\documentclass[aps,prx,twocolumn,footinbib,superscriptaddress,longbibliography]{revtex4-2}
\usepackage{graphicx}%
\usepackage{siunitx}
\usepackage{bm}
\usepackage[caption=false]{subfig}
\usepackage{mathtools}
\usepackage{multirow}
\usepackage{amssymb}
\usepackage{braket}
\usepackage{amsmath}
\usepackage{placeins}
\usepackage{cancel}
\usepackage{xr-hyper}
\usepackage{xcolor}
\usepackage{spreadtab}
\usepackage{dsfont}
\usepackage[colorlinks=true,
            linkcolor=blue,
	        citecolor=blue,
	         urlcolor=blue]{hyperref}

\newcommand\bk{{\mathbf{k}}}
\newcommand\bq{{\mathbf{q}}}
\newcommand\br{{\mathbf{r}}}
\newcommand\bR{{\mathbf{R}}}

\begin{document}

\title{In search of the electron-phonon contribution to total energy}

\author{Samuel Ponc\'e}
\email{samuel.ponce@uclouvain.be}
\affiliation{%
European Theoretical Spectroscopy Facility, Institute of Condensed Matter and Nanosciences, Université catholique de Louvain, Chemin des Étoiles 8, B-1348 Louvain-la-Neuve, Belgium. 	
}%
\affiliation{%
WEL Research Institute, avenue Pasteur, 6, 1300 Wavre, Belgium.		
}%
\author{Xavier Gonze}
\affiliation{%
European Theoretical Spectroscopy Facility, Institute of Condensed Matter and Nanosciences, Université catholique de Louvain, Chemin des Étoiles 8, B-1348 Louvain-la-Neuve, Belgium. 	
}%

\date{\today}

\begin{abstract}
The total energy is a fundamental characteristic of solids, molecules, and nanostructures.
In most first-principles calculations of the total energy, the nuclear kinetic operator is decoupled from the many-body electronic Hamiltonian and nuclear potential, and the dynamics of the nuclei is reintroduced afterward. 
This two-step procedure introduced by Born and Oppenheimer (BO) is approximate. 
Energies beyond the electronic and vibrational (or phononic) main contributions might be relevant when small energy differences are important, such as when predicting stable polymorphs or describing magnetic energy landscape.
We clarify the different flavors of BO decoupling and give an exact formulation for the total energy in the basis of BO electronic wavefunctions.
Then, we list contributions, beyond the main ones, that appear in a perturbative expansion in powers of $M_0^{-1/4}$, where $M_0$ is a typical nuclear mass, up to sixth order.
Some of these might be grouped and denoted the electron-phonon contribution to total energy, 
$E^{\textrm{elph}}$, that first appears at fourth order. 
The electronic inertial mass contributes at sixth order.
We clarify that the sum of the Allen-Heine-Cardona zero-point renormalization of eigenvalues over occupied states is not the electron-phonon contribution to the total energy but a part of the phononic contribution. 
The computation of the lowest-order 
$E^{\textrm{elph}}$ is implemented and shown to be small but non-negligible (3.8~meV per atom) in the case of diamond and its hexagonal polymorph.
We also estimate the electronic inertial mass contribution and the quasi-harmonic one. 
We confirm the size consistency of all computed terms.    
\end{abstract}

\maketitle

\section{Introduction}

The total energy is a cornerstone of modern solid-state physics and quantum chemistry. 
It governs material stability~\cite{Jones2015}, phase transition~\cite{Wentzcovitch1991}, crystal structure predictions~\cite{Payne1992}, or defect formation energy~\cite{Freysoldt2014}.
Density functional theory (DFT) is often the method of choice when computing the total energy of solids~\cite{Hohenberg1964,Kohn1965}. 
If the exact exchange-correlation functional was known, then for any system of nuclei and electrons, DFT would give the exact total
electronic energy at any chosen set of fixed nuclear positions $\{ \mathbf{R} \}$.
An open question is how to get the exact total energy (i.e. electronic plus nuclear vibrational) for the ground state of such a system.
Indeed, DFT rests on the Born-Oppenheimer (BO) decoupling of nuclear and electronic degrees of freedom~\cite{Born1927, Born1954}. 
Often, the first-principles BO nuclear contribution is added to the electronic energy.
Phononic (or vibrational) degrees of freedom are taken into account through Bose-Einstein statistics in the case of harmonic solids~\cite{Baroni2001}, or through the quasiharmonic approximation to obtain thermal expansion~\cite{Rignanese1996, Rostami2024, Rostami2025} or using effective interatomic force constants to include anharmonic (phonon-phonon) corrections~\cite{Hellman2011,Monacelli2021,Knoop2024}.
However, the BO decoupling leaves away numerous contributions to the total energy beyond the dominant electronic (static nucleus) energy and the phononic harmonic, quasi-harmonic, or anharmonic contributions. 
So, one can wonder whether electron-phonon contributions to the total energy exist, how to characterize them, and how to compute them.

As a justification for their approach, Born and Oppenheimer~\cite{Born1927} relied on a perturbation expansion, using $M_0^{-1/4}$ as a small parameter, where $M_0$ is a typical nuclear mass.
In the equilibrium BO geometry, the total energy is obtained for frozen nuclei, so it does not depend on $M_0$, while the first-order contribution vanishes.
The second-order contribution, scaling as $M_0^{-1/2}$, is the phononic contribution.
The third-order correction also vanishes.
Additional corrections come at the fourth order, with $M_0^{-1}$ scaling. 
Among these, two contributions are of the anharmonic phonon-phonon type, and can nowadays be included in first-principles calculations of total energy~\cite{Hellman2011,Monacelli2021,Knoop2024}. 

In this work, we observe that the other fourth-order BO expansion correction can be computed approximately (within DFT) on the basis of the electron-phonon matrix elements.
Thanks to recent advances in efficient calculations of electron-phonon interaction~\cite{Giustino2017}, mostly relying on density functional perturbation theory~\cite{Baroni1987,Gonze1997,Gonze1997b,Baroni2001,Gonze2005a,Gonze2024a}, electron-phonon matrix elements are readily available when computing electronic transport~\cite{Ponce2020,Lihm2026} or thermal transport~\cite{Allen1993,Lindsay2016} properties, superconductivity~\cite{Margine2013}, polaron formation~\cite{Sio2019,Vasilchenko2025}, or the temperature-dependence of the band gap and its zero-point renormalization~\cite{Ponce2015,Ponce2025a}.  
Several terms of even higher order in the small parameter $M_0^{-1/4}$ fall into that category. 
All could be grouped as ``electron-phonon contributions to the total energy''.
It is expected that such contributions are small compared to electronic and phononic contributions, but they might be crucial for the study of defects and surfaces~\cite{Freysoldt2018}, magnetic configurations~\cite{Cerny2003}, chemical reactions and bonding~\cite{Xu2020}, or accurate phase transition~\cite{Bouchet2019} where small energy differences matter. 

Given the recent advances in the computation of electron-phonon matrix elements, several recent works also discussed electron-phonon contributions to total energy.
In 2020, P.~B.~Allen~\cite{Allen2020} suggested that this contribution could be calculated using ingredients available when electron-phonon band-gap renormalization calculations are performed. 
His approach is based on the summation over occupied states of partial Fan-Migdal and Debye-Waller self-energies~\cite{Allen1976}. 
Recently, S.~Paul \textit{et al.}~\cite{Paul2023} have performed first-principles calculations based on this idea. 
They argued that such formulation violates size consistency, in which the total energy would depend on the size of the unit cell. 
In the present work, we show that the formulation of Allen is well defined and size-consistent.
Crucially, we show that the approach proposed by Allen does not deliver the fourth-order $M_0^{-1}$ electron-phonon contribution term but is already part of the second-order phononic $M_0^{-1/2}$ contribution to the total energy and should not be added to total energy calculations.   

We first present in Section~\ref{sec:BOapproximations} the BO Hamiltonian, and approximations based on BO wavefunctions.
We also derive an expression for the exact total energy that includes contributions beyond the BO wavefunctions. 
We then perform a detailed mass-scaling analysis of all the terms entering the total energy in Section~\ref{sec:scaling}.
From this mass scaling analysis, we identify which terms are phononic contributions, Section~\ref{sec:phonon}, while the remaining ones are detailed in Section~\ref{sec:remaining}.
In particular, we report, for the first time explicitly, the lowest-order electron-phonon contribution to the total energy. 
In Sec.~\ref{sec:normal_modes} of the Supplemental Information (SI), the BO perturbation expansion is derived and expressed in normal mode coordinates. 
We then show in Section~\ref{sec:ahc} the extension to solids with practical formulae for implementation in first-principles software.
In Section~\ref{sec:Allenelph}, we show that Allen's energy is not the electron-phonon contribution to the total energy but a part of the phononic contribution to the total energy.
We then recast in Section~\ref{sec:connection} Allen's formula into the standard Allen-Heine-Cardona (AHC)~\cite{Allen1976,Allen1981} theory for band renormalization thanks to an identically null term. 
In Section~\ref{sec:validation}, we present a detailed analysis of the energy terms at the phonon zone-center, comparing a frozen-phonon approach and perturbation theory. 
We find excellent agreement for all terms, below 0.5\% differences. 
We show size consistency by computing the diamond band renormalization using the \textsc{Abinit} code~\cite{Gonze2016,Gonze2020,Romero2020,Verstraete2025} and an interpolation of the perturbed potential which includes dipole and quadrupole contributions~\cite{Brunin2020,Brunin2020a}.
We show that the same results are obtained between the primitive and a supercell of diamond when no crystal symmetries are used.

Finally, in Section~\ref{sec:comparison} we compute the (fourth order) electron-phonon contribution to the total energy and approximate the effect of electron mass rescaling (sixth order), for two allotropes of carbon: the stable face-centered cubic diamond and the high-pressure metastable hexagonal diamond called lonsdaleite which is typically found in meteorites. 
For both materials, we report their converged total energy with momentum integration. 
We compute the electron-phonon forces and the new equilibrium lattice parameters of diamond and the internal parameter of lonsdaleite due to quasi-harmonic effects. 
At 0~K, we find that diamond is 52.7~meV/(2 atoms) more stable than lonsdaleite. 

In addition to clarifying Allen's work, the main findings of this manuscript are (i) an equation fulfilled by the exact total energy, Eq.~\eqref{eq:exact_Schreq_2}, (ii) the derivation of all terms in a perturbation expansion series in terms of M$^{-1/4}$ up to order 6 included, (iii) the identification of terms that do not come from the Born-Oppenheimer anharmonicity, including those coming from the vector potential, the electron-phonon coupling, mixed terms, and the renormalization of nuclei mass due to electronic inertial effects, and (iv) we derived a practical formula for the lowest-order electron-phonon contribution to the total energy, implement it and applied to study diamond and lonsdaleite.

\section{Born-Oppenheimer-based approximations}
\label{sec:BOapproximations}

We first consider finite systems and extend the theory to periodic crystals in Sec.~\ref{sec:ahc}. 
The total Hamiltonian of a system composed of electrons and nuclei can be expressed as~\cite{Born1927,Born1996}
\begin{align}\label{eq:totalH}
\hat{\mathcal{H}} \{ \bR \}  \equiv & \hat{T}^\textrm{N} + \hat{H}^{\rm BO}\{\bR \}, \\
\hat{T}^\textrm{N} \equiv& \sum_{\kappa\alpha}\frac{\hat{P}^2_{\kappa\alpha}}{2M_{\kappa}}= -\frac{1}{2}\sum_{\kappa\alpha}\frac{1}{M_{\kappa}}\frac{\partial^2}{\partial R_{\kappa\alpha}^2}, \label{eq:kineticN}
\end{align}
where $\{\bR \}$ denotes the set of nuclear positions, 
$\hat{T}^\textrm{N}$ is the nuclear kinetic operator,
$\hat{P}_{\kappa\alpha}\equiv-i\partial / \partial R_{\kappa\alpha}$ is the canonical momentum operator for nuclei, with $M_{\kappa}$ the mass of nucleus $\kappa$ and $\alpha$ an index for cartesian axes. 
Electronic coordinates are implicit, unlike the nuclear ones. 
The global Hamiltonian in Eq.~\eqref{eq:totalH} is written in calligraphic style as well as the forthcoming associated total energies, which are independent of $\{\bR \}$.
Hartree atomic units are used throughout.

The Born-Oppenheimer (BO) Hamiltonian, $\hat{H}^{\rm BO}\{\bR \}$, is defined as:
\begin{align}
\hat{H}^{\rm BO}\{\bR \} \equiv & \hat{H}^{\rm e}\{\bR \} + V^{\rm NN}\{ \bR \} \label{eq:BOHamiltonian} \\
\hat{H}^{\rm e}\{\bR \}  \equiv & \hat{T}^{\rm e} + \hat{V}^{\rm ee} + \hat{V}^{\rm eN}\{ \bR \},  
\label{eq:ElectronicHamiltonian}
\end{align}  
where $\hat{H}^{\rm e}\{\bR \}$ is the electronic Hamiltonian, $V^{\rm NN}\{\bR \}$ the bare nucleus-nucleus interaction, $\hat{T}^{\rm e}$ the electronic kinetic operator, $\hat{V}^{\rm ee}$ the electron-electron operator, and $\hat{V}^{\rm eN}\{\bR \}$ the electron-nucleus operator. 
The Schr\"odinger equation associated to the total Hamiltonian is 
\begin{align}\label{eq:Schreq}
\hat{\mathcal{H}} \{ \bR \} \ket{ \psi_i \{ \bR \} } =&  \mathcal{E}_i \ket{\psi_i \{\bR\}}, 
\end{align}
with ground state labeled $i=0$.
The Schr\"odinger equation for the BO Hamiltonian is
\begin{align}\label{eq:BO_Schreq}
\hat{H}^{\rm BO} \{ \bR \} \ket{ \phi_j \{ \bR \} } =&  E^{\rm BO}_j\{\bR\} \ket{ \phi_j \{\bR\}}, 
\end{align}
with ground state for each configuration $\{\bR\}$ labeled $j=0$. 
The wavefunction $\psi_i$ is normalized when integrated over electronic $\mathbf{r}$ and nuclear $\mathbf{R}$ coordinates, while the BO wavefunction $\phi_j$ is normalized for each nuclear configuration $\{\bR\}$ when integrated over electronic coordinates only. 
The Dirac bracket notation is used for the electronic degrees of freedom.

The denominations \emph{``Born-Oppenheimer Hamiltonian''}, \emph{``Born-Oppenheimer wavefunction''} and \emph{``Born-Oppenheimer eigenenergy''}
for quantities in Eqs.~\eqref{eq:BOHamiltonian} and \eqref{eq:BO_Schreq} are well-established and consistent in the literature, but there is some confusion
about the meaning of \emph{``Born-Oppenheimer approximation''} and \emph{``adiabatic approximation''}~\cite{Ballhausen1972, Nafie1983, Hagedorn1986, Tully2000, Scherrer2017, Cohen2025}.
In many publications, these are used interchangeably while in other they have different meanings. 
For example, in their 1972 publication, Ballhausen and Hansen~\cite{Ballhausen1972} discuss various adiabatic approximations including \emph{``Born-Oppenheimer''}, \emph{``Born-Huang''}, \emph{``Herzberg-Teller''}, and \emph{``crude''}, based on wavefunctions of the type 
\begin{equation}\label{eq:adiabwf}
\ket{\psi_i \{\bR\}}=X_{i} \{\bR\}\ket{ \zeta_i\{\bR\}},
\end{equation}
that is, a simple product of a configuration-dependent (normalized) function of the nuclear configuration only $X_{i}\{\bR\}$, with a single normalized electronic wavefunction, also possibly configuration-dependent. 
The electronic wavefunctions in these schemes change smoothly with the configuration.
However, beyond the different \emph{``adiabatic approximations''} used by Ballhausen and Hansen, the recent
\emph{``exact factorization''}~\cite{Abedi2010,Scherrer2017,Cohen2025} approach also has the form of Eq.~\eqref{eq:adiabwf} but is exact in that case.
Therefore, it is not helpful to define the \emph{``adiabatic approximation"} as being based on Eq.~\eqref{eq:adiabwf}, which is too general.

Instead, if one considers that $\ket{ \zeta_{j} \{\bR\}}$ is a BO wavefunction $\ket{ \phi_j \{\bR\}}$, a meaningful set of approximations follows.
We start from the exact relation:
\begin{equation}\label{eq:wfexact}
\ket{\psi_i \{\bR\}}=\sum_j \chi_{ij} \{\bR\}\ket{ \phi_j\{\bR\}},
\end{equation}
where the wavefunction $\chi_{ij}\{\bR\}$ defines the nuclear dynamics.
This decomposition can always be done, since for every configuration $\{\bR\}$, the set of $\ket{ \phi_j\{\bR\}}$ forms a complete orthonormalized basis.
Normalization of $\ket{\psi_i \{\bR\}}$, combined with the electronic normalization of $\ket{ \zeta_j \{\bR\}}$ implies that $\chi_{ij}\{\bR\}$ is normalized with respect to integration over $\{\bR\}$ configurations:
\begin{equation}\label{eq:chi_normalization}
\langle\chi_{ij}\{\bR\} | \chi_{ij}\{\bR\}\rangle_{\bR} \equiv \int \textrm{d}\{\bR\} \chi_{ij}^*\{\bR\} \chi_{ij}\{\bR\} = 1,
\end{equation}
where we introduce the notation $\langle .| .\rangle_{\mathbf{R}}$ to indicate that we evaluate the scalar product by integrating functions on all nuclear configurations $\{\bR\}$.
Scalar products without such a subscript are implicitly only for electronic degrees of freedom.

\begin{figure}[ht]
  \centering
\includegraphics[width=0.99\linewidth]{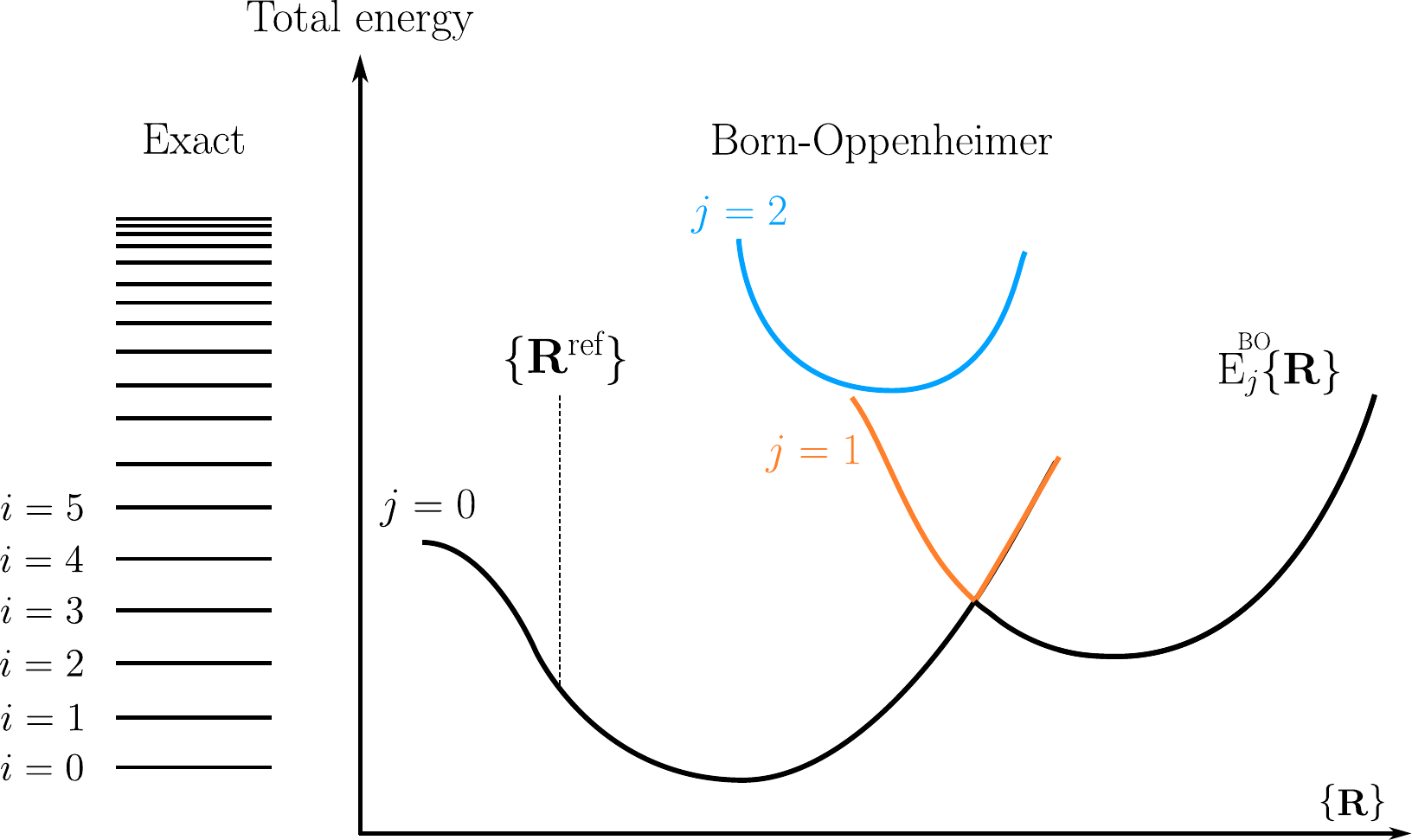}
  \caption{\label{fig:hypersurface}
Example of coordinate diagram for the BO energy hypersurfaces $E^{\rm BO}_j\{\bR\}$
of the electronic ground state ($j=0$) and first- and second-excited states ($j=1$ and $j=2$).
Eigenenergies of the total Hamiltonian are also shown, labeled with the $i$ index. 
Unlike their wavefunctions, or the BO eigenenergies, they do not depend on the configuration of nuclei.
The ground state of the total Hamiltonian has $i=0$ label.
Usually, the corresponding wavefunctions will be made predominantly of the BO
wavefunctions of one of the BO hypersurfaces, but there could be hybridized states.
The chosen reference configuration is denoted by $\{ \mathbf{R}^{\rm ref}\}$. 
In this example, we present a specific case where the $j=0$ BO hypersurface has two minima, a global one and a local one, that might both be chosen to be the reference state $\{ \mathbf{R}_{j=0}^0\}$ in Sec.~\ref{sec:scaling}.
The $j=2$ BO hypersurface has one global minimum.
}
\end{figure}

We choose a reference configuration $\{\bR^{\rm ref} \}$ and the associated reference BO wavefunction $\ket{ \phi_j \{\bR^{\rm ref}\}}$.
We impose unicity and that $\ket{ \phi_j \{\bR\}}$ changes smoothly as a function of $\{\bR\}$ around $\{\bR^{\rm ref}\}$. 
We note that such an imposition might be difficult when there are degeneracies around $\{\bR^{\rm ref} \}$ or when the BO wavefunction changes abruptly. 
Moreover, the arbitrary phase of the BO wavefunction should be taken consistently to reach smoothness.
We also use the index $j$ to label this set of BO wavefunctions, even beyond the neighborhood where the smoothness is established. 
We will assume in Sec.~\ref{sec:scaling} that
$\{\bR^{\rm ref} \}$ is the configuration of a (local or global) minimum of a BO hypersurface, denoted by $\{\bR_{j}^0 \}$, but such restriction is not yet needed.
For illustrative purposes, we present in Fig.~\ref{fig:hypersurface} a specific case of a 
one-dimensional coordinate diagram with a ground-state BO hypersurface and two BO excited surfaces.

In the literature, there are several possible approximations to Eq.~\eqref{eq:wfexact} in which only one term is kept.  
A first possible approximation consists in considering $E^{\rm BO}_j\{\bR\}$ as an effective scalar potential for nuclear dynamics:
\begin{align}\label{eq:sca_wf}
\ket{\psi_i \{\bR\}} \approx &  \chi_{ij}^{\rm sca} \{\bR\}\ket{ \phi_j\{\bR\}} \\
\label{eq:sca_Schreq}
\big(\hat{T}^\textrm{N} + E^{\rm BO}_j\{\bR\} \big) \chi_{ij}^{\rm sca} \{ \bR \} =& \mathcal{E}_{ij}^{\rm sca} \chi_{ij}^{\rm sca}\{\bR\}.
\end{align}
We refer to this approximation as the \emph{``scalar potential''} approximation.

A second possible approximation, still using one term in Eq.~\eqref{eq:wfexact}, which we refer to as the \emph{``optimal approximation''}, is based on choosing 
$\chi_{ij} \{\bR\}$ such that the expectation value of the Hamiltonian, Eq.~\eqref{eq:totalH}, is stationary (minimal in the ground-state case).
In that case, the nuclear wave function fulfills the following equation:  
\begin{multline}\label{eq:optimal_Schreq_1}
\Big(\hat{T}^\textrm{N} + E^{\rm BO}_j\{\bR\} + \sum_{\kappa\alpha}\frac{1}{2M_{\kappa}} 
\Big[B_{\kappa\alpha j}\{\bR\} + 2A_{\kappa\alpha j}\{\bR\}\hat{P}_{\kappa\alpha} \Big]\Big)   \\
\times \chi_{ij}^{\rm opt} \{ \bR \} = {\cal{E}}_{ij}^{\rm opt} \chi_{ij}^{\rm opt} \{\bR\},
\end{multline}
with
\begin{align}\label{eq:A}
A_{\kappa\alpha j}\{\bR\} \equiv & \bigg\langle \phi_j \{\bR\}  \bigg| -i \frac{\partial \phi_j}{\partial R_{\kappa\alpha}}_{|\bR} \bigg\rangle \\
B_{\kappa\alpha j}\{\bR\} \equiv & -\frac{1}{2} \bigg\langle  \phi_j \{\bR\}  \bigg| \frac{\partial^2 \phi_j}{\partial R^2_{\kappa\alpha}}_{|\bR} \bigg\rangle,
\label{eq:B}
\end{align}
where the notation $X_{|\mathbf{R}}$ indicates that we take the derivative at the nuclear configuration $\{\bR\}$.
We also note that the presence of the terms $A_{\kappa\alpha j}\{\bR\}$ and $B_{\kappa\alpha j}\{\bR\}$ was observed by Born in 1951 (see Appendix VIII of Ref.~\onlinecite{Born1954}). 
However, $A_{\kappa\alpha j}\{\bR\}$ was then long ignored because it disappears if the BO wavefunctions $\ket{ \phi_j \{\bR\}}$ can be chosen to be real. 
Additionally, even if the BO wavefunctions are not real, there might be systems for which a choice of phase (gauge choice) in the relevant smoothness region yields vanishing $A_{\kappa\alpha j}\{\bR\}$. 
This is usually possible for finite systems, except in certain cases. 
The first such case is related to the so-called \emph{conical intersections} detailed in 1979 by Mead and Truhlar~\cite{Mead1979}.
A conical intersection can appear if the reference BO hypersurface $E^{\rm BO}_j\{\bR\}$ is not separated from the other BO hypersurfaces.
However, if there is such a separation in the relevant region in the neighborhood of $\{\bR^{\textrm{ref}}\}$, this problem can be neglected.
A second exception arises if the Hamiltonian breaks the time-reversal symmetry. 
In this case, $A_{\kappa\alpha j}\{\bR\}$ will generally not vanish in a finite region around $\{\bR^{\textrm{ref}}\}$,
although it can made to vanish for the chosen reference configuration.
There has recently been interest in the presence of such a term, inducing chirality and/or angular momentum of phonons~\cite{Qin2012,Saito2019, Bistoni2021,Saparov2022,Bonini2023,Ren2024, Royo2025}.
The $A_{\kappa\alpha j}\{\bR\}$ term is actually a Berry connection.

If $A_{\kappa\alpha j}\{\bR\}$ does not disappear, it is nevertheless possible to reformulate Eq.~\eqref{eq:optimal_Schreq_1} such that $A_{\kappa\alpha j}\{\bR\}$ plays the role of a vector potential that modifies the nuclear momenta, giving
\begin{equation}\label{eq:optimal_Schreq_2}
\Big(\hat{T}^\textrm{vec}_{j}\{\bR\} + E^{\rm opt}_j\{\bR\}\Big) \chi_{ij}^{\rm opt} \{ \bR \}
= {\cal{E}}_{ij}^{\rm opt} \chi_{ij}^{\rm opt} \{\bR\},
\end{equation}
where $\hat{T}^\textrm{N}$ is replaced by
\begin{equation}\label{eq:Tvec}
\hat{T}^\textrm{vec}_{j}\{\bR\} \equiv \sum_{\kappa\alpha} \frac{\Big(\hat{P}_{\kappa\alpha} + A_{\kappa\alpha j}\{\bR\}\Big)^2}{2M_{\kappa}},
\end{equation}
and $E^{\rm BO}_j\{\bR\}$ is replaced by an ``optimal" potential,
\begin{equation}\label{eq:E_opt} 
E^{\rm opt}_j\{\bR\}\equiv 
E^{\rm BO}_j\{\bR\}+ E^{\rm Born}_j\{\bR\}.
\end{equation}
The Born contribution $E^{\rm Born}_j\{\bR\}$ is defined as
\begin{equation}\label{eq:Born}
E^{\rm Born}_j\{\bR\} \equiv \sum_{\kappa\alpha}\frac{1}{M_{\kappa}} B_{\perp j\kappa\alpha}\{\bR\}, 
\end{equation}
with
\begin{equation}\label{eq:Bperp}
B_{\perp j\kappa\alpha}\{\bR\} \equiv
\frac{1}{2} \bigg\langle \frac{\partial \phi_j}{\partial R_{\kappa\alpha}}_{|\bR} \bigg| \hat{\cal{P}}^{\textrm{BO}}_{\perp j} \bigg|
\frac{\partial \phi_j}{\partial R_{\kappa\alpha}}_{|\bR} \bigg\rangle, 
\end{equation}
where the projector on the $j$-BO manifold is defined by its matrix elements for two electronic and nuclear wavefunctions as follows, 
\begin{multline}\label{eq:P_jBO_matrix_elm}
\bra{ \psi' \{\bR'\}} \hat{\cal{P}}^{\textrm{BO}}_j 
\ket{ \psi \{\bR\}} \equiv \int \mathrm{d}\{\bR''\} 
\langle \psi' \{\bR'\} \ket{ \phi_j \{\bR''\}} \\
\times \bra{ \phi_j \{\bR''\}} \psi \{\bR\} \rangle,
\end{multline}
or, in a more compact notation,
\begin{equation}\label{eq:P_jBO}
\hat{\cal{P}}^{\textrm{BO}}_j \equiv \int \mathrm{d}\{\bR\} \ket{ \phi_j \{\bR\}} \bra{ \phi_j \{\bR\}}, 
\end{equation}
while its orthogonal is
\begin{equation}
\hat{\cal{P}}^{\textrm{BO}}_{\perp j} \equiv \,\, \hat{1}-\hat{\cal{P}}^{\textrm{BO}}_{j}. \label{eq:PperpBO}
\end{equation}
The presence of the projector in Eq.~\eqref{eq:Bperp} removes the phase arbitrariness that is present in Eq.~\eqref{eq:B}.
In the literature dealing with phonon chirality~\cite{Bonini2023}, a simplified version of Eq.~\eqref{eq:optimal_Schreq_2} is often encountered
where the vector potential is considered but not the Born term, see Sec.~\ref{sec:hybrid} of SI.

Both approximations, Eqs.~\eqref{eq:sca_Schreq} and \eqref{eq:optimal_Schreq_2}, to the exact Schrödinger equation, Eq.~\eqref{eq:totalH}, have been referred to as \emph{``adiabatic approximation''} and/or as \emph{``Born-Oppenheimer approximation''} in the literature, with the following approximate total wavefunctions:
\begin{subequations}
\begin{align}
   \ket{\psi_i^{\rm sca} \{\bR\}} =& \chi_{ij}^{\rm sca} \{\bR\}\ket{ \phi_j\{\bR\}} 
   \label{eq:psi_sca}\\ 
   \ket{\psi_i^{\rm opt} \{\bR\}} =& \chi_{ij}^{\rm opt} \{\bR\}\ket{ \phi_j\{\bR\}}. 
   \label{eq:optwf} 
\end{align}
\end{subequations}
The introduction of the reference set of BO wavefunctions, and its use in Eqs.~\eqref{eq:wfexact}, \eqref{eq:psi_sca}, and \eqref{eq:optwf},
already allows one to obtain an exact representation of the total wavefunction, as well as to introduce different adiabatic approximations, but even yields a nuclear wavefunction equation that must be fulfilled by the exact eigenenergy.

For this purpose, we come back to the exact wavefunction $\ket{\psi_i \{\bR\}}$. 
The Hamiltonian Eq.~\eqref{eq:totalH} can be block-decomposed (2$\times$2 decomposition) in the subspace spanned by the set of all $\ket{ \phi_j \{\bR\}}$ for a chosen $j$ value (but varying nuclear configurations) and its orthogonal. 
The first is left invariant upon application of $\hat{\mathcal{P}}^{\textrm{BO}}_j$ and the second is left invariant upon application of $\hat{\mathcal{P}}^{\textrm{BO}}_{\perp j}$. 
This corresponds to the exact decomposition of the wavefunction,
\begin{equation}\label{eq:WFdecomposition}
\ket{\psi_i \{\bR\}}=\chi_{ij} \{\bR\}\ket{ \phi_j\{\bR\}}+\ket{{\cal{F}}_i\{\bR\}},
\end{equation}
where $\ket{{\cal{F}}_i\{\bR\}}$ fulfills the constraint
\begin{equation}\label{eq:conditiononF}
\hat{\cal{P}}^{\textrm{BO}}_{ j} \ket{{\cal{F}}_i\{\bR\}}=\ket{0}.
\end{equation}
The left-hand side of the Schrödinger equation, Eq.~\eqref{eq:BO_Schreq}, can be moved to the right-hand side and the Hamiltonian can be replaced by its expression in Eq.~\eqref{eq:totalH}.
The BO energy and Hamiltonian being diagonal in this decomposition, the block decomposition delivers two equations:
\begin{widetext}
\begin{align}
\Bigg\{
\begin{matrix}
\hat{\cal{P}}^{\textrm{BO}}_{j}(\hat{T}^\textrm{N}+E^{\rm BO}_j\{\bR\} - \mathcal{E}_i)
    \chi_{ij} \{\bR\}\ket{ \phi_j\{\bR\}}
& + &
\hat{\cal{P}}^{\textrm{BO}}_{j}\hat{T}^\textrm{N}\ket{{\cal{F}}_i\{\bR\}}&=\ket{0}, \\ 
\hat{\cal{P}}^{\textrm{BO}}_{\perp j}\hat{T}^\textrm{N}\chi_{ij} \{\bR\}\ket{ \phi_j\{\bR\}}
& + &
\hat{\cal{P}}^{\textrm{BO}}_{\perp j}(\hat{T}^\textrm{N}+\hat{H}^{\rm BO}\{\bR \}- \mathcal{E}_i)
\hat{\cal{P}}^{\textrm{BO}}_{\perp j}\ket{{\cal{F}}_i\{\bR\}}
& = \ket{0}. 
\end{matrix}
\label{eq:block_decomposition}
\end{align}
\end{widetext}

Let us first consider the second equation of this system. 
The action of $\hat{\cal{P}}^{\textrm{BO}}_{\perp j}\hat{T}^\textrm{N}$ on a product of a nuclear wavefunction and a BO wavefunction can be rewritten in terms of a non-hermitian operator, $\hat{\cal{T}}^\textrm{N}\{\bR\}$, acting on the nuclear wavefunction only, which delivers its result in the full nuclear and electronic space:
\begin{equation}
\hat{\cal{P}}^{\textrm{BO}}_{\perp j}\hat{T}^\textrm{N}
\chi_{ij} \{\bR\}\ket{ \phi_j\{\bR\}} = \hat{\cal{T}}^\textrm{N}_j\{\bR\} \chi_{ij} \{\bR\},
\end{equation}
where
\begin{multline}\label{eq:TN_R}
\hat{\cal{T}}^\textrm{N}_j\{\bR\} \equiv \\
\sum_{\kappa\alpha}\frac{\hat{\cal{P}}^{\textrm{BO}}_{\perp j}}{2M_{\kappa}} \Bigg[ -\bigg| \frac{\partial^2 \phi_j}{\partial R_{\kappa\alpha}^2}_{|\bR} \bigg\rangle -2i \bigg| \frac{\partial \phi_j}{\partial R_{\kappa\alpha}}_{|\bR} \bigg\rangle \hat{P}_{\kappa\alpha} \Bigg].
\end{multline}
To treat the second term of the second equation in Eq.~\eqref{eq:block_decomposition}, one introduces the Green's function of the full Hamiltonian, that is a function of a complex variable $z$,
\begin{equation}\label{eq:Green}
\hat{\mathcal{G}}(z,\{\bR\})\equiv \Big(z-\hat{\mathcal{H}} \{ \bR \}\Big)^{-1}.
\end{equation}
Its restriction in the space orthogonal to the BO-wavefunctions is defined as
\begin{equation}\label{eq:Green_perp}
\hat{\mathcal{G}}_{\perp j}(z,\{\bR\}) \equiv  \hat{\cal{P}}^{\textrm{BO}}_{\perp j}
\Big(z-\hat{\mathcal{H}} \{ \bR \}\Big)^{-1} \hat{\cal{P}}^{\textrm{BO}}_{\perp j},
\end{equation}
with the term between parentheses defining a pseudo-inverse, limited to that subspace.
Although Green's function might not be defined for specific real $z$ values above the ground-state energy $z=\mathcal{E}_0$,
the restriction to the space orthogonal to the BO-wavefunctions removes this problem up to $z$ being equal to the lowest excited-state BO eigenenergy.
We also assume that $z=\mathcal{E}_i$ is a value for which the projected Green's function $\hat{\mathcal{G}}_{\perp j}(z,\{\bR\})$ exists.

Finally, we single out $\ket{{\cal{F}}_i\{\bR\}}$ in the second line of Eq.~\eqref{eq:block_decomposition} by multiplying by $\hat{\mathcal{G}}_{\perp j}(\mathcal{E}_i,\{\bR\})$
and inject it in the first line of Eq.~\eqref{eq:block_decomposition}.
After some algebraic manipulations, similar to those needed to go from Eq.~\eqref{eq:Schreq} to Eq.~\eqref{eq:optimal_Schreq_2},
one deduces
\begin{multline}\label{eq:exact_Schreq_2}
\Big(\hat{T}^\textrm{vec}_{j}\{\bR\} + E^{\rm opt}_j\{\bR\} \\ 
+ \hat{\Sigma}_j(\mathcal{E}_i,\{\bR\})\Big) \chi_{ij} \{ \bR \} = \mathcal{E}_i \chi_{ij} \{\bR\},
\end{multline}
where
\begin{equation}\label{eq:self-energy}
\hat{\Sigma}_j(z,\{\bR\}) \equiv \hat{\cal{T}}^{\textrm{N},\dagger}_j\{\bR\} \hat{\mathcal{G}}_{\perp j}(z,\{\bR\}) \hat{\cal{T}}^\textrm{N}_j\{\bR\}
\end{equation}
is an operator similar to self-energy operators found in the context of many-body perturbation theory~\cite{Mahan1990}.

Importantly, Eq.~\eqref{eq:exact_Schreq_2} is an exact version of the approximate Eqs.~\eqref{eq:sca_Schreq} and \eqref{eq:optimal_Schreq_2}, and focuses only on nuclear dynamics where all electronic degrees of freedom are condensed in the self-energy operator. 
It is a key result of this work. 

The Green's function of the full Hamiltonian fulfills a Dyson-type equation:
\begin{equation} \label{eq:Dyson}
\hat{\mathcal{G}}_{\perp j}(z,\{\bR\}) \!=\! \hat{G}_{\perp j}^{\rm BO}(z,\{\bR\}) \Big(\hat{1} \!+\! \hat{T}^\textrm{N}\hat{\mathcal{G}}_{\perp j}(z,\{\bR\})\Big),
\end{equation}
where $\hat{G}_{\perp j}^{\rm BO}(z,\{\bR\})$ is the BO Green's function defined as:
\begin{equation}\label{eq:GreenBO_perp}
\hat{G}_{\perp j}^{\rm BO}(z,\{\bR\}) \equiv  \hat{\cal{P}}^{\textrm{BO}}_{\perp j}
\Big(z-\hat{H}^{\rm BO}\{\bR \}\Big)^{-1}\hat{\cal{P}}^{\textrm{BO}}_{\perp j}.
\end{equation}

Inserting Eq.~\eqref{eq:Dyson} into Eq.~\eqref{eq:self-energy} can yield further approximate treatments. 
For example, motivated by the inverse mass of nuclei that is present in $\hat{T}^\textrm{N}$, one might neglect $\hat{T}^\textrm{N}
\hat{\mathcal{G}}_{\perp j}(z,\{\bR\})$ with respect to $\hat{1}$ 
in Eq.~\eqref{eq:Dyson}, 
and set
\begin{equation} \label{eq:Greenperp_approx_GreenBO}
\hat{\mathcal{G}}_{\perp j}(z,\{\bR\})  \approx \hat{G}_{\perp j}^{\rm BO}(z,\{\bR\}).
\end{equation}
We will study this approximation in Sec.~\ref{sec:scaling} on the basis of its mass-scaling behavior.

As a final comment in this section, let us now focus on the ground state. 
The variational freedom of the wavefunction Eq.~\eqref{eq:WFdecomposition} is complete, producing the lowest possible value of the energy of the Hamiltonian Eq.~\eqref{eq:totalH}. 
So, restricting the trial wavefunction to the BO-wavefunction subspace, which corresponds to the solution of Eq.~\eqref{eq:optimal_Schreq_2}, or correspondingly to the neglect of the self-energy in Eq.~\eqref{eq:exact_Schreq_2}, must come with an increase of the energy such that $\mathcal{E}_0 \leq \mathcal{E}_0^{\textrm{opt}}$.

\section{Mass-scaling analysis}\label{sec:scaling}
In the original BO work~\cite{Born1927}, it is shown how to formulate a perturbation expansion around a minimum (global or local) of a chosen BO energy hypersurface with index $j$, taken as reference configuration, when the nuclei masses are much larger than the electronic masses. 
The small parameter of this expansion is here denoted $\lambda$.
Taking $M_0$ as a characteristic nuclear mass, expressed in units of electronic mass, one defines relative nuclear masses
\begin{equation}\label{eq:m_kappa}
m_\kappa \equiv M_{\kappa}/M_0.
\end{equation}
Then, the masses become $\lambda$-dependent and are scaled together,
\begin{equation}\label{eq:M_kappa_lambda}
M_{\kappa}(\lambda)=m_\kappa \lambda^{-4}.
\end{equation}
For the value 
$\lambda = \lambda_0 \equiv M_0^{-1/4}$ the physical masses are recovered.
We note that $M_0^{-1/4}$ is small and at most 0.153 in the case of the hydrogen nucleus.
The choice of Eq.~\eqref{eq:M_kappa_lambda} will be explained later by the observation that the average fluctuations of nuclear displacements with respect to a reference configuration chosen to be a local BO energy minimum $\{\bR_{j}^0 \}$, explicitly
\begin{equation}
\Delta R_{\kappa\alpha j} = R_{\kappa\alpha} - R_{\kappa\alpha j}^0,
\end{equation}
are proportional to $\lambda$ at the lowest order.
Similarly, the average momenta are inversely proportional to $\lambda$.
Also, when $\lambda$ tends to zero, all masses become infinite so that there is no nuclear dynamics.

The Schrödinger equation, Eq.~\eqref{eq:Schreq}, becomes
\begin{equation}\label{eq:Schreq_lambda}
\hat{\mathcal{H}}_\lambda \{ \bR \} \ket{ \psi_{i\lambda} \{ \bR \} } =  \mathcal{E}_{i\lambda} \ket{\psi_{i\lambda} \{\bR\}},
\end{equation}
with a $\lambda$-dependent Hamiltonian replacing Eqs.~\eqref{eq:totalH}-\eqref{eq:kineticN},
\begin{equation}\label{eq:hamiltonian_with_lambda}
\hat{\mathcal{H}}_{\lambda}\{ \bR \} = \sum_{\kappa\alpha} \frac{\lambda^4\hat{P}^2_{\kappa\alpha}}{2m_\kappa} + \hat{H}^{\rm BO}\{ \bR\}.
\end{equation}

Quantities that depend on $\lambda$ (wavefunctions, Hamiltonian, total energy, and Green's function) can be expanded in a Taylor series in $\lambda$ as follows, generically: 
\begin{subequations}
\begin{align}\label{eq:XRl_Taylor}
X(\lambda) =& X^{(0)} + \lambda X^{(1)} + \lambda^2 X^{(2)} + \mathcal{O}( \lambda^3), \\
X^{(n)} =& \frac{1}{n!} \frac{ \partial^n X}{\partial \lambda^n}_{|\lambda=0}. \label{eq:coeff_l_Taylor}
\end{align} 
\end{subequations}
For $\lambda=\lambda_0$, the scaled Hamiltonian is equal to the original Hamiltonian, so that the series 
\begin{align}\label{eq:ETaylor_lambda}
\mathcal{E}_{i\lambda_0} = \mathcal{E}_{i}^{(0)} + \lambda_0 \mathcal{E}_{i}^{(1)} + (\lambda_0)^2 \mathcal{E}_{i}^{(2)} + {\cal{O}}(\lambda_0^3)
\end{align} 
delivers the exact total energy.

Quantities that depend on the nuclei coordinates (the wavefunctions, different potentials, Hamiltonians, and Green's functions) can also be expanded in another series, now considering atomic displacements, generically:
\begin{multline}\label{eq:XR_Taylor}
X\{\bR\} = X_{j|\mathbf{0}} + \sum_{\kappa\alpha} \frac{\partial X}{\partial R_{\kappa\alpha }}_{j|\mathbf{0}}\Delta R_{\kappa\alpha j} \\
+ \frac{1}{2} \sum_{\substack{\kappa\alpha \\ \kappa'\alpha'}} \frac{\partial^2 X}{\partial R_{\kappa\alpha}\partial R_{\kappa'\alpha'}}_{j|\mathbf{0}} \!\! \Delta R_{\kappa\alpha j} \Delta R_{\kappa'\alpha' j} \!+\! \mathcal{O} \big( \Delta \bR^3_j\big),
\end{multline} 
where the $j|\mathbf{0}$ subscript of $X_{j|\mathbf{0}}$ indicates that the quantity $X$ is evaluated at $\{\bR\}=\{\bR_{j}^0 \}$.
For conciseness, we also introduce a compact notation for the terms of such Taylor series:
\begin{subequations}
\begin{align}\label{eq:XR_Taylor_concise}
X\{\bR\} =& X_{j|\mathbf{0}} +X^{\{\mathbf{1}\}} \cdot  \Delta \bR_j + X^{\{\mathbf{2}\}}  \cdot \Delta \bR_j^2 \nonumber \\
& + \mathcal{O}\big( \Delta \bR^3_j\big), 
\\
X^{\{ \mathbf{n} \}} \equiv & \frac{1}{n!} \frac{\partial^n X}{\partial R_{\kappa_{s_1}\alpha_{s_1}}\dots \partial R_{\kappa_{s_n}\alpha_{s_n}}}_{j|\mathbf{0}}, 
\\
\Delta \bR_{j}^n \equiv & \prod_{s=1}^{n} \Delta R_{\kappa_s\alpha_s j}, 
\end{align} 
\end{subequations}
where the dot product in Eq.~\eqref{eq:XR_Taylor_concise} is defined as $\cdot \equiv \sum_{\kappa_1 \alpha_1} \dots \sum_{\kappa_n \alpha_n}$.
Curly brackets are used as superscripts in this notation, instead of the superscript parentheses of Eq.~\eqref{eq:XRl_Taylor}.
Being at the BO equilibrium positions, the first-order derivatives of the BO energy vanish, so that
\begin{equation}\label{eq:EBO_0R_lambda}
E_j^{\rm BO}\{\bR\} = E^{\rm BO}_{j|\mathbf{0}}
+E^{\rm BO \{\mathbf{2}\}}_{j}\cdot \Delta \bR_j^2 +\mathcal{O} \big( \Delta \bR_j^3\big).
\end{equation}  

\begin{figure}[t]
  \centering
\includegraphics[width=0.99\linewidth]{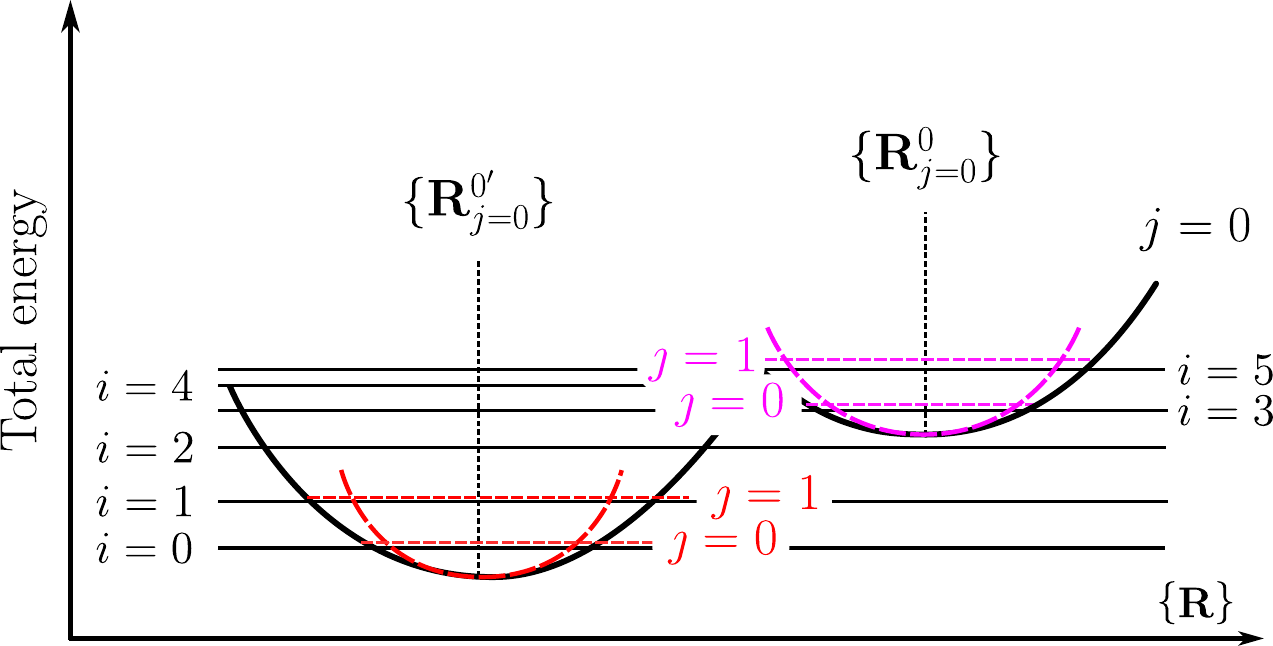}
  \caption{\label{fig:hypersurface2}
Same coordinate diagram as Fig.~\ref{fig:hypersurface} where we focus on the specific $j=0$ BO hypersurface (black curved line). 
This BO hypersurface presents here two minima denoted $\{\mathbf{R}_{j=0}^0\}$ and $\{\mathbf{R}_{j=0}^{0'}\}$.
The exact total energy $\mathcal{E}_i$ (black horizontal line) of some level $i$ is obtained from a series expansion considering one or the other reference configuration as starting point, in the neighborhood of which the BO hypersurface is expanded to quadratic order (dashed parabola),
defining an harmonic oscillator Hamiltonian.
An eigenstate of this harmonic oscillator Hamiltonian is labeled with the index $\jmath$ (dotless-$j$). 
The corresponding eigenenergy is added to the reference BO energy giving the red dashed horizontal line.
The series expansion (red dash) maps the corresponding black energy level if the  series converges.
}
\end{figure}

When considering only the second-order term in this expansion (curvature), $E^{\rm BO \{\mathbf{2}\}}_{j}$, the nuclear dynamics is described by the following auxiliary multidimensional harmonic oscillator Hamiltonian, 
\begin{equation} \label{eq:H_E_vib_first}
\hat{\mathfrak{H}}_j^{\textrm{ho}}\{ \Delta \bR\} \equiv \sum_{\kappa\alpha}\frac{\hat{P}^2_{\kappa\alpha}}{2m_\kappa} 
+E^{\rm BO \{\mathbf{2}\} }_j \cdot \Delta\mathbf{R}_j^2,
\end{equation}
with associated nuclear wavefunctions, $\vartheta_{\jmath j}$, and eigenenergies $\mathfrak{E}_{\jmath j}^{\textrm{ho}}$,
\begin{equation} \label{eq:Hvib_Schreq}
\hat{\mathfrak{H}}_j^{\textrm{ho}}\{ \Delta\bR\} \vartheta_{\jmath j} \{ \Delta\bR\} =  
\mathfrak{E}_{\jmath j}^{\textrm{ho}} \vartheta_{\jmath j} \{ \Delta\bR\},
\end{equation}
where the subscript $\jmath$ (dotless $j$) characterizes the nuclear state, considering the $j$ BO hypersurface energy and curvature in the chosen reference configuration.
The expression of the energy in terms of the nuclear state and the expansion around a (local or global) minima is depicted in Fig.~\ref{fig:hypersurface2} for the specific $j=0$ BO hypersurface, for which one has to choose among two possible reference configurations.
Importantly, the nuclear state index $\jmath$ starts at zero around each minimum $\{\mathbf{R}_j^{0}\}$ of the $j$ hypersurface. 
Both $\hat{\mathfrak{H}}_j^{\textrm{ho}}\{ \Delta\bR\}$ and $\mathfrak{E}_{\jmath j}^{\textrm{ho}}$
are denoted with a gothic calligraphy to indicate that they are related only to nuclear dynamics.
Because $\hat{\mathfrak{H}}_j^{\textrm{ho}}$ is even with respect to a sign change of $\Delta\bR_j$, its eigenfunctions are even or odd under such a sign change.

In Sec.~\ref{App:wf_factorizes} of the SI, we show that the Schrödinger equation, Eq.~\eqref{eq:Schreq_lambda}, is satisfied up to the second order in the displacements by the following wavefunction,
\begin{equation}\label{eq:trialwf}
\ket{\Psi_{i\lambda} \{ \bR \}} 
\equiv N_\lambda\vartheta_{\jmath j} \{ \Delta\bR_j /\lambda \} \ket{ \phi_j \{\bR\}},
\end{equation}
that has the form of Eq.~\eqref{eq:optwf} and where we have associated the index $i=(\jmath,j)$ to a specific pair $\jmath$ and $j$ where the index $\jmath$ is associated with the local minimum ${\mathbf{R}_j^0}$ which best describes $i$. 
This association is possible when the $\lambda$-based expansion starting 
from the auxiliary harmonic oscillator Hamiltonian converges and is exemplified by Fig.~\ref{fig:hypersurface2}.
The nuclear displacements are scaled by the inverse of $\lambda$ in the first factor, describing the nuclear dynamics, while they are not scaled in the second factor, describing the electronic BO wavefunctions.
In Sec.~\ref{App:wf_factorizes} of the SI, it is also shown that $\mathcal{E}^{(2)}_{i}= \mathfrak{E}_{\jmath j}^{\textrm{ho}}$, so that
\begin{equation}\label{eq:Eilambda_Taylor_up2}
\mathcal{E}_{i\lambda}=E^{\rm BO}_{j|\mathbf{0}} + \lambda^2 \mathfrak{E}_{\jmath j}^{\textrm{ho}}+\mathcal{O}( \lambda^3),
\end{equation}
where the first term is independent of the characteristic mass and the second term scales as $M_0^{-1/2}$.
In Eq.~\eqref{eq:Eilambda_Taylor_up2}, the second-order energy is the vibrational correction with respect to a zero-order energy that depends on the index $j$. 

Born and Oppenheimer~\cite{Born1927} have analyzed anharmonic corrections, expanding $E_j^{\rm BO}\{\bR\}$
beyond the second order in the nuclear displacements, up to the fourth order in $\lambda$.
We proceed with the analysis of the mass scaling of other corrections beyond those examined by them, up to the sixth order in $\lambda$.
In order to do this, inspired by Eq.~\eqref{eq:trialwf} we define scaled nuclear displacements with respect to the reference configuration, and express the nuclear dynamics in terms of these.
Explicitly,
\begin{align}
\rho_{\kappa\alpha j} \equiv & \Delta R_{\kappa\alpha j}/\lambda \\
\hat{\pi}_{\kappa\alpha} \equiv & \lambda \hat{P}_{\kappa\alpha}.
\end{align}
The associated notation for functions of the configuration argument is
\begin{equation}
 X[\lambda\boldsymbol\rho_j] \equiv X\{ \bR_{j}^0+\lambda \boldsymbol\rho_j  \}  ,
\end{equation}
with the shift by $\bR_{j}^0$ being taken into account and square brackets being used.
The square bracket notation for functions that depend separately and explicitly on $\lambda$ and $\{ \bR \}$ does not follow the above simple correspondence and will be specified in their definition. 
In particular, we will favor a convention in which the dominant $\lambda$ power is factored outside of such functions.

Importantly, the \emph{derivatives} of the \emph{electronic BO wavefunctions} present in $A_{\kappa\alpha j}\{\bR\}$, Eq.~\eqref{eq:A}, in $B_{\perp \kappa\alpha j}\{\bR\}$, Eq.~\eqref{eq:Bperp},
as well as in $\hat{\cal{T}}^\textrm{N}_j\{\bR\}$, Eq.~\eqref{eq:TN_R}, remain derivatives in terms of nuclear configurations. 
This is due to the expression of the second factor in Eq.~\eqref{eq:trialwf}, which is still expressed in nuclear configurations, not in scaled displacements.
However, these functions are evaluated in the configuration expressed equivalently by $\{ \bR \}$ or by $[\lambda\boldsymbol\rho]$.
So, one has
\begin{align}\label{eq:A_lambda}
A_{\kappa\alpha j}\{ \bR \} =& A_{\kappa\alpha j}[\lambda\boldsymbol\rho_j] \\
B_{\perp \kappa\alpha j}\{ \bR \} =& B_{\perp \kappa\alpha j}[\lambda\boldsymbol\rho_j].\label{eq:B_lambda}
\end{align}
Also, referring to Eq.~\eqref{eq:TN_R}, we define
\begin{multline} \label{eq:TN_lambda}
\lambda^3 \hat{\cal{T}}^\textrm{N}_\lambda[\boldsymbol\rho_j] 
\equiv  \lambda^3\sum_{\kappa\alpha} \frac{\hat{\cal{P}}^{\textrm{BO}}_{\perp j}}{2m_{\kappa}} \\
\times \Bigg[ -\lambda \bigg| \frac{\partial^2 \phi_j}{\partial R_{\kappa\alpha}^2}_{|[\lambda \boldsymbol\rho_j]} \bigg\rangle - 2i \bigg| \frac{\partial \phi_j}{\partial R_{\kappa\alpha}} 
_{|[\lambda \boldsymbol\rho_j]}
\bigg\rangle \hat{\pi}_{\kappa\alpha} \Bigg].
\end{multline}
Transformed to the mass-scaled coordinates, the $\lambda$-dependent Hamiltonian, Eq.~\eqref{eq:hamiltonian_with_lambda}, becomes
\begin{align}\label{eq:reduced_hamiltonian_with_lambda}
\hat{\mathcal{H}}_\lambda [ \boldsymbol\rho_j ] =& \lambda^2 \hat{T}^\textrm{N} [ \boldsymbol\rho_j ] + \hat{H}^{\rm BO}[\lambda\boldsymbol\rho_j] \\
\hat{T}^\textrm{N}[ \boldsymbol\rho_j ] =& \sum_{\kappa\alpha} \frac{\hat{\pi}^2_{\kappa\alpha}}{2m_\kappa}.
\end{align}
The corresponding Schrödinger equation is
\begin{equation}\label{eq:Schreq_lambda_rho}
\hat{\mathcal{H}}_\lambda [ \boldsymbol\rho_j ] \ket{ \psi_{i\lambda} [ \boldsymbol\rho_j ]} =  \mathcal{E}_{i\lambda} \ket{\psi_{i\lambda} [ \boldsymbol\rho_j ]}.
\end{equation}

We now examine the mass-scaling behavior of the different approximations presented in Sec.~\ref{sec:BOapproximations}.
Similarly to Eqs.~\eqref{eq:psi_sca}-\eqref{eq:optwf}, we express the total wavefunction using scaled displacements $[ \boldsymbol\rho_j ]$ instead of real space coordinates:
\begin{equation}\label{eq:BOwf_lambda_rho}
\ket{\psi_{i\lambda} [ \boldsymbol\rho_j ]} = \chi_{\jmath j \lambda}[\boldsymbol\rho_j ]\ket{ \phi_j [ \lambda \boldsymbol\rho_j ]}.
\end{equation}
The optimal approximation of Eq.~\eqref{eq:optimal_Schreq_2} becomes
\begin{equation}\label{eq:optimal_Schreq_2_lambda}
\Big(\lambda^2\hat{T}^\textrm{vec}_{j\lambda}[ \boldsymbol\rho_j ] + E^{\rm  opt}_{j \lambda}[\boldsymbol\rho_j ]\Big) \chi_{\jmath j\lambda}^{\rm opt}[\boldsymbol\rho_j ] = {\mathcal{E}}^{\rm opt}_{i\lambda} \chi_{\jmath j\lambda}^{\rm opt} [\boldsymbol\rho_j ],
\end{equation}
with
\begin{subequations}
\begin{align}\label{eq:Tvec_lambda}
\hat{T}_{j\lambda}^\textrm{vec}[\boldsymbol{\rho}_j] =& \sum_{\kappa\alpha} \frac{\Big(\hat{\pi}_{\kappa\alpha} 
+ \lambda A_{\kappa\alpha j}[ \lambda \boldsymbol\rho_j ]\Big)^2}{2m_{\kappa}} \\
E^{\rm opt}_{j\lambda}[\boldsymbol\rho_j ] =& E^{\rm BO}_j [\lambda \boldsymbol\rho_j ]
+ \lambda^4 E^{\rm Born}_j[\lambda\boldsymbol\rho_j ] \label{eq:E_opt_lambda} \\
E^{\rm Born}_j [\lambda\boldsymbol\rho_j ] =& \sum_{\kappa\alpha}
\frac{1}{m_{\kappa}} B_{\perp \kappa\alpha j}[\lambda\boldsymbol\rho_j ]. \label{eq:Born_lambda} 
\end{align}
\end{subequations}

In the scalar approximation, one sets $\lambda A_{\kappa\alpha j}[ \lambda \boldsymbol\rho_j ]$ and $E^{\rm Born}_j [\lambda\boldsymbol\rho_j ]$
to zero, while keeping either of them leads to alternative approximations, see Sec.~\ref{sec:hybrid} of SI.
The exact equation for the energy $\mathcal{E}_{i\lambda}$ of the full Hamiltonian, Eq.~\eqref{eq:Schreq_lambda_rho}, is now expressed in terms of the self-energy, Eq.~\eqref{eq:exact_Schreq_2}:
\begin{multline}\label{eq:exact_Schreq_2_lambda}
\Big(\hat{T}^\textrm{vec}_{j\lambda}[\boldsymbol\rho_j ] + E^{\rm opt}_{j\lambda}[\boldsymbol\rho_j ]+\lambda^6 \hat{\Sigma}_{j\lambda}(\mathcal{E}_{i\lambda},[\boldsymbol\rho_j ])\Big) \chi_{\jmath  j\lambda} [\boldsymbol\rho_j ] \\
= \mathcal{E}_{i\lambda} \chi_{\jmath j \lambda} [\boldsymbol\rho_j ],
\end{multline}
where
\begin{align}\label{eq:self-energy_lambda}
\hat{\Sigma}_{j\lambda}(z,[\boldsymbol\rho_j ]) =& \hat{\cal{T}}^{\textrm{N},\dagger}_{j\lambda}[\boldsymbol\rho_j ] \hat{\mathcal{G}}_{\perp j}(z,[\boldsymbol\rho_j ]) \hat{\cal{T}}^\textrm{N}_{j\lambda} [\boldsymbol\rho_j ], \\
\hat{\mathcal{G}}_{\perp j\lambda}(z,[ \boldsymbol\rho_j ]) =& \hat{\cal{P}}^{\textrm{BO}}_{\perp j} \Big(z - \hat{\mathcal{H}}_\lambda [\boldsymbol\rho_j ])\Big)^{-1} \hat{\cal{P}}^{\textrm{BO}}_{\perp j}. \label{eq:Green_perp_lambda}
\end{align}

Referring to Eq.~\eqref{eq:Dyson}, the latter Green's function can equivalently  be expressed in terms of the BO Green's function as
\begin{align} \label{eq:Dyson_lambda}
\hat{\mathcal{G}}_{\perp j \lambda}(z,[ \boldsymbol\rho_j ]) =& \hat{G}_{\perp j}^{\rm BO}(z,[ \lambda \boldsymbol\rho_j ]) \nonumber \\
& \times \Big(\hat{1} + \lambda^2 \hat{T}^\textrm{N}_j
[ \boldsymbol\rho_j ] \hat{\mathcal{G}}_{\perp j \lambda}(z,[ \boldsymbol\rho_j ]\Big), \\
\hat{G}_{\perp j}^{\rm BO}(z,[ \lambda\boldsymbol\rho_j ]) =&  \hat{\cal{P}}^{\textrm{BO}}_{\perp j} \Big(z-\hat{H}^{\rm BO}[ \lambda\boldsymbol\rho_j ]\Big)^{-1}\hat{\cal{P}}^{\textrm{BO}}_{\perp j}. \label{eq:GreenBO_perp_lambda}
\end{align}

In Eq.~\eqref{eq:exact_Schreq_2_lambda}, the reference BO energy $E^{\rm BO}_{j|\textbf{0}}$ can be subtracted from both $E^{\rm opt}_{j\lambda}[\boldsymbol\rho_j ]$ and $\mathcal{E}_{i\lambda}$, leaving quantities that are at least second-order in $\lambda$. 
Such second-order component can be used as a zeroth-order Hamiltonian to find the remaining expansion for $\mathcal{E}_{i\lambda}$ and to find the wavefunction $\chi_{\jmath j\lambda} [\boldsymbol\rho_j ]$.

Explicitly, from Eq.~\eqref{eq:exact_Schreq_2_lambda},
one deduces 
\begin{equation} \label{eq:Hchi_split}
\hat{\mathfrak{H}}_{j\lambda}[ \boldsymbol\rho_j ] \chi_{\jmath j\lambda} [\boldsymbol\rho_j ] 
= \mathfrak{E}_{\jmath j\lambda}
\chi_{\jmath j\lambda} [\boldsymbol\rho_j ],
\end{equation}
with
\begin{equation} \label{eq:Hamiltonian_split}
\hat{\mathfrak{H}}_{j\lambda}[ \boldsymbol\rho_j ]=
\hat{\mathfrak{H}}^{\textrm{ho}}_j 
[ \boldsymbol\rho_j ] + \Delta\hat{\mathfrak{H}}_{j\lambda}[ \boldsymbol\rho_j ] ,
\end{equation}
that includes the harmonic oscillator Hamiltonian
\begin{equation}\label{eq:Hvib_[0]}
\hat{\mathfrak{H}}^{\textrm{ho}}_j[ \boldsymbol\rho_j ] \equiv \sum_{\kappa\alpha}\frac{\hat{\pi}^2_{\kappa\alpha}}{2m_\kappa} 
+E^{\rm BO \{\mathbf{2}\}}_{j} \cdot \boldsymbol\rho_j^2,
\end{equation}
and corrections to this harmonic oscillator Hamiltonian
\begin{align}\label{eq:Hvib_[1]}
\Delta\hat{\mathfrak{H}}_{j\lambda}[ \boldsymbol\rho_j ] \equiv & \lambda
\hat{T}_{j}^\textrm{vec(1)} [ \lambda\boldsymbol\rho_j ] + \lambda^2 \hat{T}_{j}^\textrm{vec(2)} [ \lambda\boldsymbol\rho_j ] \nonumber\\
& + \sum_{n=1} \lambda^n E^{\rm BO\{\mathbf{n+2}\}}_j \cdot \boldsymbol\rho_j^{n+2} \nonumber\\
& + \lambda^2 E^{\rm Born}_j [\lambda\boldsymbol\rho_j ] + \lambda^4 \hat{\Sigma}_{j\lambda} (\mathcal{E}_{i\lambda},[\boldsymbol\rho_j ]),
\end{align}
where
\begin{align}\label{eq:Tvec1}
\hat{T}_{j}^\textrm{vec(1)}[ \lambda \boldsymbol\rho_j ] \equiv & \sum_{\kappa\alpha}
\frac{ \hat{\pi}_{\kappa\alpha} A_{\kappa\alpha j}[ \lambda \boldsymbol\rho_j ]
+ A_{\kappa\alpha j}[ \lambda \boldsymbol\rho_j ]\hat{\pi}_{\kappa\alpha}}{m_\kappa} \\
\hat{T}_{j}^\textrm{vec(2)}[ \lambda \boldsymbol\rho_j ] \equiv & 
\sum_{\kappa\alpha}\frac{(A_{\kappa\alpha}[ \lambda \boldsymbol\rho_j ])^2}{m_\kappa}. \label{eq:Tvec2}
\end{align}
The eigenvalues in Eq.~\eqref{eq:Hchi_split} are related to those in Eq.~\eqref{eq:exact_Schreq_2_lambda} by 
\begin{equation}
\mathcal{E}_{i\lambda} - E^{\rm BO}_{j|\mathbf{0}} = \lambda^{2}\mathfrak{E}_{\jmath j\lambda}.
\end{equation}
In Eq.~\eqref{eq:Hvib_[1]} and in later equations, the functions of the argument $\lambda \boldsymbol\rho_j$ are expanded using the notation of Eq.~\eqref{eq:XR_Taylor_concise} adapted to this argument:
\begin{equation}\label{eq:Xrho_Taylor_concise}
X[\boldsymbol\rho_j] = X_{j|\mathbf{0}} +X^{\{\mathbf{1}\}} \cdot  \boldsymbol\rho_j +X^{\{\mathbf{2}\}} \cdot \boldsymbol\rho_j^2 + \mathcal{O}\big( \boldsymbol\rho_j^3\big).
\end{equation} 
Without the corrections $\Delta\hat{\mathfrak{H}}_{j\lambda}$, Eq.~\eqref{eq:Hchi_split} reduces to the $\lambda$-independent equation
\begin{equation}\label{eq:Schro_[0]}
\hat{\mathfrak{H}}^{\textrm{ho}}_j [ \boldsymbol\rho_j ] \chi_{\jmath j}^{\textrm{ho}} [\boldsymbol\rho_j ] = \mathfrak{E}_{\jmath j}^{\textrm{ho}} \chi_{\jmath j}^{\textrm{ho}} [\boldsymbol\rho_j ],
\end{equation}
where identification with Eq.~\eqref{eq:Hvib_Schreq} shows that
\begin{equation}
\chi_{\jmath j}^{\textrm{ho}} [\boldsymbol\rho_j ] = \vartheta_{\jmath j}[\boldsymbol\rho_j ].
\end{equation}

Contributions with different powers of $\lambda$ in $\Delta\hat{\mathfrak{H}}_{j\lambda}[ \boldsymbol\rho_j ]$ can be ordered to understand the effect of
${\cal{O}}(\lambda)$ and ${\cal{O}}(\lambda^2)$ contributions on $\mathfrak{E}_{\jmath j\lambda}$. 
Moreover, it is possible to fix the gauge of the BO wavefunction at the reference configuration, as well as its first-order derivative with respect to atomic displacements, such that $A_{\kappa \alpha j}[\boldsymbol 0]=0$.
Hence $\hat{T}_{j}^\textrm{vec(1)}$ is ${\cal{O}}(\lambda)$ and $\hat{T}_{j}^\textrm{vec(2)}$ is ${\cal{O}}(\lambda^2)$.
The corrections to the harmonic oscillator Hamiltonian, Eq.~\eqref{eq:Hvib_[1]} becomes
\begin{align}\label{eq:Hvib_[1]_lambda}
&\Delta\hat{\mathfrak{H}}_{j\lambda}[ \boldsymbol\rho_j ] 
\equiv  
\lambda E^{\rm BO\{\mathbf{3}\}}_j \cdot  \boldsymbol\rho_j^{3} 
\nonumber\\
& + \lambda^2 
\Big( \hat{T}_{j}^ \textrm{vec(1)\{{\bf{1}}\}} \cdot \boldsymbol\rho_j
+ E^{\rm BO\{\mathbf{4}\}}_j \cdot \boldsymbol\rho_j^{4}
+E^{\rm Born}_j [\lambda\boldsymbol\rho_j ] \Big) \nonumber\\
& 
+ \sum_{n=3} \lambda^n \Big(
\hat{T}_{j}^ \textrm{vec(1)\{{\bf{n-1}}\}}
\cdot \boldsymbol\rho_j^{n-1}
+E^{\rm BO\{\mathbf{n+2}\}}_j \cdot \boldsymbol\rho_j^{n+2}\Big) \nonumber\\
&+ \sum_{n=4} \lambda^n \Big(
\hat{T}_{j}^ \textrm{vec(2)\{{\bf{n-2}}\}}
\cdot \boldsymbol\rho_j^{n-2}\Big)
+ \lambda^4 \hat{\Sigma}_{j\lambda} (\mathcal{E}_{i\lambda},[\boldsymbol\rho_j ]).
\end{align}
In this definition, the $\lambda$ behavior of the first line is  ${\cal{O}}(\lambda)$, ${\cal{O}}(\lambda^2)$ for the second line, ${\cal{O}}(\lambda^3)$ for the third line, and ${\cal{O}}(\lambda^4)$ for the fourth line.
Note that the second, third and fourth lines contain also contributions
of higher orders because of the $[\lambda\boldsymbol\rho_j ]$ argument of some of their terms or due to explicit higher $n$ values in the summation.

As previously observed, $\vartheta_{\jmath j}$ is even or odd with respect to a change of sign of $\boldsymbol\rho_j$ (also changing sign of the associated momentum). 
Moreover, all terms of $\Delta\hat{\mathfrak{H}}_{j\lambda}$ that are even with respect to such a sign change have an even $\lambda$ exponent, while all terms that are odd with respect to such a sign change have an odd $\lambda$ exponent. 
Due to these symmetry considerations, the $\lambda$-expansion of the total energy turns out to have only even power terms.

The eigenenergy present in Eq.~\eqref{eq:Hchi_split} is studied in the SI, Sec.~\ref{app:auxH4}, and gives
\begin{equation}\label{eq:E^opt4}
\mathcal{E}_{i}^{\rm(4)} = \mathcal{E}_{i}^{\rm (4), anh} + \mathcal{E}_{i}^{(4), \text{vec}} + E^{\rm Born}_{j}[\textbf{0}],
\end{equation}
with three components, first the anharmonic term,  
\begin{multline}\label{eq:E^sca4}
\mathcal{E}_{i}^{\rm (4), anh} = \langle \vartheta_{\jmath j} | E^{\rm BO\{\mathbf{4}\}}_j \cdot \boldsymbol\rho_j^{4} | \vartheta_{\jmath j} \rangle_{\boldsymbol\rho_j} \\ 
\!+\! \langle \vartheta_{\jmath j} | \big(E^{\rm BO\{\mathbf{3}\}}_j \cdot \boldsymbol\rho_j^{3}\big)
\hat{\mathfrak{G}}_{\perp \jmath j} ^{\textrm{ho}}[\boldsymbol\rho_j]\big(E^{\rm BO\{\mathbf{3}\}}_j \cdot \boldsymbol\rho_j^{3}\big)
| \vartheta_{\jmath j}\rangle_{\boldsymbol\rho_j},
\end{multline}
that is present in the expansion of the scalar approximation, then also two terms that disappear in this approximation, the vector one,
\begin{equation}\label{eq:vec4term}
   \mathcal{E}_{i}^{(4), \text{vec}} = \langle \vartheta_{\jmath j} | \hat{T}_{j}^ \textrm{vec(1)\{{\bf{1}}\}} \cdot \boldsymbol\rho_j  | \vartheta_{\jmath j} \rangle_{\boldsymbol\rho_j},
\end{equation}
and the Born one,
\begin{equation}\label{eq:Eijl[1]2_Born}
  E^{\rm Born}_{j}[\textbf{0}] = \langle \vartheta_{\jmath j} | E^{\rm Born}_{j} [\textbf{0}]| \vartheta_{\jmath j} \rangle_{\boldsymbol\rho_j}.
\end{equation}

In Eq.~\eqref{eq:E^sca4}, we have used the Green's function of the auxiliary harmonic oscillator Hamiltonian
$\hat{\mathfrak{H}}_j^{\textrm{ho}}[\boldsymbol\rho_j]$, see Eq.~\eqref{eq:H_E_vib_first}.
More precisely, we denote  
\begin{equation}\label{eq:Green[0]_perp}
\hat{\mathfrak{G}}_{\perp \jmath j}(z,[\boldsymbol\rho_j]) \equiv  \hat{\cal{P}}_{\perp \vartheta_{\jmath j}}\Big(z-\hat{\mathfrak{H}}^{\textrm{ho}}_j [\boldsymbol\rho_j]\Big)^{-1} \hat{\cal{P}}_{\perp \vartheta_{\jmath j}},
\end{equation}
with the term between parenthesis being a pseudo-inverse, limited to the subspace perpendicular to $\vartheta_{\jmath j}[\boldsymbol\rho_j ]$.
The Green's function evaluated at $z=\mathfrak{E}_{\jmath j}^{\textrm{ho}}$ is labeled with a `ho' superscript, and used in Eq.~\eqref{eq:E^sca4},
\begin{equation}\label{eq:Green_ho_perp}
\hat{\mathfrak{G}}_{\perp \jmath j}^{\textrm{ho}}
[\boldsymbol\rho_j] \equiv \hat{\mathfrak{G}}_{\perp \jmath j}
(\mathfrak{E}_{\jmath j}^{\textrm{ho}},[\boldsymbol\rho_j]).
\end{equation}

Now that we have established the mass scaling of the optimal and scalar approximations up to order five, we can carry on our search for the electron-phonon contribution to the total energy, although there is an intermediate question first: what is the \textit{phonon} contribution to the total energy?

\section{What is the phonon contribution to the total energy ?}\label{sec:phonon}

To identify terms that can be called electron-phonon contributions to the total energy, we first identify terms that can be called phonon contributions to the total energy.
At first sight, the second-order contribution $\mathcal{E}^{(2)}_{i}=\mathfrak{E}_{\jmath j}^{\textrm{ho}}$ is precisely the phonon contribution, solution of the auxiliary harmonic oscillator Hamiltonian for the nuclear dynamic, with
the well-known $M_0^{-1/2}$ scaling, in which case higher-order terms should not be labeled as phonon contributions. 

However, this choice has three drawbacks.
The first concerns the electronic inertial mass, especially the one of the core electrons.
The core electrons follow rigidly the nuclei movement. 
Therefore, the mass of core electrons should be added to the bare nuclear mass in Eq.~\eqref{eq:kineticN}. 
The second drawback comes from the vector potential in Eq.~\eqref{eq:Tvec}.
In fact, in the classical equation of motion that stems from the harmonic
approximation, a spatially homogeneous Berry curvature simply generates a force-velocity coupling, which does not prevent one to obtain normal modes of vibrations, with well-characterized frequencies~\cite{Qin2012,Saito2019, Bistoni2021,Saparov2022,Bonini2023,Ren2024, Royo2025}.
The third drawback stems from the fact that the optimal geometry (the average expectation of the nuclear positions for the exact wavefunction) is not exactly the one of the minimum of the BO hypersurface. 
The BO curvature at the optimal geometry will differ from the one at the BO geometry and would be better suited to define the phonon contribution to the total energy.

In these three cases, the contribution from phonons will not have a pure $M_0^{-1/2}$ scaling.
For example, the electronic inertial mass adds to the bare nuclear mass
$M_\kappa$ a $n_{e\kappa}m_e$ correction, where $n_{e\kappa}$ is the number of electrons moving with the nucleus $\kappa$, while $m_e$ is the electronic mass (so, 1 in atomic units). 
One can take such modification into account by updating accordingly the value of $m_\kappa$, with a factor $M_0^{-1}$, that is $\lambda^4$. 
Coupled with the $\lambda^2$ scaling of the phonon energy, one sees that this contribution should appear in a $\lambda^6$ term in the perturbation expansion of Sec.~\ref{sec:scaling}. 

Indeed, the lowest-order $\lambda$ contribution to the total energy by the self-energy operator $\hat{\Sigma}_{j\lambda}$ contains such an inertial electronic mass correction.
Using Eqs.~\eqref{eq:TN_lambda}, \eqref{eq:self-energy_lambda}, \eqref{eq:Green_perp_lambda}, \eqref{eq:Dyson_lambda}, and  \eqref{eq:GreenBO_perp_lambda}, we find
\begin{equation}\label{eq:self-energy_lowest_lambda}
\hat{\Sigma}_{j\lambda}(z,[\boldsymbol\rho_j ]) =
\sum_{\substack{\kappa\alpha \\ \kappa'\alpha'}}
\frac{1}{2}\hat{\pi}_{\kappa\alpha}
\Delta m_{\kappa\alpha,\kappa'\alpha'}^{\textrm{elm},-1}(z)
\hat{\pi}_{\kappa'\alpha'} +\mathcal{O}(\lambda),
\end{equation}
where
\begin{equation}\label{eq:inertial_mass_correction}
\!\!\!\!\Delta m_{\kappa\alpha,\kappa'\alpha'}^{\textrm{elm},-1}(z)
= \bigg\langle  \frac{\partial \phi_j}{\partial R_{\kappa\alpha}}_{|\mathbf{0}} 
\bigg|
\frac{2\hat{G}_{\perp j}^{\rm BO}(z,[\mathbf{0}])}{m_{\kappa}m_{\kappa'}}
\bigg| \frac{\partial \phi_j}{\partial R_{\kappa'\alpha'}}_{|\mathbf{0}} \bigg\rangle
\end{equation}
is the correction to the inverse electronic mass.
Taking into account that this is only the lowest-order correction to the electronic mass, this expression matches the one obtained using the exact factorization technique, see Eqs.~(13), (14), (15) and (B10) of  Ref.~\onlinecite{Scherrer2015}.
We note that Eq.~\eqref{eq:self-energy_lowest_lambda} is quadratic in the scaled momenta, and does not depend on the scaled displacements $\boldsymbol\rho_j$.
Such a quadratic contribution can be transferred from 
$\Delta\hat{\mathfrak{H}}_{j\lambda}[ \boldsymbol\rho_j ]$ as defined in Eq.~\eqref{eq:Hvib_[1]} or \eqref{eq:Hvib_[1]_lambda} to 
$\hat{\mathfrak{H}}^{\textrm{ho}}_j [ \boldsymbol\rho_j ]$ defined in Eq.~\eqref{eq:Hvib_[1]_lambda} without changing the fact that the latter is an harmonic oscillator Hamiltonian.
It simply becomes $\lambda$-dependent.

Similarly, quadratic contributions in the scaled displacements $\boldsymbol\rho_j$
can be transferred from $\Delta\hat{\mathfrak{H}}_{j\lambda}[ \boldsymbol\rho_j ]$   
to $\hat{\mathfrak{H}}^{\textrm{ho}}_j [ \boldsymbol\rho_j ]$, as well as mixed contributions that include the scaled momenta and scaled displacements such as
\begin{equation}
\hat{T}_{j}^ \textrm{vec(1)\{{\bf{1}}\}} \cdot \boldsymbol\rho_j
= \sum_{\kappa\alpha}\frac{1}{m_\kappa} A_{\kappa\alpha j}^{\{\bf{1}\}} \cdot
(\hat{\pi}_{\kappa\alpha}  \boldsymbol\rho_j+\boldsymbol\rho_j \hat{\pi}_{\kappa\alpha}).
\end{equation} 

Moreover, some contributions might be extracted from the terms in $\Delta\hat{\mathfrak{H}}_{j\lambda}[ \boldsymbol\rho_j ]$ that are higher-order in the combined scaled displacements and scaled momenta, such as $\lambda E^{\rm BO\{\mathbf{3}\}}_j \cdot  \boldsymbol\rho_j^{3}$, and
might be transferred to $\hat{\mathfrak{H}}^{\textrm{ho}}_j [ \boldsymbol\rho_j ]$, in connection to a $\lambda$-dependent change of reference configuration. 
These are so-called ``quasi-harmonic'' contributions.
Let us consider Eq.~\eqref{eq:E^sca4}. 
Changing the reference configuration from $\{\bR_{j}^0 \}$ to some nearby configuration $\{\bR^{\rm ref} \}$ will increase the BO energy contribution to the total energy, with a dominant effect proportional to the square of the change of configuration.
However, such a change of configuration might induce a linear change of $\lambda E^{\rm BO\{\mathbf{3}\}}_j \cdot  \boldsymbol\rho_j^{3} | \vartheta_{\jmath j}\rangle_{\boldsymbol\rho_j}$ that is present twice in Eq.~\eqref{eq:E^sca4}.
The net result might be a gain in energy.
This possibility is especially relevant when one examines the behavior of the optimized volume of the primitive cell.
The standard BO approach starts from the volume that minimizes the BO energy.
The derivative of the BO energy with respect to a volume change is zero.
However, the zero-point energy of phonons creates a modification of this energy, which is incorporated in the total energy. 
The gradient of the zero-point energy of phonons with respect to a volume change will usually not vanish at the BO optimized configuration. 
Hence, the re-optimized volume will be linear in the zero-point energy, that is $\mathcal{\lambda}^2$.
A $\mathcal{\lambda}^2$ change in volume will create a $\mathcal{\lambda}^4$ change in total energy.

It is possible to establish a BO expansion both with the BO optimized configuration as reference, as we have done until now, or with a $\lambda$-dependent optimized configuration as reference as obtained in such quasiharmonic approximation. 
Since the reference configuration changes, the terms of both series will not be identical but the series summation gives the same final result.

However, some terms cannot be eliminated by a modification of the parameters of the auxiliary harmonic oscillator,
those of order lower than quadratic in scaled displacements and momenta, coming from the Born energy and from the self-energy. 
They will be examined in the next section.

\section{Remaining contributions to the total energy}\label{sec:remaining}

The Hamiltonian $\Delta\hat{\mathfrak{H}}_{j\lambda}[ \boldsymbol\rho_j ]$
includes the Born potential, that is order $\mathcal{O}(\lambda^2)$ when developed in powers of the scaled displacement:
\begin{equation}\label{eq:EBorn_expansion}
\lambda^2 E^{\rm Born}_j [\lambda\boldsymbol\rho_j ] = \sum_{n=0} \lambda^{n+2} 
\Big(E_j^{\rm Born\{{\bf{n}}\}}\cdot \boldsymbol\rho_j^{n}\Big).
\end{equation}

The displacement-independent contribution to the total energy of Eq.~\eqref{eq:EBorn_expansion} at the lowest order in $\lambda$ is given in Eq.~\eqref{eq:Eijl[1]2_Born} and can be further evaluated since it does not depend on displacement.
Using Eqs.~\eqref{eq:Born_lambda}, \eqref{eq:B_lambda}, and \eqref{eq:Bperp}:
\begin{equation}\label{eq:Eijl[1]2_Born_2}
  E^{\rm Born}_{j}[\textbf{0}] = \sum_{\kappa\alpha}
\bigg\langle \frac{\partial \phi_j}{\partial R_{\kappa\alpha}}_{|\bf{0}} \bigg| \frac{\hat{\cal{P}}^{\textrm{BO}}_{\perp j}}{2m_\kappa} \bigg|
\frac{\partial \phi_j}{\partial R_{\kappa\alpha}}_{|\bf{0}} \bigg\rangle.
\end{equation}

The derivative of the wavefunction with respect to the displacement of a nucleus is a typical quantity present in the first-principles calculations of dynamical matrices and phonons for solids~\cite{Gonze1997,Gonze1997b,Gonze2024a}.
Similarly, for molecules, the interatomic force constants are found from such calculations, as well as vibrational frequencies.
The scalar product of such a quantity with excited electronic states delivers the electron-phonon (or electron-vibration) matrix elements, as shown later.
As such, among all the contributions to the total energy that appear in the BO expansion, Eq.~\eqref{eq:Eijl[1]2_Born_2} is the lowest-order one that can be called an \emph{``electron-phonon contribution to the total energy''} (or electron-vibration).

Accordingly, one defines
\begin{equation}\label{eq:elph4_main}
\mathcal{E}^{(4),\textrm{elph}}_{i} \equiv E^{\rm Born}_{j}[\bf{0}]
\end{equation}
that is independent of the vibrational state.
The next Born lowest-order contribution is linear in the scaled displacement, $n=1$ in Eq.~\eqref{eq:EBorn_expansion}.
Being odd with respect to the inversion of $\boldsymbol\rho_j$, its expectation value for the vibrational wavefunction $\vartheta_{\jmath j}$ vanishes.
It will not contribute to first order in $\Delta\hat{\mathfrak{H}}_{j\lambda}[ \boldsymbol\rho_j ]$ but will contribute at the next order.
This contribution might then be called an electron-phonon contribution to total energy as well, albeit of higher order.

Finally, we have seen that the self-energy $\hat{\Sigma}_{j\lambda}(z,[\boldsymbol\rho_j ])$ yields Eq.~\eqref{eq:self-energy_lowest_lambda} at lowest-order and corresponds to a mass renormalization of the Hamiltonian.
From the expression of $\hat{\cal{T}}^\textrm{N}_\lambda[\boldsymbol\rho_j]$, 
Eq.~\eqref{eq:TN_lambda}, and its double use in the self-energy, Eq.~\eqref{eq:self-energy_lambda}, one sees that first- and second-order derivatives of the BO wavefunctions play an important role in the self-energy, which might also be named
an electron-phonon contribution to the total energy.
Beyond the lowest-order contribution, Eq.~\eqref{eq:self-energy_lowest_lambda}, the next order brings one momentum operator multiplied by $\lambda$.
Being odd with respect to the inversion of $\boldsymbol\rho_j$, its expectation value for the dynamical wavefunction $\vartheta_{\jmath j}$ vanishes. 
The order 6 energy associated with the mass renormalization from Eq.~\eqref{eq:inertial_mass_correction} is $\mathcal{E}_i^{(6),\rm elm}$ and is given in Eq.~\eqref{eq:6el} of Sec.~\ref{App:wf_factorizes} of the SI.

Altogether, there are five groups of energy terms at order 6, see Sec.~\ref{app:auxH4} of the SI, including an electron-phonon energy contribution, $\mathcal{E}_i^{(6),\rm elph}$, terms associated with the vector potential (which is zero in materials with time-reversal symmetry), $\mathcal{E}_i^{(6),\rm vec}$, anharmonic energy terms, $\mathcal{E}_i^{(6),\rm anh}$, and mixed terms which involve anharmonic and vector potential contributions, as well as one anharmonic and an electron-phonon term, $\mathcal{E}_i^{(6),\rm mix}$.
This concludes our listing of all the energy contributions to the total energy up to order 6 in the mass scaling and we present a summary of the terms in Table~\ref{tab:table0}.
We now turn to the practical evaluation of the Born contribution to the total energy, which we will now refer to as the \emph{electron-phonon contribution} to the total energy.
\begin{table}[t]
  \caption{\label{tab:table0}
Summary of all the energy terms up to order 6 in the mass scaling, that contribute to the total energy. 
The main text contains expressions up to order 4, while Sec.~\ref{app:auxH4} of the SI presents those of order 6.
}
\begin{tabular}{ l r r }
  \toprule\\
Energy  & Equation & Mass-scaling  \\
\hline
$\mathcal{E}_i^{(0)} = E_{j|\mathbf{0}}^{\rm BO}$ & \eqref{eq:EBO_0R_lambda}  & $\mathcal{O}(\lambda^0)$   \\
$\mathcal{E}_i^{(2)} = \mathfrak{E}_{\jmath j}^{\textrm{ho}}$ & \eqref{eq:Hvib_Schreq}  & $\mathcal{O}(\lambda^2)$ \\
$\mathcal{E}_i^{(4),\rm vec} $ & \eqref{eq:vec4term}  & $\mathcal{O}(\lambda^4)$ \\
$\mathcal{E}_i^{(4),\rm anh} $ & \eqref{eq:E^sca4}  & $\mathcal{O}(\lambda^4)$ \\
$\mathcal{E}_i^{(4),\rm elph} $ & \eqref{eq:elph4_main}  & $\mathcal{O}(\lambda^4)$ \\
$\mathcal{E}_i^{(6),\rm elph} $ & \eqref{eq:6elph}  & $\mathcal{O}(\lambda^6)$ \\
$\mathcal{E}_i^{(6),\rm elm} $ & \eqref{eq:6el}  & $\mathcal{O}(\lambda^6)$ \\
$\mathcal{E}_i^{(6),\rm vec} $ & \eqref{eq:6vec}  & $\mathcal{O}(\lambda^6)$ \\
$\mathcal{E}_i^{(6),\rm anh} $ & \eqref{eq:6anh}  & $\mathcal{O}(\lambda^6)$ \\
$\mathcal{E}_i^{(6),\rm mix} $ & \eqref{eq:6hyb}  & $\mathcal{O}(\lambda^6)$ \\

  \botrule
\end{tabular}
\end{table}

\section{Semiconductors and insulators treated within density-functional theory}
\label{sec:ahc}

In Sec.~\ref{sec:BOapproximations}, we introduced $\phi_j \{ \bR \}$ as the many-body electronic wavefunction for a fixed nuclear configuration, in the Born-Oppenheimer approximation, see Eq.~\eqref{eq:BO_Schreq}.
Such a many-body wavefunction can only be computed in the simplest systems.
In this section, one approximates it by a Slater determinant formed from the combination of density-functional theory (DFT) one-electron Kohn-Sham wavefunctions~\cite{Martin2004}.
Moreover, the application to gapped periodic solids is considered.
As mentioned in the previous section, the obtained results rely on non-degenerate perturbation theory, with an energy separation between the ground-state energy and the first excited-state energy. 
This justifies our consideration of gapped solids only.
Modifications would be needed to deal with metallic periodic solids.

For periodic solids, eigenfunctions are Bloch waves, made of a periodic part times a phase.
A band index $n$ and a crystalline momentum $\bk$  characterize each wavefunction $\phi_{n\bk}$.
For simplicity, the spin index is not explicitly noted and the summations over bands are to be understood as being on bands and spin channels if the two spin channels are treated separately.
The dependence on the nuclear configurations is not explicitly noted for the wavefunctions of the charge density, or for entities that depend on such configurations only indirectly.

The ground-state Born-Oppenheimer DFT energy for solids includes~\cite{Martin2004}
a kinetic energy term, a one-electron atomic potential energy term, the Hartree-exchange-correlation (Hxc) energy term, and a nucleus-nucleus interaction energy term, as follows:
\begin{equation}
E^{\rm BO}\{ \bR \} = T + E^{\textrm{e-N}}\{ \bR \} + E^{\textrm{Hxc}} + E^{\textrm{NN}}\{ \bR \},
\end{equation}
where we now assume to be in the ground state $j=0$ and therefore omit the $j$ index. 
The DFT Kohn-Sham eigenenergies are obtained from the expectation value of the kinetic, one-electron nuclear potential and Hxc potential operators for the Kohn-Sham wavefunctions,
\begin{equation}
    \varepsilon_{n\bk}\{\bR \}= \braket{ \phi_{n\bk} | \hat{T} \!+\! \hat{V}^{\textrm{e-N}}\{ \bR \} \!+\! \hat{V}^{\textrm{Hxc}}| \phi_{n\bk}},
\end{equation}
where $\hat{V}^{\textrm{e-N}}$ and $\hat{V}^{\textrm{Hxc}}$ are periodic, and the Kohn-Sham wavefunctions are normalized to one in the unit cell.
The one-electron nuclear potential can be considered in an all-electron, pseudopotential, or projector-augmented wave framework. 
In the latter two frameworks, the potential operator has local and non-local contributions.
Also, the nucleus-nucleus energy includes an Ewald energy term from periodic positively charged particles placed in a negatively charged homogeneous background, but also possibly a pseudo-core energy~\cite{Hamann1979} for the pseudopotential and projector-augmented wave frameworks.

The ground-state Born-Oppenheimer DFT energy per unit cell can be written in terms of its eigenenergy component, $E^{\textrm e}$, plus other contributions, denoted $E^{\textrm{other}}$,
\begin{equation}\label{eq:decomposition}
E^{\rm BO} \{\bR \} = E^{\rm e} \{\bR \} + E^{\rm other} \{\bR \},
\end{equation}
with the eigenenergy component per unit cell being
\begin{equation}\label{eq:epart}
E^{\rm e}\{\bR \} = \int_{\rm BZ} \frac{\rm d \bk}{\Omega^{\rm BZ}} \sum_n f_{n\bk} \varepsilon_{n\bk}\{\bR \},
\end{equation}
$\Omega^{\rm BZ}$ is the Brillouin-Zone volume and $f_{n\bk}$ is the Fermi-Dirac occupation function.
In this work, zero temperature is considered, so that the occupation function is one or zero.
Moreover, for gapped solids, the occupation function for all energy bands below the gap is one and zero for bands above the gap.
We therefore write Eq.~\eqref{eq:epart} as:
\begin{equation}\label{eq:epart2}
E^{\rm e}\{ \bR \} = \int_{\rm BZ} \frac{\rm d \bk}{\Omega^{\rm BZ}} \sum_n^{\textrm{occ}} \varepsilon_{n\bk}\{\bR \},
\end{equation}
and the other contributions are
\begin{subequations}
\begin{align}
E^\textrm{other}\{\bR \} =& E^{\textrm{Hxc}} - V^{\textrm{Hxc}} \!+\! E^{\textrm{NN}}\{ \bR \}, \\
V^{\textrm{Hxc}} =& \int_{\rm BZ} \frac{\rm d \bk}{\Omega^{\rm BZ}} \sum_n^{\textrm{occ}}\braket{ \phi_{n\bk} | \hat{V}^{\textrm{Hxc}} | \phi_{n\bk}}. 
\end{align}
\end{subequations}

Returning to the different orders of the Taylor expansion of the total ground-state energy, Eq.~\eqref{eq:ETaylor_lambda} becomes:
\begin{equation}
\mathcal{E}_\lambda = E^{\rm BO}_{|\mathbf{0}} + \lambda^2 \mathcal{E}^{(2)} 
+ \lambda^4 \mathcal{E}^{(4)} + \mathcal{O}(\lambda^5),
\end{equation}
where one has to write expressions for the second-order and fourth-order energies in the case of DFT for a solid.
The physical total energy is obtained with $\lambda=\lambda_0$.
We then perform a transformation to normal mode coordinates, see Sec.~\ref{sec:normal_modes} of the SI, and express the second-order contribution to the total energy per unit cell resulting from atomic vibrations (or phonons), Eq.~\eqref{eq:Eph}, as:
\begin{equation}\label{eq:E2qalternative}
E^{\textrm{ph}} \equiv \lambda_0^2 \mathcal{E}^{(2)} = \lambda_0^2 \mathfrak{E}_0^{\rm ho} = \int_{\rm BZ} \frac{\rm d \bq}{\Omega^{\rm BZ}}\sum_\nu \frac{ \omega_{\nu\bq}}{2}, 
\end{equation} 
where  $\omega_{\nu\bq}$ is the phonon frequency for mode $\nu$ and momentum $\bq$.

The fourth-order contribution to the total energy due to phonons from Eq.~\eqref{eq:E4elphfinal} can be written in different ways.
Considering normal mode derivatives of the wavefunctions yields
\begin{subequations}
\begin{align}\label{eq:fourthorder}
E^{\rm elph} \equiv & \lambda_0^4 \mathcal{E}^{(4), \rm elph} \\
=& \frac{1}{2} 
\iint_{\rm BZ} \! \frac{\rm d \bk \rm d \bq }{(\Omega^{\rm BZ})^2}
\sum_{n}^{\rm occ}
\sum_{\nu}  \bigg\langle  \frac{\partial\phi_{n\bk}}{\partial \eta_{\nu\bq}} \bigg|  \hat{\mathcal{P}}_{\perp\textrm{occ}, \mathbf{k}}^{\rm BO} \bigg|
\frac{\partial\phi_{n\bk}}{\partial \eta_{\nu\bq}} \bigg\rangle, 
\label{eq:fourthorder_explicit}
\end{align}
\end{subequations}
with a possible spin doubling in the unpolarized spin case and where $\hat{\mathcal{P}}_{\perp\textrm{occ}, \mathbf{k}}^{\rm BO}$ is the projector on the unoccupied states.

The relation between nuclei displacements $\Delta R_{l\kappa\alpha}$ and normal mode amplitude
$\eta_{\nu\bq}$, in the case of periodic solids is given by
\begin{equation}\label{eq:deltaRl}
\Delta R_{l\kappa\alpha} = \sum_{\nu} \eta_{\nu\bq} e^{i\bq \cdot \mathbf{R}_l }U_{\nu\kappa\alpha \bq}, 
\end{equation} 
where $\mathbf{R}_l$ is the lattice vector of the cell $l$, or alternatively by
\begin{equation}\label{eq:deltaRl_alt}
\Delta R_{l\kappa\alpha} = \sum_{\nu} \eta_{\nu\bq} e^{i\bq \cdot \mathbf{R}_{l\kappa} }U_{\nu\kappa\alpha\bq},
\end{equation} 
where  $\mathbf{R}_{l\kappa}$ is the position vector of the atom $\kappa$ in cell $l$.
The phase difference between these two possibilities does not impact observables as they cancel in the scalar product present in Eq.~\eqref{eq:fourthorder_explicit}.

Direct derivatives of wavefunctions with respect to the atomic displacements of a specific sublattice can also be used. 
Defining
\begin{equation}\label{eq:deltaRl_fromR}
\Delta R_{l\kappa\alpha} \equiv R_{\kappa\alpha\bq} e^{i\bq \cdot \mathbf{R}_l }, 
\end{equation} 
where $R_{\kappa\alpha\bq} = \sum_\nu \eta_{\nu\bq} U_{\nu\kappa\alpha\bq}$, gives
\begin{multline}\label{eq:fourthorder2}
E^{\rm elph} = \frac{1}{2} 
\iint_{\rm BZ} \frac{\rm d \bk}{\Omega^{\rm BZ}}\frac{\rm d \bq}{\Omega^{\rm BZ}}\\
\times \sum_{n}^{\rm occ} \sum_{\kappa\alpha}
\frac{1}{M_{\kappa}}
 \bigg\langle \frac{\partial\phi_{n\bk}}{\partial R_{\kappa\alpha\bq}} \bigg| \hat{\mathcal{P}}_{\perp\textrm{occ}, \mathbf{k}}^{\rm BO} \bigg|
 \frac{\partial\phi_{n\bk}}{\partial R_{\kappa\alpha\bq}} \bigg \rangle. 
\end{multline}

A third possibility to express the fourth-order contribution to the total energy due to electron-phonon interaction relies on the usual electron-phonon matrix elements and a sum over unoccupied states. This expression reads
\begin{multline}\label{eq:fourthorder3}
E^{\rm elph} = \frac{1}{2} 
\iint_{\rm BZ} \frac{\rm d \bk}{\Omega^{\rm BZ}}\frac{\rm d \bq}{\Omega^{\rm BZ}}\\
\times \sum_{n}^{\rm occ}\sum_{m}^{\rm unocc} \sum_{\nu}
 \frac{2\omega_{\nu\mathbf{q}}|g_{mn\nu}(\bk,\bq)|^2  }{(\varepsilon_{m\bk+\bq} - \varepsilon_{n\bk} )^2},
\end{multline}
where   
$g_{mn\nu}(\bk,\bq) = \braket{ \phi_{m\bk+\bq}  |  \Delta V_{\bq\nu} | \phi_{n\bk} }$ are the electron-phonon matrix elements in which 
$\Delta V_{\bq\nu}$ is the first-order change to the Kohn-Sham potential induced by a phonon mode $\nu$.
The first-order potential in the mode basis is expressed in terms of the movement of atoms $\kappa$ in the Cartesian direction $\alpha$ as
\begin{multline}
\label{eq:V(1)}
\Delta V_{\bq\nu}(\mathbf{r},\mathbf{r}') = \Big[  \frac{1}{2\omega_{\nu\mathbf{q}}}\Big]^{\frac{1}{2}} \sum_{\kappa\alpha l}U_{\nu\kappa\alpha\bq} \\
\times \frac{\partial V_{\kappa\alpha}(\mathbf{r}+{\mathbf{R}_l},\mathbf{r}'+{\mathbf{R}_l})}{\partial R_{\kappa\alpha\bq}} e^{i\mathbf{q}\cdot (\mathbf{r}'+\mathbf{R}_l)}.
\end{multline}
In the \textsc{Abinit} software, for the purpose of the present work, Eqs.~\eqref{eq:fourthorder2} and \eqref{eq:fourthorder3} have been implemented.
As mentioned earlier, the spin index is omitted and the summations over bands are to be understood as being on bands and spin channels if the two spin channels are treated separately.
In the spin-unpolarized case, the summation might be done over the (non-spin-polarized) wavefunctions, and a factor of two might be included afterward.

\section{Allen's electron-phonon contribution to the total energy}
\label{sec:Allenelph}

In Ref.~\onlinecite{Allen2022}, Allen proposed that the contribution from the electron-phonon interaction to the total energy could be computed from the same ingredients as for the zero-point renormalization of the electronic eigenenergies due to the electron-phonon interaction. 
Using the notation of the present paper, Eq.~(4) of Ref.~\onlinecite{Allen2022} becomes
\begin{multline}\label{eq:EAllen}
E^{\rm Allen} = \iint_{\rm BZ} \frac{\rm d \bk}{\Omega^{\rm BZ}} \frac{\rm d \bq}{\Omega^{\rm BZ}} \sum_n f_{n\bk} \sum_{\nu} \Big[
 \braket{ \phi_{n\bk} | \Delta_2 V_{\nu\bq} | \phi_{n\bk} } \\
+ \sum_m \frac{|\braket{\phi_{n\bk} | \Delta V_{\nu\bq} | \phi_{m\bk+\bq} }|^2}{\varepsilon_{n\bk} - \varepsilon_{m\bk+\bq} } (1-f_{m\bk+\bq}) \Big],
\end{multline}
where the expression for the second-order potential $\Delta_2 V_{\nu\bq}$ can be found in Ref.~\onlinecite{Giustino2017} (there, it is denoted $V^{(2)}_{\nu\bq}$).
It is asserted in Ref.~\onlinecite{Allen2022} that this expression has the same structure as the sum of the electron-phonon energy shifts $\Delta \varepsilon_{n\bk}$ of the occupied states, but that it differs from the latter because of the presence of the
$(1-f_{m\bk+\bq})$ factor in Eq.~\eqref{eq:EAllen}.
For comparison, such shifts write
\begin{multline} \label{eq:Deig}
\Delta \varepsilon_{n\bk} = \int_{\rm BZ} \frac{\textrm{d} \bq}{\Omega^{\rm BZ}} \sum_{\nu} \Big[
 \braket{ \phi_{n\bk} | \Delta_2 V_{\nu\bq} | \phi_{n\bk} } \\
 + {\sum_{m}}' \frac{|\braket{\phi_{n\bk} | \Delta V_{\nu\bq} | \phi_{m\bk+\bq} }|^2}{\varepsilon_{n\bk} - \varepsilon_{m\bk+\bq} }
 \Big],
\end{multline}
where the prime indicates that the intraband contribution ($m=n$) is omitted for $\bq=\mathbf{0}$.

Interestingly, the $(1-f_{m\bk+\bq})$ factor can be suppressed from Eq.~\eqref{eq:EAllen}. 
Indeed, due to the change of sign in the denominator $\varepsilon_{n\bk} - \varepsilon_{m\bk+\bq}$
upon exchange between $n\bk$ and $m\bk+\bq$, one has
\begin{equation} \label{eq:EAllen_vanish}
  \iint_{\rm BZ} \frac{\textrm{d} \bk}{\Omega^{\rm BZ}} \frac{\textrm{d} \bq}{\Omega^{\rm BZ}} {\sum_{\nu nm}}'  f_{n\bk} \frac{|\braket{\phi_{n\bk} | \Delta V_{\nu\bq} | \phi_{m\bk+\bq}  }|^2}{\varepsilon_{n\bk} - \varepsilon_{m\bk+\bq} }
 f_{m\bk+\bq} = 0.
\end{equation}
Thus, without any approximation,
\begin{equation}\label{eq:EAllen2}
E^{\rm Allen} = \int_{\rm BZ} \frac{\rm d \bk}{\Omega^{\rm BZ}} \sum_n f_{n\bk} \Delta \varepsilon_{n\bk},
\end{equation}
which at 0~K reduces to
\begin{equation}\label{eq:EAllen3}
E^{\rm Allen}  = \sum_{n}^{\rm occ} \int_{\rm BZ} \frac{\text{d}\bk}{\Omega^{\rm BZ}} \Delta \varepsilon_{n\bk}.
\end{equation}
In a subsequent work~\cite{RamasimhaVarma2023}, Varma~\textit{et al.} computed the right-hand side of Eq.~\eqref{eq:EAllen3}, and argued that it approximates well Eq.~\eqref{eq:EAllen}, the electron-phonon contribution to the total energy.
Like Allen, they did not mention that both quantities are equal.
We also remark that the eigenvalue renormalization in Eq.~\eqref{eq:Deig} is expressed using the adiabatic approximation, but the nonadiabatic version of $\Delta \varepsilon_{n\bk}$ should be used for infra-red active materials~\cite{Ponce2015,Miglio2020}. 

Let us proceed with the mass-scaling analysis of $E^{\textrm{Allen}}$.
Relying on the analysis presented in Sec.~\ref{sec:scaling}, in particular the fact that the fluctuations of nuclear displacements are proportional to $\lambda$, one finds that $\Delta \varepsilon_{n\bk}$ scales as $\lambda^2$. 
Actually, $\Delta V_{\nu\bq}$ scales as $\lambda$, and is present twice in Eq.~\eqref{eq:Deig}, while $\Delta_2 V_{\nu\bq}$ scales like $\lambda^2$, see Eqs.~(34) and (40) of Ref.~\onlinecite{Giustino2017}.
Hence, $E^{\textrm{Allen}}$ also scales as $\lambda^2$. 
It is of second order in the perturbation, not of fourth order. 
However, the perturbation approach of the previous sections shows that the second-order term gives the phonon contribution to the total energy, no second-order contribution in 
$\lambda$ can be called an electron-phonon contribution.
As we will show, this alleged electron-phonon contribution to the total energy is actually a part of the phonon contribution to the total energy, previously mentioned, $E^{\rm ph}$. 
Hence $E^{\textrm{Allen}}$ should not be added to $E^{\rm BO}_{|\mathbf{0}}$ and $E^{\rm ph}$. 
The correct expressions for the lowest-order electron-phonon contribution to the total energy, if any, $E^{\rm elph}$, are given by Eqs.~\eqref{eq:fourthorder}, \eqref{eq:fourthorder2}, and \eqref{eq:fourthorder3}.

To prove that $E^{\textrm{Allen}}$ is a contribution to $E^{\rm ph}$ and not to $\lambda_0^4\mathcal{E}^{(4)}$, we further start from Eqs.~(1) and (2) of Ref.~~\onlinecite{Ponce2015}:
\begin{multline}\label{eq:Deig2}
\Delta \varepsilon_{n\bk} = \frac{1}{2}\sum_{\nu} \int_{\rm BZ} \frac{\rm d \bq}{\Omega^{\rm BZ}} \frac{1}{2\omega_{\nu\bq}}   \\
 \times \sum_{\kappa\alpha\kappa'\beta l} \frac{\partial^2 \varepsilon_{n\bk}}{\partial R_{0\kappa\alpha}\partial R_{l\kappa'\beta}} e^{i\bq \cdot \mathbf{R}_l } U_{\nu\kappa\alpha\bq}^*U_{\nu\kappa'\beta\bq}.
\end{multline}
This expression is inserted in Eq.~\eqref{eq:EAllen3} and combined with Eq.~\eqref{eq:epart2} to give
\begin{multline} \label{eq:EAllen4}
E^{\rm Allen} = \frac{1}{2} \sum_{\nu} \int_{\rm BZ} \frac{\rm d \bq}{\Omega^{\rm BZ}} \frac{1}{2\omega_{\nu\bq}}  \\
 \times \sum_{\kappa\alpha\kappa'\beta l} \frac{\partial^2 E^{\rm e}}{\partial R_{0\kappa\alpha}\partial R_{l\kappa'\beta}} e^{i\bq \cdot \mathbf{R}_l } U_{\nu\kappa\alpha\bq}^*U_{\nu\kappa'\beta\bq},
\end{multline}
where the dependence of the electronic energy $E^{\rm e}$ on $\phi$ and $\{\bR \}$ is omitted for conciseness.
In addition, the second-order derivative of the BO total energy is the interatomic force constant, see Eq.~\eqref{eq:secular}, which extends for the periodic case to
\begin{equation}\label{eq:omega2}
\omega_{\nu\bq}^2 = \sum_{\kappa\alpha\kappa'\beta l} \frac{\partial^2 E^{\rm BO}}{\partial R_{0\kappa\alpha}\partial R_{l\kappa'\beta}} e^{i\bq \cdot \mathbf{R}_l } U_{\nu\kappa\alpha\bq}^*U_{\nu\kappa'\beta\bq}.
\end{equation}
This is the same expression as the second summation in Eq.~\eqref{eq:EAllen4} when $E^{\rm e}$ is replaced by $E^{\rm BO}$.
Dividing Eq.~\eqref{eq:omega2} by $2\omega_{\nu\bq}$ and inserting it into Eq.~\eqref{eq:E2qalternative} gives
\begin{multline}\label{eq:E2qalternative_2}
E^{\rm ph} = \sum_\nu \int_{\rm BZ} \frac{\rm d \bq}{\Omega^{\rm BZ}} \frac{1}{2\omega_{\nu\bq}}\\
 \times \sum_{\kappa\alpha\kappa'\beta l}
\frac{\partial^2 E^{\rm BO}}{\partial R_{0\kappa\alpha}\partial R_{l\kappa'\beta}} e^{i\bq \cdot \mathbf{R}_l } U_{\nu\kappa\alpha\bq}^*U_{\nu\kappa'\beta\bq}.
\end{multline} 
The decomposition of the BO energy had been introduced in Eq.~\eqref{eq:decomposition} and allows one to split the second-order phononic contribution to the total energy, delivering
\begin{equation} \label{E2ph2=EAllen+}
E^{\rm ph} = 2 E^{\textrm{Allen}} + E^{\textrm{ph,other}},
\end{equation} 
with
\begin{multline}
E^{\textrm{ph,other}} = \sum_\nu \int_{\rm BZ} \frac{\textrm{d} \bq}{\Omega^{\rm BZ}} \frac{1}{2\omega_{\nu\bq}} \\
 \times \sum_{\kappa\alpha\kappa'\beta l}
 \frac{\partial^2 E^{\rm other}}{\partial R_{0\kappa\alpha}\partial R_{l\kappa'\beta}} e^{i\bq \cdot \mathbf{R}_l } U_{\nu\kappa\alpha\bq}^*U_{\nu\kappa'\beta\bq}.
\end{multline} 
As announced, $E^{\textrm{Allen}}$ is a contribution to $E^{\rm ph}$.
The factor two in front of $E^{\textrm{Allen}}$ in Eq.~\eqref{E2ph2=EAllen+} might seem surprising.
It comes from the fact that the zero-point energy of phonons includes both a kinetic energy and a potential energy contribution, which are actually equal, while the zero-point renormalization of eigenvalues, Eq.~\eqref{eq:Deig2}, does not include a kinetic contribution but contains the curvature of $\varepsilon_{n\bk}$, which acts as a potential energy.

Therefore, Allen's energy defined in Ref.~\onlinecite{Allen2022} is not an electron-phonon energy but a part of the phonon contribution to the total energy.
We can inquire about the origin of the problem in Ref.~\onlinecite{Allen2022}. 
Apparently, the issue stems from the use of an effective Hamiltonian, with separate electronic and phononic terms fixed \emph{a priori}, to which an electron-phonon interaction is added, changing the total energy. 
As derived in the present work, following the BO approach, the first-principles BO-DFT phonons come from the BO curvature, \textit{which already includes the electronic energy curvature}, in addition to nuclear-nuclear energy curvature, and Hartree and exchange-correlation corrections.
Hence, phonons cannot be separated from electrons \emph{a priori}, unlike in the effective Hamiltonian of Ref.~\onlinecite{Allen2022}.

\section{Connection with the Allen-Heine-Cardona theory}
\label{sec:connection}

Despite $E^{\textrm{Allen}}$ not being the desired electron-phonon interaction energy, one can nonetheless study it and examine its connection with the Allen-Heine-Cardona (AHC) theory of zero-point renormalization.
Using Ref.~\onlinecite{Ponce2015}, $\Delta \varepsilon_{n\bk}$ in Eq.~\eqref{eq:EAllen3} can be decomposed as: 
\begin{equation}\label{eq:EtotNRIA}
\Delta \varepsilon_{n\bk} = \Delta \varepsilon_{n\bk}^{\rm AHC} + \Delta \varepsilon_{n\bk}^{\rm NRIA},  
\end{equation}
where $\Delta \varepsilon_{n\bk}^{\rm AHC}$ is obtained from the AHC theory~\cite{Allen1976,Allen1981} with the rigid-ion approximation~\cite{Ponce2014a} to describe the Debye-Waller term. 
$\Delta \varepsilon_{n\bk}^{\rm AHC}$ has the following sum-over-state expression~\cite{Gonze2011, Ponce2015}
\begin{multline}\label{eq:ahceq}
\Delta \varepsilon_{n\bk}^{\rm AHC} =  \frac{1}{2}\sum_{\nu} 
\int_{\rm BZ} \frac{\rm d \bq}{\Omega^{\rm BZ}}  \sum_{\kappa\alpha\kappa'\beta} \bigg[ U_{\kappa\alpha\nu\bq}^*U_{\kappa'\beta\nu\bq} \\
\times  \sum_{m}{}^{'} \frac{g_{mn\kappa\alpha}^*(\bk,\bq) g_{mn\kappa'\beta}(\bk,\bq)}{\varepsilon_{n\bk} - \varepsilon_{m\bk+\bq} + i\eta}  \\
- \frac{1}{2} [ U_{\kappa\alpha\nu\bq}^* U_{\kappa\beta\nu\bq} + U_{\kappa'\alpha\nu\bq}^*U_{\kappa'\beta\nu\bq}]  \\
\times  \sum_{m}{}^{'} \frac{ g_{mn\kappa\alpha}^*(\bk,\boldsymbol{\Gamma}) g_{mn\kappa'\beta}(\bk,\boldsymbol{\Gamma})}{\varepsilon_{n\bk} - \varepsilon_{m\bk} + i\eta}\bigg],
\end{multline}
where $\eta$ should be a positive infinitesimal in the AHC theory. 
Beyond AHC, $\eta$ would effectively describe the lifetime of the quasiparticle state~\cite{Lihm2024}. 
In this work, without computing $\eta$, it is chosen to be 5~meV.

The non-rigid-ion approximation (NRIA) $\Delta \varepsilon_{n\bk}^{\rm NRIA}$ terms in Eq.~\eqref{eq:EtotNRIA} have been shown to be small in solids~\cite{Ponce2014a,Ponce2025a}. 
Computing them requires the second-order derivative of the Hamiltonian, not available with current DFPT implementation, that reads: 
\begin{multline}\label{eq:NRIAterm}
\Delta \varepsilon_{n\bk}^{\rm NRIA} = \sum_{\nu}  \int_{\rm BZ} \frac{\rm d \bq}{\Omega^{\rm BZ}} \! \sum_{l\alpha\beta,\kappa \neq \kappa'} \! \frac{e^{-i\bq \cdot \mathbf{R}_l }}{4 \omega_{\nu\bq}} U_{\nu\kappa'\beta\bq}^*U_{\nu\kappa\alpha\bq}   \\
 \times \int  \textrm{d} \br \, \phi_{n\bk}^{*}(\br)\phi_{n\bk}(\br)  \frac{\partial^2 V^{\rm Hxc}(\br)}{\partial R_{0\kappa\alpha}\partial R_{l\kappa'\beta}}  ,
\end{multline}
where $V^{\rm Hxc}(\br)$ is the Hartree and exchange-correlation potential. 
Therefore, at the DFPT level, one cannot currently compute $E^{\textrm{Allen}}$, but only its AHC approximation,
\begin{equation}\label{eq:Allen(AHC)}
E^{\rm Allen,AHC}  \equiv \sum_{n}^{\rm occ} \int_{\rm BZ} \frac{\text{d}\bk}{\Omega^{\rm BZ}}  \Delta \varepsilon_{n\bk}^{\rm AHC},
\end{equation}
with possible spin doubling in the unpolarized case.

\section{Validation for diamond at the phonon zone-center}
\label{sec:validation}

In this section, we compute different quantities mentioned in Secs.\ref{sec:ahc}-\ref{sec:connection} using diamond as an example, and compare results from  perturbation theory with those from finite differences (FD - frozen phonon approach). 
Although the numbers will first be obtained for a single wavevector at the zone center, the procedure
is defined for a grid of wavevectors compatible with a supercell of the primitive cell.

Let $N$ be the size ratio between the primitive cell and the supercell, the latter being commensurate with the former, and consider similarly that there are $N$ allowed $\bq$ wavevectors that match periodic boundary conditions in the supercell.
We define $E^{\rm BO,sup}$, the BO energy of the supercell, that will be evaluated with atoms collectively displaced from their equilibrium position. 
Precisely, the atom labeled $\kappa$ in the cell $l$, with equilibrium
position $\mathbf{R}_{l\kappa}^0$, will be displaced by $h$ times the eigendisplacement
$\boldsymbol{U}_{\nu\kappa\bq}$, modulated by the phase $e^{-i\bq \cdot \mathbf{R}_l}$, giving the displaced position
\begin{equation}
\mathbf{R}_{l\kappa}(h)=\mathbf{R}_{l\kappa}^0+ h\boldsymbol{U}_{\nu\kappa\bq}e^{-i\bq \cdot \mathbf{R}_l }.
\end{equation}
We refer the reader to Ref.~\onlinecite{Ponce2014}, Eqs.~(70) and (71),
for the treatment of complex displacements, using linear combinations of such displacements.
The phonon frequencies are obtained using the following FD formulation:
\begin{multline}\label{eq:omega2FD}
\omega_{\nu\bq}^2 = \frac{1}{Nh^2} \Big( E^{\rm BO,sup} [\{\mathbf{R}_{l\kappa}(h)\}] \\
+ E^{\rm BO,sup}[\{\mathbf{R}_{l\kappa}(-h)\}] - 2 E^{\rm BO,sup}[\{\mathbf{R}_{l\kappa}^0\}] \Big).
\end{multline}
This FD value can be introduced in a discretized version of Eq.~\eqref{eq:E2qalternative} to obtain $E^{\rm ph}$.
Alternatively, 
\begin{multline}\label{eq:E2BOFD}
E^{\rm ph} =  \sum_{\nu\bq} \frac{1}{2\omega_{\nu\bq}N^2h^2} \Big( E^{\rm BO,sup}
[\{\mathbf{R}_{l\kappa}(h)\}] \\
+E^{\rm BO,sup}[\{\mathbf{R}_{l\kappa}(-h)\}]- 2 E^{\rm BO,sup}
[\{\mathbf{R}_{l\kappa}^0\}] \Big). 
\end{multline} 
In this expression, one of the factors $N$ comes from the number of $\bq$ vectors, while the other comes from the reduction of $E^{\rm BO,sup}$ to the primitive cell.
Similarly, the change of eigenvalues due to electron-phonon interaction can be computed with FD as
\begin{multline}
\Delta \varepsilon_{n\bk} =  \frac{1}{2N}\sum_{\nu\bq} \sum_{n}^{\rm occ} \frac{1}{2\omega_{\nu\bq}}  \frac{1}{h^2} \Big( \varepsilon_{n\bk}[\{\mathbf{R}_{l\kappa}(h)\}]  \\
+ \varepsilon_{n\bk}[\{\mathbf{R}_{l\kappa}(-h)\}] -2 \varepsilon_{n\bk}[\{\mathbf{R}_{l\kappa}^0\}] \Big). 
  \label{eq:FDvarepsilon}
\end{multline} 

Allen's total energy has two equivalent formulations using FD:
\begin{align}\label{eq:EAllenFD1}
E^{\rm Allen} =&  \frac{1}{2}\sum_{ \nu\bq} \frac{1}{2\omega_{\nu\bq}N^2h^2} \Big( E^{\rm e,sup}[\{\mathbf{R}_{l\kappa}(-h)\}]  \nonumber\\
  &+  E^{\rm e,sup}[\{\mathbf{R}_{l\kappa}(h)\}] - 2 E^{\rm e,sup}[\{\mathbf{R}_{l\kappa}^0\}] \Big) \\
  =& \sum_{n}^{\rm occ} \int \frac{\text{d}\bk}{\Omega^{\rm BZ}}  \Delta \varepsilon_{n\bk}, \label{eq:EAllenFD2}
\end{align} 
with a possible spin doubling in the unpolarized case for the last equation.
The remaining contribution is:
\begin{multline}\label{eq:E2otherFD}
E^{\rm ph,other} = \sum_{\nu\bq} \frac{1}{2\omega_{\nu\bq}N^2h^2}
\Big( E^{\rm other,sup}[\{\mathbf{R}_{l\kappa}(h)\}]  \\
  +  E^{\rm other,sup}[\{\mathbf{R}_{l\kappa}(-h)\}] - 2 E^{\rm other,sup}[\{\mathbf{R}_{l\kappa}^0\}] \Big). 
\end{multline}

Finally, Eq.~\eqref{eq:NRIAterm} can also be computed using finite differences. 
We first define
\begin{multline}
\mathbf{R}_{l\kappa}(h_1,\kappa_1,h_2,\kappa_2) \equiv \mathbf{R}^0_{l\kappa} \\
+ h_1\delta_{\kappa\kappa_1}\boldsymbol{U}_{\nu\kappa_1\bq}
+ h_2\delta_{\kappa\kappa_2}\boldsymbol{U}_{\nu\kappa_2\bq}e^{-i\bq \cdot \mathbf{R}_l }, 
\end{multline}
and obtain
\begin{align}\label{eq:NRIAtermFD}
\Delta \varepsilon_{n\bk}^{\rm NRIA} = \sum_{\nu\bq} & \frac{1}{4 \omega_{\nu\bq}4N^2h^2} 
\sum_{l\kappa_1\neq\kappa_2} \int_{\textrm{sup}} \! \textrm{d} \br \phi_{n\bk}^{*}
(\br) \phi_{n\bk}(\br) \nonumber\\
 \times  \Big\{ \,\,\,\, & V^{\rm Hxc}[\mathbf{R}_{l\kappa}(h,\kappa_1,h,\kappa_2)](\br)  \nonumber \\
-&V^{\rm Hxc}[\mathbf{R}_{l\kappa}(h,\kappa_1,-h,\kappa_2)](\br) \nonumber \\
-& V^{\rm Hxc}[\mathbf{R}_{l\kappa}(-h,\kappa_1,h,\kappa_2)](\br) \nonumber \\
+&V^{\rm Hxc}[\mathbf{R}_{l\kappa}(-h,\kappa_1,-h,\kappa_2)](\br) \Big\},
\end{align}
where the two arguments of $V^{\rm Hxc}$ indicate that two nuclei are moved in the FD approach, and where the integral is performed in the supercell.
This term gives the corresponding total energy:
\begin{equation}\label{eq:Allen(NRIA)}
E^{\rm Allen,NRIA} \equiv \sum_{n}^{\rm occ} \int_{\rm BZ} \frac{\text{d}\bk}{\Omega^{\rm BZ}}  \Delta \varepsilon_{n\bk}^{\rm NRIA},
\end{equation}
with possible spin doubling in the unpolarized case.

In this section, we consider the $\mathbf{q}$=$\Gamma$ case so that finite difference calculations can be performed only in the primitive cell. 
We will use the finite $\mathbf{q}$ formulation in Sec.~\ref{sec:comparison}. 
Calculations are performed with the \textsc{Abinit} v10.4.5 software~\cite{Gonze2016,Gonze2020,Verstraete2025} with a 40~Ha plane-wave energy cutoff and a zone-centered 8$\times$8$\times$8 $\bk$-point grid.
We rely on the  Generalized Gradient Approximation (GGA) PBE~\cite{Perdew1996} exchange-correlation functional, with 
the carbon scalar-relativistic standard PBE pseudopotential from \textsc{PseudoDojo}~\cite{vanSetten2018} v0.5. 
We compute a relaxed 6.7513~Bohr lattice parameter, slightly overestimating the 6.74~Bohr experimental value~\cite{Surratt1973}.   
Diamond is a non-polar material with a long-range quadrupole value of 2.46~$e$Bohr~\cite{Ponce2021}. 
We note that the calculation of the quadrupole tensor in solids from perturbation theory has so far only been implemented in the \textsc{Abinit} software~\cite{Royo2019,Gonze2020}. 
Since \textsc{Abinit} v10.4.5, the calculations of quadrupole tensor supports nonlinear core correction (NLCC) of pseudopotentials and has been used here.
We obtain a quadrupole value of 2.502~$e$Bohr using a 16$\times$16$\times$16 $\bk$-point grid.

This recent development is crucial because the use of a model core charge has been shown to be important~\cite{Louie1982} in the case of generalized gradient approximation pseudopotentials such as PBE. 
In fact, when the valence and core densities overlap, neglecting the core density results in an artificially large gradient of the density.
We have numerically tested this effect for diamond and found that ignoring the NLCC leads to numerically unstable finite-difference calculations but is acceptable for perturbation theory. 
Unless specified otherwise, we use the above-mentioned pseudopotential which includes NLCC.

\begin{table}[ht]
  \caption{\label{tab:table1}
Comparison for diamond between the finite difference (FD) and density functional perturbation theory (DFPT) for different quantities at $\mathbf{q}$=$\boldsymbol{\Gamma}$ and T=0~K, at the optimized BO lattice parameters.
The $\Delta \varepsilon_{\boldsymbol{\Gamma}}$ refers to the band zero-point renormalization of the triply degenerate valence band maximum, $\Delta \varepsilon_{\boldsymbol{\Gamma}}^{\rm NRIA} $ is the non-rigid ion approximation (NRIA), and
$\Delta \varepsilon_{\boldsymbol{\Gamma}}^{\rm AHC} $ refers to the Allen-Heine-Cardona (AHC) theory using the rigid-ion approximation. 
The 0$^{\rm th}$-order Born-Oppenheimer (BO) total energy is given by $E^{\rm BO}$ and include the electronic part is given by $E^{\rm e}$, and the \emph{other} terms $E^{\rm other}$. 
The second-order contribution to total energy due to phonons is $E^{\rm ph}$.
It can be decomposed into half the electronic contribution $E^{\rm Allen}$, and the rest $E^{\rm ph,other}$. 
The electron-phonon part of the fourth-order contribution to total energy $E^{\rm elph}$ can be computed using DFPT. 
Energies are given per unit cell (2 atoms).
For total energies, we use a 8$\times$8$\times$8 $\mathbf{k}$-point grid with symmetry, detailed in Table~\ref{tab:table2}.
Absolute total energy values depend on the choice of pseudopotential but relative energy differences are observables.  
}
\begin{tabular}{ l c c  r }
  \toprule\\
  & Evaluation  & Equation & Energy  \\
 &   &  &  (meV)  \\
\hline
$\omega_{\boldsymbol{\Gamma}}$ & DFPT     & \eqref{eq:secular}               & 159.989 \\
$\omega_{\boldsymbol{\Gamma}}$ & FD       & \eqref{eq:omega2FD}               & 160.002 \\
$\Delta \varepsilon_{\boldsymbol{\Gamma}}$ & FD & \eqref{eq:FDvarepsilon}    & 27.695 \\
$\Delta \varepsilon_{\boldsymbol{\Gamma}}^{\rm NRIA} $ & FD & \eqref{eq:NRIAtermFD} & 3.894 \\
$\Delta \varepsilon_{\boldsymbol{\Gamma}} $-$ \Delta \varepsilon_{\boldsymbol{\Gamma}}^{\rm NRIA} $ & FD  & \eqref{eq:FDvarepsilon} \!-\! \eqref{eq:NRIAtermFD}  &  23.801  \\
$\Delta \varepsilon_{\boldsymbol{\Gamma}}^{\rm AHC} $ & DFPT & \eqref{eq:ahceq}    & 23.806 \\
\hline
$E^{\rm BO}$         & DFT           & \eqref{eq:decomposition} & -327559.404  \\
$E^{\rm e}$  & DFT  & \eqref{eq:epart}         &   10119.176 \\
$E^{\rm other}$             & DFT             & \eqref{eq:decomposition} \!-\! \eqref{eq:epart} & -337678.580 \\
$E^{\rm ph}$  & DFPT                     & (\ref{eq:E2qalternative}, \ref{eq:secular}) &  240.040   \\
$E^{\rm ph}$  & FD                       & (\ref{eq:E2qalternative}, \ref{eq:omega2FD}) &  240.023   \\
$E^{\rm ph}$  & FD                       & (\ref{eq:omega2FD}, \ref{eq:E2BOFD}) &  240.003   \\
$E^{\rm Allen}$          & FD             & \eqref{eq:EAllenFD1}    &  -84.214   \\
$E^{\rm Allen}$          & FD             & \eqref{eq:EAllenFD2}    &  -84.214   \\
$E^{\rm ph,other}$      & FD             & (\ref{eq:E2qalternative}, \ref{eq:omega2FD}) \!-\! 2\eqref{eq:EAllenFD2} &  408.451  \\
$E^{\rm ph,other}$   & FD             & \eqref{eq:E2otherFD} &   408.466  \\
$E^{\rm Allen,NRIA}$         & FD        & \eqref{eq:Allen(NRIA)}      &   17.086    \\
$E^{\rm Allen}  $-$ E^{\rm Allen,NRIA}$ & FD & \eqref{eq:EAllenFD2} \!-\! \eqref{eq:Allen(NRIA)}   & -101.300    \\
$E^{\rm Allen,AHC}$      & DFPT           & \eqref{eq:Allen(AHC)}      & -100.783    \\
$E^{\rm elph}$     & DFPT             & \eqref{eq:fourthorder2}   &   3.639   \\
$E^{\rm elph}$     & DFPT             & \eqref{eq:fourthorder3}   &   3.638   \\
  \botrule
\end{tabular}
\end{table}

We start by verifying phonon frequencies computed by perturbation theory and finite difference (FD) in Table~\ref{tab:table1} using a 8$\times$8$\times$8 $\bk$-grid, since phonon frequencies will be used in later quantities.  
All finite-difference calculations are performed using a Richardson extrapolation with Romberg order 4, starting from $h$=2. 
We find that the triply degenerate optical modes agree within 0.013~meV between the density functional perturbation theory (DFPT)~\cite{Gonze1997,Baroni2001} and FD values. 
We then compute the phonon zone-center contribution to the zero-point renormalization (ZPR) of the valence band maximum (VBM) computed using FD and obtain 27.695~meV, close to the value of 28.428~meV of a prior study~\cite{Ponce2014a}.
Since the VBM is triply degenerate in band indices and any linear combination is a valid one, we report the average value over the three bands.  
The same value cannot be directly computed using DFPT as it requires a higher-order perturbation currently not available in first-principles software. 
However, it can be approximated using the RIA which gives 23.806~meV, again in close agreement with the 24.830~meV value of a previous study~\cite{Ponce2014a}.
The neglected NRIA term can be computed using FD, and then added to the RIA term giving 27.700~meV, in very close agreement with the direct FD value. 
We note that the RIA is a good approximation for solids when integrated over $\mathbf{q}$-points~\cite{Ponce2014a,Ponce2025a}.
As a technical note, it is possible in \textsc{Abinit} to reduce the computation cost by using a reduced ratio between the radius of the $\mathbf{G}$-vector sphere that can be inserted in the real-space fast Fourier transform box and the $\mathbf{G}$-vector sphere used to represent the wavefunction.
However, we find here that it is crucial to use a ratio of 2 to have an exact density, see Sec.~\ref{app:improvement} of the SI.

Next, we report in Table~\ref{tab:table1} the zeroth-order BO total energy.
Such total energy, provided as an output of a standard self-consistent DFT cycle, contains a core electron energy, that is assumed constant, and ignored via the use of pseudopotential.  
Therefore, the total energy depends on the arbitrary choice of the pseudopotential. 
In contrast, the contribution from its eigenenergy component $E^{\rm e}$ is only weakly dependent on the choice of pseudopotential, in a way similar to other predicted observables with DFT. 
We compute a value of 10.119~eV directly from the output of the \textsc{Abinit} software and verify that it gives the same result as the direct sum over the occupied eigenvalues $\sum_{n\bk}\varepsilon_{n\bk}f_{n\bk}$.
We then compute the second-order contribution to the total energy due to phonons and find a value of 240.040~meV using DFPT, which is less than 0.04~meV larger than the two FD results.  
We see that such a contribution is significantly lower than the $E^{\rm e}$ energy contribution. 

To assess the eigenenergy contribution to that second-order energy, we compute the ZPR for all $\mathbf{k}$-point of an homogeneous 8$\times$8$\times$8 $\bk$-grid using crystal symmetry, and report their values in Sec.~\ref{sec:zpr} of the SI.
We find a close agreement between perturbation theory and FD, with less than 1.5~meV difference. 
From its weighted summation, we find that the eigenenergy contribution to the second-order energy is -84.214~meV, which amounts to over one-third the total second-order contribution.
This value can be approximated via DFPT and the RIA which gives $-100.783$~meV, overestimating by 20\% the true value. 
As mentioned earlier, we expect the RIA to improve when summing over dense $\mathbf{q}$-points grids.

\begin{figure}[t]
  \centering
  \includegraphics[width=0.99\linewidth]{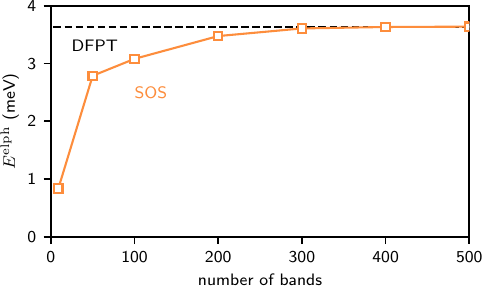} %
  \caption{\label{fig:sos}
Diamond fourth-order total energy contribution (per unit cell) computed using DFPT, Eq.~\eqref{eq:fourthorder2}, and with a sum-over-state (SOS) expression, Eq.~\eqref{eq:fourthorder3}.  
The $\mathbf{k}$-point grid used is 8$\times$8$\times$8 with only a single $\mathbf{q}=\boldsymbol{\Gamma}$ point. 
}
\end{figure}

Furthermore, we compute the electron-phonon contribution to the fourth-order total energy using DFPT, $E^{\rm elph}$, Eq.~\eqref{eq:fourthorder2}, and obtain 3.639~meV.
Alternatively, we can use Eq.~\eqref{eq:fourthorder3} to compute the same quantity via the electron-phonon interaction matrix elements. 
As shown in Fig.~\ref{fig:sos}, both result in the same value but Eq.~\eqref{eq:fourthorder3} requires a sum over empty states which converges slowly.
In the case of diamond, we need at least 300 bands to converge that formulation. 
Such a small value is nearly two orders of magnitude smaller than the second-order contribution, which is expected given the $M_0^{-1/2}$ scaling between each higher-even order terms.  
Although small, this contribution could play a role in stabilizing an allotrope over another, especially in the case of magnetic materials that can have a very shallow energy landscape.

Finally, it was argued in Ref.~\onlinecite{Paul2023} that the DFPT formulation of AHC zero-point renormalization, $\Delta \varepsilon_{n\mathbf{k}}^{\rm AHC}$  violates size-consistency, whereby the total energy $E^{\rm Allen, AHC}$ would depend on the unit cell size.
To show the size consistency of Eq.~\eqref{eq:Allen(AHC)}, we simply need to prove the size consistency of $\Delta \varepsilon_{n\bk}^{\rm AHC}$.
For this, we compute the adiabatic zero-point renormalization using the rigid-ion approximation for the Debye-Waller term~\cite{Ponce2014a} and a Fourier interpolation of the perturbed potential~\cite{Eiguren2008,Gonze2020}.
We evaluate the diagonal~\cite{Lihm2020} self-energy at the DFT eigenvalues (on-the-mass shell) and use a Sternheimer expression to replace the sum-over-state expression~\cite{Gonze2011,Ponce2015}, and report the details in Sec.~\ref{sec:size} of the SI. 
We find excellent agreement, validating size consistency.

\section{Diamond and Lonsdaleite allotropes}\label{sec:comparison}

Having analyzed and compared the behavior of the theory and implementation in different cases, we now converge the various contributions to the total energies with phonon momentum $\bq$.
In this case, we can no longer use FD methods as supercell size quickly becomes prohibitively larger. 
First, we remark that the various total energy calculations require dense phonon momentum $\bq$ grid samplings of the electron-phonon interaction, which can be computationally expansive. 
Two interpolation methods are often used to accurately and efficiently interpolate $\bq$-dependent properties: the Wannier-Fourier interpolation of the electron-phonon matrix elements~\cite{Giustino2007,Ponce2016} and the Fourier interpolation of the perturbed potential~\cite{Gonze2020,Brunin2020,Brunin2020a}.
Importantly, long-range electrostatics should analytically be removed and added back to ensure the accuracy of the interpolation~\cite{Gonze1997,Verdi2015,Sjakste2015,Stengel2016,Royo2019,Royo2020,Brunin2020,Brunin2020a,Jhalani2020,Royo2021,Ponce2021,Ponce2023,Ponce2023a}. 

\begin{table}[b]
  \caption{\label{table4}
Total energy contributions for diamond and lonsdaleite, at the BO-DFT optimized geometries, with a 40~Ha cutoff.
For diamond, we use a 8$\times$8$\times$8 $\bk$-point grids and a 8$\times$8$\times$8 $\bq$-point grids interpolated to a fine 64$\times$64$\times$64 $\bq$-grid.
For londsdaleite, we use a 6$\times$6$\times$4 $\bk$-point grids and a 6$\times$6$\times$4 $\bq$-point grids interpolated to a fine 60$\times$60$\times$40 $\bq$-grid. 
The energies are reported per 2 atoms (diamond primitive cell). 
We also provide the band-by-band decomposition of $E^{\rm elph}$. 
The total energy presented are computed at the BO atomic positions $\{ \bR_0^0 \}$ while $\Delta E^{\rm QH}$ is the total energy difference
between the atomic position corresponding to the energy minimum of $E^{\rm BO} \!+\! E^{\rm ph} \!+\! E^{\rm elph} \!+\! E^{\rm elm}$ and 
the same energy at $\{ \bR_0^0 \}$.
}
\begin{tabular}{ l r r }
  \toprule\\
Total energy term & diamond (meV) & londsdaleite (meV)  \\
\hline                                      
$E^{\rm BO}$                             &  -327559.216 & -327505.244 \\
$E^{\rm e}$                              &    10119.176 &   13468.413  \\
$E^{\rm other}$                          &  -337678.580 & -340973.863 \\
$E^{\rm ph}$                             &      358.685 &     357.381 \\
$E^{\rm Allen, AHC}$                     &     -104.189 &    -112.906 \\
$E^{\textrm{elph}}$                      &        7.600 &       7.608 \\
$E^{\textrm{elm}}$                       &       -0.049 &      -0.049  \\
\hline
$E_1^{\rm elph}$                     &        0.509 &       0.517 \\
$E_2^{\rm elph}$                     &        1.510 &       1.476 \\
$E_3^{\rm elph}$                     &        2.516 &       2.419 \\
$E_4^{\rm elph}$                     &        3.066 &       3.197 \\
\hline
$E^{\rm BO} \!+\! E^{\rm ph} \!+\! E^{\rm elph} \!+\! E^{\rm elm}$ & -327192.969 & -327140.304  \\
Relative difference                  & -52.691 & 0 \\
$\Delta E^{\rm QH}$                  & -2.103 & - \\
  \botrule
\end{tabular}
\end{table}

To analyze the importance of the different terms, we compute these for another allotrope of diamond called \emph{lonsdaleite} or \emph{hexagonal diamond}~\cite{Bundy1967} which forms naturally under high-pressure and high-temperature and is typically found in meteorite impacts. 
It crystallizes in the P6$_3$/mmc space group with 4 atoms per primitive cell.
We computed a DFT-BO crystal structure of a=4.7477~Bohr and c=7.9011~Bohr, slightly above the experimental lattice values of a=4.743~Bohr and c=7.786~Bohr~\cite{Bundy1967}.
Our computed fractional internal parameter is u=0.374535.
The volume per atom is similar for diamond and lonsdaleite, with 38.465~Bohr$^3$/atom and 38.560~Bohr$^3$/atom, respectively.
We compute the quadrupole tensor of lonsdaleite integrated on a 18$\times$18$\times$12 $\textbf{k}$-point grid and obtain the non-equivalent values: 
Q$_{\kappa y xx}$ = -1.643~e Bohrs, 
Q$_{\kappa z xx}$ = -1.663~e Bohrs,
Q$_{\kappa x xz}$ = -1.018~e Bohrs, and
Q$_{\kappa z zz}$ = 2.775~e Bohrs.

\begin{figure}[t]
  \centering
  \includegraphics[width=0.99\linewidth]{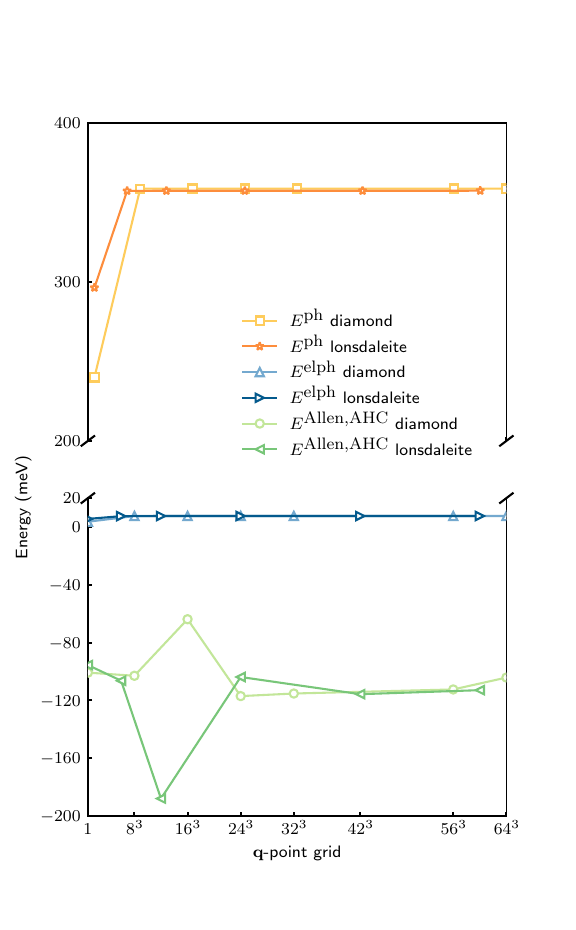} 
  \caption{\label{fig:convergence}
Diamond and lonsdaleite convergence rates for the second- and electron-phonon fourth-order contributions to the total energy (per 2 atoms) with respect to $\mathbf{q}$-point grid integration. 
The $\mathbf{k}$-point grid used is 8$\times$8$\times$8 and 6$\times$6$\times$4 for lonsdaleite. 
In the case of lonsdaleite the reported grid size correspond to the first two numbers, e.g. 24$^3$ means a $\mathbf{q}$ grid of 24$\times$24$\times$16.
}
\end{figure}

Prior DFT calculations have reported that diamond is more stable than lonsdaleite by 2.6 kcal/mol~\cite{Jones2016}.
As reported in Table~\ref{table4}, we also find diamond to be more stable but only by -53.964~meV/(2 atoms), which corresponds to 0.6222~kcal/mol.
%
 We then study the second (phononic) and fourth-order (electron-phonon only) contributions to the total energy using DFPT and in particular the formulation of Eq.~\eqref{eq:fourthorder2} for $E^{\rm elph}$. 
We report in Fig.~\ref{fig:convergence} the values of $E^{\rm ph}$, $E^{\rm Allen,AHC}$, and 
$E^{\rm elph}$ as a function of momentum $\mathbf{q}$-grid integration. 
We also integrate on $\mathbf{k}$-points using crystal symmetries and a 8$\times$8$\times$8 and 6$\times$6$\times$4 grids for diamond and lonsdaleite, respectively. 
We verify that increasing the $\mathbf{k}$-point grids changes the results by less than 1\%. 
Both $E^{\rm ph}$ and $E^{\rm elph}$ converge quickly with $\mathbf{q}$-point grids but $E^{\rm Allen,AHC}$ requires very dense grids. 
This is not surprising, as this term is obtained by integrating the eigenvalue renormalization, a quantity that is known to converge slowly with phonon momentum grids~\cite{Ponce2015,Ponce2025a}.

We find the second-order energy 
$E^{\rm ph}$ per 2 atoms to be 358.7~meV and 357.4~meV for diamond and londsdaleite, respectively. 
Interestingly, these values are very close, which is also the case for their $E^{\rm Allen,AHC}$ decomposition, with diamond having a value of -104.2~meV while londsdalite is -112.9~meV.  
Finally, we look at the fourth-order contribution and find that $E^{\rm elph}$ is similar in both cases, finding a value of 7.600~meV for diamond and 7.608~meV for lonsdaleite.
In addition, and as expected, we see in Table~\ref{table4} that the band decomposition of $E^{\rm elph}$ yields the largest contribution from the highest energy bands. 
However, interestingly, the difference in the band decomposition between diamond and londsdaleite is greater than their sum, indicating bandstructure differences between these two materials.

\begin{figure}[t]
  \centering
  \includegraphics[width=0.99\linewidth]{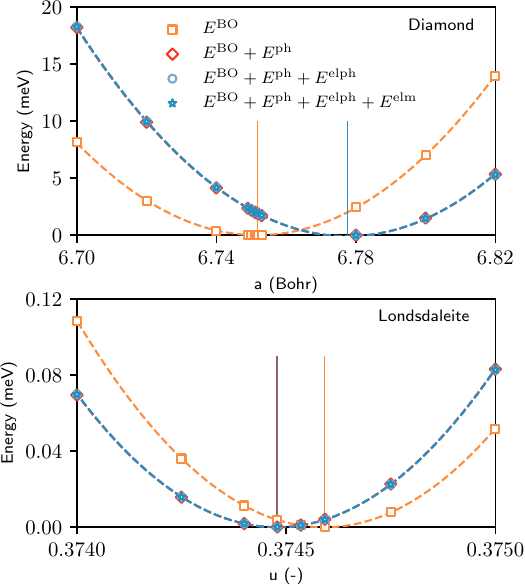} 
  \caption{\label{fig:forces}  
Diamond energy dependence on the lattice parameter $a$ and lonsdaleite energy dependence on the internal parameter $u$.  
All the energies are presented with their minimal value adjusted to 0.
Vertical lines indicates the position of the energy minimum.
}
\end{figure}

Overall, when accounting for phononic and electron-phonon interaction up to the 4$^{\rm th}$ order, we find that londsdalelite becomes a little more stable with respect to diamond than the straight BO value, going from -54.015~meV/(2 atoms) to -52.691~meV/(2 atoms), which contributes marginally to explaining why lonsdalelite is stable and can be naturally observed. 
We remark that this stability will be modified at finite temperature, which is outside the scope of this study.
Also, we did not compute the anharmonic fourth-order contribution to the total energy.
We can still expect the relative difference to remain similar given the similarity in $E^{\rm ph}$ between the two materials. 

We note that among all the $\lambda^6$ order energy terms presented in Table~\ref{tab:table0}, the effect of the $\mathcal{E}^{\rm (6),elm}$ term given in Eq.~\eqref{eq:6el} can be easily computed, approximately, by renormalizing the $E^{\rm ph}$ term as:
\begin{equation}
    E^{\rm elm} \equiv E^{\rm ph} \bigg( \sqrt{\frac{M_\kappa}{M_\kappa + n_{e\kappa}m_e}} -1 \bigg).
\end{equation}
In this approximation, we consider the $\Delta m_{\kappa\alpha,\kappa'\alpha'}^{\textrm{elm},-1}$
tensor to be diagonal, isotropic for each nucleus, with the value needed to fulfill the sum rule mentioned in Eq.~(17) of Ref.~\onlinecite{Scherrer2017}, namely,
$n_{e\kappa}=6$ for carbon nuclei.
We report it in Table~\ref{table4} and find the very small rounded value of -0.049~meV for both allotropes, which therefore does not affect their relative stability.

We conclude this work by computing the forces and strain resulting from the addition of the phonon energy, $E^{\rm ph}$, electron-phonon energy $E^{\rm elph}$ and electronic mass renormalization energy $E^{\rm elm}$
when the atoms are located at $\{ \mathbf{R}^0 \}$.

Indeed, as already discussed at the end of Sec.~\ref{sec:phonon}, the $\{ \mathbf{R}^0 \}$ atomic positions are at their BO minimum, but might no longer be at their minimum once the additional energy terms are included. 
This is indeed the case as seen in Fig.~\ref{fig:forces}, and the diamond lattice parameter increases from 6.7518~Bohr to 6.7776~Bohr when including the three additional forces, which represent a 0.4\% increase. 
This term is known as zero-point lattice expansion~\cite{BrousseauCouture2022}.
We emphasize that this is a small but non-negligible increase that occurs at zero Kelvin and will be much larger with the temperature increase (thermal expansion). 
For londsdaleite, a similar increase would occur, and we do not report it. 
However, in contrast to diamond, londsdaleite has an internal parameter $u$ and we find in Fig.~\ref{fig:forces} a reduction of the internal parameter from  0.37459 to 0.37448.
Moreover, the lowering of the energy between the optimized atomic position and $\{ \mathbf{R}^0 \}$ is the quasiharmonic energy, defined (for the terms that we address in the present study) as:
\begin{multline}
\Delta E^{\rm QH} \equiv \text{min}_{\{ \bR \}} ( E^{\rm BO} + E^{\rm ph} + E^{\rm elph} + E^{\rm elm} )_{\{ \bR \}} \\
- ( E^{\rm BO} + E^{\rm ph} + E^{\rm elph} + E^{\rm elm} )_{\{ \bR^0 \}},  
\end{multline}
which gives a decrease of 2.1~meV for the volume of diamond and a negligible decrease for the optimization of the optimal internal parameter $u$ of londsdaleite.

Overall, we expect that computing these additional forces will bring crystal structure predictions closer to those obtained from cryogenic temperature measurements.

\section{Conclusions}
\label{sec:conclusions}

In this work, we have examined which terms in the BO expansion might be called \emph{``electron-phonon contribution to the total energy''}.
The lowest-order possibility appears at fourth order in the small parameter of the Born-Oppenheimer expansion and scale as the inverse of the characteristic nuclear mass.
This contribution, not yet computed from first-principles to our knowledge, might become relevant when addressing small energy differences between competing phases, magnetic configurations, defects, or surfaces. 
We actually derive all the contributions in the BO expansion to the sixth order, even including the effect of time-reversal symmetry breaking with the appearance of a vector potential.
We clarified that the expression proposed by Allen in 2020, albeit called the electron-phonon contribution to the total energy, is in fact part of the second-order phonon contribution and should not be added independently. 
Building on this understanding, we derived and implemented a correct expression for the fourth-order electron-phonon correction, which we show to be size-consistent and compatible with standard density-functional theory workflows. 
Our implementation is validated against finite-difference calculations or the sum-over-states formulation and applied to the case of diamond and its hexagonal polymorph, lonsdaleite. 
We find that although the fourth-order electron-phonon contribution is small in absolute terms, it is non-negligible and on the order of several meV per atom. 
However, there is only a marginal difference between its values for the two allotropes. 
Overall, this work provides a practical and accurate route to include the lowest-order electron-phonon contribution to the total energy in first-principles simulations, enabling more precise energy comparisons in systems where quantum lattice effects play a significant role.


\begin{acknowledgments}
The authors acknowledge useful discussions with Dr. Matteo Giantomassi, Prof. Matthieu Verstraete, and participants to the online \emph{ALPS} meetings.  
S. P. acknowledges support from the Fonds de la Recherche Scientifique de Belgique (FRS-FNRS). 
This work was supported by the Fonds de la Recherche Scientifique - FNRS under Grants number T.0183.23 (PDR) and  T.W011.23 (PDR-WEAVE).
It is an outcome of the Shapeable 2D magnetoelectronics by design project (SHAPEme, EOSProjectNo. 560400077525) that has received funding from
the FWO and FRS-FNRS under the Belgian Excellence of Science (EOS) program. 
This publication was supported by the Walloon Region in the strategic axe FRFS-WEL-T.
Computational resources have been provided by the PRACE award granting access to MareNostrum4 at Barcelona Supercomputing Center (BSC), Spain and Discoverer in SofiaTech, Bulgaria (OptoSpin project id. 2020225411), by the Consortium des Équipements de Calcul Intensif (CÉCI), funded by the FRS-FNRS under Grant No. 2.5020.11, by the Tier-1 supercomputer of the Walloon Region (Lucia) with infrastructure funded by the Walloon Region under the grant agreement n°1910247, and by the Belgian share of the EuroHPC LUMI supercomputer.
\end{acknowledgments}

\bibliography{Bibliography}

\end{document}


\title{SI: In search of the electron-phonon contribution to total energy}

\author{Samuel Ponc\'e}
\email{samuel.ponce@uclouvain.be}
\affiliation{%
European Theoretical Spectroscopy Facility, Institute of Condensed Matter and Nanosciences, Université catholique de Louvain, Chemin des Étoiles 8, B-1348 Louvain-la-Neuve, Belgium. 	
}%
\affiliation{%
WEL Research Institute, avenue Pasteur, 6, 1300 Wavre, Belgium.		
}%
\author{Xavier Gonze}
\affiliation{%
European Theoretical Spectroscopy Facility, Institute of Condensed Matter and Nanosciences, Université catholique de Louvain, Chemin des Étoiles 8, B-1348 Louvain-la-Neuve, Belgium. 	
}%

\date{\today}

\maketitle

\section{Hybrid approximations}\label{sec:hybrid}

As discussed in the main text, the differences
between the scalar BO approximation and the optimal BO approximation consist in (i) adding the vector potential inside the kinetic energy operator and (ii) adding the Born contribution to the BO hypersurface. 
In the literature, in addition to these two approximations, two other BO based approximations had been used, and sometimes not distinguished
from the scalar or the optimal ones.

Indeed, before the work of Mead and Truhlar~\cite{Mead1979}, the Born
term was sometimes taken into account, while the vector potential was ignored~\cite{Born1954}, because it was supposed that a gauge choice could always make the vector potential vanish.
When the possibility of a vector potential was recognized, the reverse choice was then made by some: ignoring the Born term, while retaining the vector potential.
This gives two approximations that we term ``hybrid approximations". 
Explicitly,
\begin{multline}\label{eq:hyb1_Schreq}
\Big(\hat{T}^\textrm{N} + E^{\rm opt}_j\{\bR\}\Big) \chi_{ij}^{\rm hyb1} \{ \bR \}
= {\cal{E}}_{ij}^{\rm hyb1} \chi_{ij}^{\rm hyb1} \{\bR\}, 
\end{multline}
and  
\begin{equation}\label{eq:hyb2_Schreq}
\Big(\hat{T}^\textrm{vec}_{j}\{\bR\} \!+\! E^{\rm BO}_j\{\bR\}\Big) \chi_{ij}^{\rm hyb2} \{ \bR \}
= {\cal{E}}_{ij}^{\rm hyb2} \chi_{ij}^{\rm hyb2} \{\bR\},
\end{equation}
with the following approximate total wavefunctions:
\begin{align}\label{eq:psi_hyb1}
   \ket{\psi_i^{\rm hyb1} \{\bR\}} =& \chi_{ij}^{\rm hyb1} \{\bR\}\ket{ \phi_j\{\bR\}}  \\
   \ket{\psi_i^{\rm hyb2} \{\bR\}} =& \chi_{ij}^{\rm hyb2} \{\bR\}\ket{ \phi_j\{\bR\}}.
   \label{eq:psi_hyb2}
\end{align}

Different approximations based on the BO separation, beyond the four ``adiabatic" approximations, Eqs.~\eqref{eq:psi_sca},
\eqref{eq:optwf}, \eqref{eq:psi_hyb1}, and \eqref{eq:psi_hyb2}, have also been explored by Nafie~\cite{Nafie1983}, although the vector potential was not considered in that work.

\section{The total wavefunction factorizes up to second order in $\lambda$}\label{App:wf_factorizes}

In this section, we will show that 
\begin{align}
\label{eq:trialwf_app}
\ket{\Psi_{i\lambda} \{ \bR \} }  =
N_\lambda\vartheta_{\jmath j} \{ \Delta\bR_j / \lambda \}
\ket{ \phi_j \{\bR\}},
\end{align}
that has the factorized form of Eq.~\eqref{eq:adiabwf}, satisfies the Schrödinger equation 
\begin{equation}\label{eq:Schreq_lambda_app}
\hat{\mathcal{H}}_\lambda \{ \bR \} \ket{ \psi_{i\lambda} \{ \bR \} } =  \mathcal{E}_{i\lambda} \ket{\psi_{i\lambda} \{\bR\}},
\end{equation}
see Eq.~\eqref{eq:Schreq_lambda}, up to second-order in $\lambda$ and $\Delta\bR$ included. 
Explicitly,
\begin{equation}\label{eq:Schreq_lambda_app}
\hat{\mathcal{H}}_\lambda \{ \bR \} \ket{ \Psi_{i\lambda} \{ \bR \} } =  \mathcal{E}_{i\lambda} \ket{\Psi_{i\lambda} \{\bR\}}
+\mathcal{O}( \lambda^3,\Delta\bR^3),
\end{equation}
where
\begin{equation}\label{eq:Eilambda_Taylor_up2_app}
\mathcal{E}_{i\lambda}=E^{\rm BO}_{j|\mathbf{0}} + \lambda^2 \mathfrak{E}_{\jmath j}^{\textrm{ho}}+\mathcal{O}( \lambda^3).
\end{equation}
Referring to the main text, the index $i$ is associated to the pair of indices $\jmath$ (dotless $j$) and $j$, the first one characterizing the nuclear state and the second one characterizing the BO hypersurface (with the chosen reference minimum).
The nuclear wavefunction, $\vartheta_{\jmath j}$, is the eigenfunction of the standard multidimensional harmonic oscillator Hamiltonian, based on the curvature of the BO hypersurface (order 2 in the displacements), see Eqs.
~\eqref{eq:H_E_vib_first}-\eqref{eq:Hvib_Schreq},
\begin{equation} \label{eq:Hvib_Schreq_2}
\hat{\mathfrak{H}}_j^{\textrm{ho}}\{ \Delta\bR_j\} 
\vartheta_{\jmath j} \{ \Delta\bR_j\} = \mathfrak{E}_{\jmath j}^{\textrm{ho}} \vartheta_{j} \{ \Delta\bR\},
\end{equation}
with
\begin{equation} \label{eq:H_E_vib_first_2}
\hat{\mathfrak{H}}_j^{\textrm{ho}}\{\Delta \bR_j\} \equiv \sum_{\kappa\alpha}\frac{\hat{P}^2_{\kappa\alpha}}{2m_\kappa} 
+E^{\rm BO\{\mathbf{2}\}}_j\!\cdot\! \Delta\mathbf{R}_j^2.
\end{equation}

Indeed, neglecting terms of $\mathcal{O}( \lambda^3,\Delta\bR^3)$,
\begin{align}
 \hat{\mathcal{H}}_\lambda \{ \bR \} & \ket{ \Psi_{i\lambda} \{ \bR \} } \nonumber\\
=& \Bigg(\hat{H}^{\rm BO} \{ \bR \} + \sum_{\kappa\alpha} \frac{\lambda^4\hat{P}^2_{\kappa\alpha}}{2m_\kappa} \Bigg) N_\lambda \vartheta_{\jmath j} \bigg\{ \frac{\Delta\bR_j} {\lambda} \bigg\} \ket{ \phi_j \{\bR\}} \nonumber\\
=& \Bigg(E^{\rm BO}_j \{ \bR \} + \sum_{\kappa\alpha} \frac{\lambda^4\hat{P}^2_{\kappa\alpha}}{2m_\kappa}
\Bigg) N_\lambda \vartheta_{\jmath j} \bigg\{ \frac{\Delta\bR_j} {\lambda} \bigg\} \ket{ \phi_j \{\bR\}} \nonumber\\
\approx & \Bigg(E^{\rm BO}_{j|\mathbf{0}} + E^{\rm BO\{\mathbf{2}\}}_j\!\cdot\! 
(\Delta\mathbf{R}_j)^2 + \sum_{\kappa\alpha}\frac{\lambda^4\hat{P}^2_{\kappa\alpha}}{2m_\kappa} 
\Bigg) \nonumber \\
& \times N_\lambda \vartheta_{\jmath j} \bigg\{ \frac{\Delta\bR_j} {\lambda} \bigg\} \ket{ \phi_j \{\bR\}} 
\end{align}
which simplifies further as
\begin{align}
 \hat{\mathcal{H}}_\lambda \{ \bR \} & \ket{ \Psi_{i\lambda} \{ \bR \} } \nonumber\\
\approx & \Bigg( E^{\rm BO}_{j|\mathbf{0}} + \lambda^2 \bigg( E^{\rm BO\{\mathbf{2}\}}_j\!\cdot\! 
\Big(\frac{\Delta\bR_j} {\lambda}\Big)^2 + \sum_{\kappa\alpha}\frac{\lambda^2\hat{P}^2_{\kappa\alpha}}{2m_\kappa} 
\bigg) \Bigg) \nonumber\\ 
& \times N_\lambda \vartheta_{\jmath j} \bigg\{ \frac{\Delta\bR_j} {\lambda} \bigg\} \ket{ \phi_j \{\bR\}} \nonumber\\
\approx & \Bigg(E^{\rm BO}_{j|\mathbf{0}} + \lambda^2 \hat{\mathfrak{H}}^{\textrm{ho}}_j
\bigg\{ \frac{\Delta\bR_j} {\lambda} \bigg\} \Bigg) N_\lambda \vartheta_{\jmath j} \bigg\{ \frac{\Delta\bR_j} {\lambda} \bigg\}
\ket{ \phi_j \{\bR\}} \nonumber \\
\approx &  \Bigg(E^{\rm BO}_{j|\mathbf{0}} + \lambda^2 \mathfrak{E}_{\jmath j}^{\textrm{ho}}
\Bigg) N_\lambda \vartheta_{\jmath j} \bigg\{ \frac{\Delta\bR_j} {\lambda} \bigg\} \ket{ \phi_j \{\bR\}} \nonumber\\
\approx & \Bigg(E^{\rm BO}_{j|\mathbf{0}} + \lambda^2 \mathfrak{E}_{\jmath j}^{\textrm{ho}}
\Bigg) \ket{ \Psi_{i\lambda} \{ \bR \} },
\end{align}
where the second equality comes from Eq.~\eqref{eq:BO_Schreq} and is exact, 
going from the second line to the third one comes from neglecting terms of $\mathcal{O}(\Delta\bR^3)$,
while the sixth one comes from Eq.~\eqref{eq:Hvib_Schreq_2}, noting that third- and higher-order terms in $\lambda$
and $\Delta\bR$ can be neglected. 
Note that $\mathfrak{H}_{j}^{\textrm{ho}}$ also acts on $\ket{ \phi_j \{\bR\}}$, but this generates contributions of $\mathcal{O}( \lambda^3)$.

The normalization factor $N_\lambda$ is such that $N_\lambda\vartheta_{\jmath j}\{ \Delta\bR / \lambda \}$ is normalized in the space of nuclei configuration for all values of $\lambda$.
In this wavefunction, the limit $\lambda \rightarrow 0$ for $N_\lambda \vartheta_{\jmath j}(\Delta\bR/\lambda) \}$ is the square root of a normalized Dirac delta function for vanishing nuclei displacements.

\section{Eigenenergies of the auxiliary Hamiltonian up to fourth order}\label{app:auxH4}

In this section, we exhaustively evaluate the eigenenergies of the auxiliary Hamiltonian 
$\hat{\mathfrak{H}}_{j\lambda}[ \boldsymbol\rho_j ]$ up to $\mathcal{O}(\lambda^4)$ included, which provides the total energy up to $\mathcal{O}(\lambda^6)$. 
The harmonic oscillator part of the auxiliary Hamiltonian, $\hat{\mathfrak{H}}^{\textrm{ho}}_j[ \boldsymbol\rho_j ]$, is given by Eq.~\eqref{eq:Hvib_[0]} and the correction to it corresponds to Eq.~\eqref{eq:Hvib_[1]_lambda} up to $\mathcal{O}(\lambda^4)$ included.

In order to derive such perturbation expansion with respect to the small parameter $\lambda$, we use once more perturbation theory, considering a new small parameter $\Lambda$, with the following $\Lambda$-dependent Hamiltonian,
\begin{equation} \label{eq:Hchi_Lambda}
\hat{\mathfrak{H}}_{j\lambda} ([ \boldsymbol\rho_j ], \Lambda) = \hat{\mathfrak{H}}^{[0]}_j [ \boldsymbol\rho_j ] + \Lambda \hat{\mathfrak{H}}^{[1]}_{j\lambda} [ \boldsymbol\rho_j ],
\end{equation}
and generic perturbation series
\begin{equation}\label{eq:XL_serie}
X(\Lambda) \equiv X^{[0]} + \Lambda X^{[1]} + \Lambda^2 X^{[2]} + \mathcal{O}( \Lambda^3).
\end{equation}
If one chooses 
$\hat{\mathfrak{H}}^{[0]}_j [ \boldsymbol\rho_j ]=
 \hat{\mathfrak{H}}^{\textrm{ho}}_j [ \boldsymbol\rho_j ]$
and
$\hat{\mathfrak{H}}_{j\lambda}^{[1]}[ \boldsymbol\rho_j ]=
\Delta\hat{\mathfrak{H}}_{j\lambda}[ \boldsymbol\rho_j ]$, then
for $\Lambda=1$, one recovers the auxiliary Hamiltonian $\hat{\mathfrak{H}}_{j\lambda}[ \boldsymbol\rho_j ]$.
Explicitly,
\begin{align}\label{eq:Hvib_[1]_lambda4}
&\hat{\mathfrak{H}}_{j\lambda}^{[1]}[ \boldsymbol\rho_j ] = \lambda E^{\rm BO\{\mathbf{3}\}}_j \cdot  \boldsymbol\rho_j^{3} \nonumber\\
& + \lambda^2 \Big( \hat{T}_{j}^ \textrm{vec(1)\{{\bf{1}}\}} \cdot \boldsymbol\rho_j + E^{\rm BO\{\mathbf{4}\}}_j \cdot \boldsymbol\rho_j^{4}
+E^{\rm Born}_{j|\mathbf{0}} \Big) \nonumber\\
& + \lambda^3 \Big( E_{j}^{\textrm{Born} \{ \mathbf{1}\}} \cdot \boldsymbol\rho_j + \hat{T}_{j}^ \textrm{vec(1)\{{\bf{2}}\}} \cdot \boldsymbol\rho_j^{2} + E^{\rm BO\{\mathbf{5}\}}_j \cdot \boldsymbol\rho_j^{5}\Big) \nonumber\\
&+ \lambda^4 \Big( E_{j}^{\textrm{Born} \{ \mathbf{2}\}} \cdot \boldsymbol\rho_j^2 + 
\hat{T}_{j}^ \textrm{vec(1)\{{\bf{3}}\}} \cdot \boldsymbol\rho_j^{3} + \hat{T}_{j}^ \textrm{vec(2)\{{\bf{2}}\}} \cdot \boldsymbol\rho_j^{2} \nonumber \\
& \qquad + E^{\rm BO\{\mathbf{6}\}}_j \cdot  \boldsymbol\rho_j^{6} +
\hat{\Sigma}_{j\lambda} (E^{\rm BO}_{j|\mathbf{0}},[\boldsymbol\rho_j ])
\Big) + \mathcal{O}(\lambda^5).
\end{align}

For the energy, one has 
\begin{multline}\label{eq:EL_serie}
\mathcal{E}_{i \lambda}-E_{j|\textbf{0}}^{\textrm{BO}} =
 \lambda^2 \Big(\mathfrak{E}_{\jmath j}^{[0]} + \Lambda \mathfrak{E}_{\jmath j\lambda}^{[1]} + \Lambda^2 \mathfrak{E}_{\jmath j\lambda}^{[2]} \\
 + \Lambda^3 \mathfrak{E}_{\jmath j\lambda}^{[3]} + \Lambda^4 \mathfrak{E}_{\jmath j\lambda}^{[4]}+ \mathcal{O}( \Lambda^5)\Big),
\end{multline} 
where 
$\mathcal{E}_{i}^{(2)}=\mathfrak{E}_{\jmath j}^{[0]}=\mathfrak{E}_{\jmath j}^{\textrm{ho}}$ is independent of $\Lambda$. 

As mentioned in the main text, $\vartheta_{\jmath j}$ is even or odd with respect to a change of sign of $\boldsymbol\rho_j$ (also changing sign of the associated momentum). 
All terms of $\hat{\mathfrak{H}}^{[1]}_{j\lambda}$ that are even with respect to such a sign change have an even $\lambda$ exponent, while all terms that are odd with respect to such a sign change have an odd $\lambda$ exponent. 
Hence, the $\lambda$-expansion of the total energy turns out to have only even power terms.

At this stage, let us examine the contributions to $\mathcal{E}_{i \lambda}-E_{j|\textbf{0}}^{\textrm{BO}} $ up to order five in $\lambda$.
We already have the zero- and second-order contributions, see Eqs.~\eqref{eq:Hvib_Schreq}-\eqref{eq:Eilambda_Taylor_up2}, while the odd-$\lambda$ terms vanish.
So, we should compute $\mathfrak{E}_{\jmath j}^{(2)}$, which amounts to treating the effect of $\hat{\mathfrak{H}}_{j\lambda}^{[1]}$ up to second order in $\Lambda$ at most.
As usual in perturbation theory, the Hellmann-Feynman theorem allows one to compute the first-order contribution of $\hat{\mathfrak{H}}_{j\lambda}^{[1]}$
in $\Lambda$ by taking the expectation value of the Hamiltonian correction for the
unperturbed wavefunction,
\begin{equation}\label{eq:Eijl[1]}
\mathfrak{E}_{\jmath j\lambda}^{[1]} = \langle \vartheta_{\jmath j} | \hat{\mathfrak{H}}^{[1]}_{j\lambda}|\vartheta_{\jmath j} \rangle_{\boldsymbol\rho_j} = \lambda^2 \mathfrak{E}_{\jmath j}^{[1](2)} +\mathcal{O}( \lambda^4), 
\end{equation}
with
\begin{align}\label{eq:Eijl[1]2}
\mathfrak{E}_{\jmath j}^{[1](2)} =& 
\mathcal{E}_{i}^{(4), \text{vec}} + \langle \vartheta_{\jmath j} | E^{\rm BO\{\mathbf{4}\}}_j \cdot \boldsymbol\rho_j^{4} | \vartheta_{\jmath j} \rangle_{\boldsymbol\rho_j} \nonumber\\
& + \langle \vartheta_{\jmath j} | E^{\rm Born}_{j}[\textbf{0}] | \vartheta_{\jmath j} \rangle_{\boldsymbol\rho_j}.
\end{align}
where 
\begin{align}\label{eq:vec4term_SI}
   \mathcal{E}_{i}^{(4), \text{vec}} = \langle \vartheta_{\jmath j} | \hat{T}_{j}^ \textrm{vec(1)\{{\bf{1}}\}} \cdot \boldsymbol\rho_j  | \vartheta_{\jmath j} \rangle_{\boldsymbol\rho_j}.
\end{align}

The second-order contribution of $\hat{\mathfrak{H}}_{j\lambda}^{[1]}$ in $\Lambda$ can be obtained by a sum-over-state formulation, using a Sternheimer equation, or using a Green's function approach, as for the case of Eqs.~\eqref{eq:Green_perp}-\eqref{eq:self-energy}. 
We will use the latter.
The definition of the Green's function of the auxiliary harmonic oscillator Hamiltonian
$\hat{\mathfrak{H}}_j^{\textrm{ho}}[\boldsymbol\rho_j]$,
evaluated at 
$\mathfrak{E}_{\jmath j}^{\textrm{ho}}$, denoted
$\hat{\mathfrak{G}}_{\perp \jmath j}^{\textrm{ho}}$,
is given in the main text, see Eqs.~\eqref{eq:Green[0]_perp} and \eqref{eq:Green_ho_perp}.
Then,
\begin{equation}\label{eq:Eijl[2]}
\mathfrak{E}_{\jmath j\lambda}^{[2]} = \langle \vartheta_{\jmath j} | \hat{\mathfrak{H}}^{[1]}_{j\lambda}
\hat{\mathfrak{G}}_{\perp \jmath j}^{\textrm{ho}}
[\boldsymbol\rho_j] \hat{\mathfrak{H}}^{[1]}_{j\lambda} |\vartheta_{\jmath j}
\rangle_{\boldsymbol\rho_j}.
\end{equation}
Considering this expression up to second order in $\lambda$ gives
\begin{subequations}
\begin{align}\label{eq:Eijl[2]_lambda}
\mathfrak{E}_{\jmath j\lambda}^{[2]} =& \lambda^2 \mathfrak{E}_{\jmath j}^{[2](2)} + \mathcal{O}( \lambda^4), \\
\mathfrak{E}_{\jmath j}^{[2](2)} =& \langle \vartheta_{\jmath j} | \hat{\mathfrak{H}}^{[1](1)}_{j} 
\hat{\mathfrak{G}}_{\perp \jmath j} ^{\textrm{ho}}[\boldsymbol\rho_j] 
\hat{\mathfrak{H}}^{[1](1)}_{j} | \vartheta_{\jmath j}
\rangle_{\boldsymbol\rho_j}, \label{eq:Eijl[2]2} \\
\hat{\mathfrak{H}}^{[1](1)}_{j} =& E^{\rm BO\{\mathbf{3}\}}_j \cdot \boldsymbol\rho_j^{3}. \label{eq:H11}
\end{align}
\end{subequations}

With the explicit expressions Eqs.~\eqref{eq:Eijl[1]2} and \eqref{eq:Eijl[2]2}, the mass-scaling expansion of the total eigenenergies obtained from the different approximations presented in Sec.~\ref{sec:BOapproximations} can be established. 
They all share the same zero- and second-order terms, see Eq.~\eqref{eq:Eilambda_Taylor_up2}.
The fourth-order terms are the sum of contributions from Eqs.~\eqref{eq:Eijl[1]2} and \eqref{eq:Eijl[2]2}, with all terms being present for the optimal approximation,
\begin{equation}\label{eq:E^opt4_SI}
\mathcal{E}_{i}^{\rm(4)} = \mathfrak{E}_{\jmath j}^{[1](2)} + \mathfrak{E}_{\jmath j}^{[2](2)},
\end{equation}
which is the same as the expansion of the exact energy $\mathcal{E}_i$ up to order 5 included.
This gives Eqs.~\eqref{eq:E^opt4}-\eqref{eq:Eijl[1]2_Born}.

For the scalar approximation, the vector contribution and the Born contribution are neglected, so that only the BO anharmonic contribution is left,
\begin{multline}\label{eq:E^sca4_SI}
\mathcal{E}_{i}^{\rm (4), anh} = \langle \vartheta_{\jmath j} | E^{\rm BO\{\mathbf{4}\}}_j \cdot \boldsymbol\rho_j^{4} | \vartheta_{\jmath j} \rangle_{\boldsymbol\rho_j} \\ 
+ \langle \vartheta_{\jmath j} | E^{\rm BO\{\mathbf{3}\}}_j \cdot \boldsymbol\rho_j^{3}
\hat{\mathfrak{G}}_{\perp \jmath j} ^{\textrm{ho}}[\boldsymbol\rho_j]
E^{\rm BO\{\mathbf{3}\}}_j \cdot \boldsymbol\rho_j^{3} | \vartheta_{\jmath j}
\rangle_{\boldsymbol\rho_j}.
\end{multline} 
For the ``hyb1" approximation, the Born contribution in 
$\mathfrak{E}_{\jmath j}^{[1](2)}$ is neglected, while for the ``hyb2", the vector potential contribution in $\mathfrak{E}_{\jmath j}^{[1](2)}$ is neglected.

Now that we have established the mass scaling of the adiabatic approximations up to order five, we can carry on with the next order, obtaining  
the total energy up to sixth order in $\lambda$ (even up to seventh order, since it vanishes).
Considering the $\Lambda$ expansion of the total energy, 
Eq.~\eqref{eq:EL_serie}, one needs to go to the fourth order in $\Lambda$
\begin{equation}\label{eq:Ei6}
    \mathcal{E}_i^{(6)} = \mathfrak{E}_{\jmath j}^{[1](4)} + \mathfrak{E}_{\jmath j}^{[2](4)} + \mathfrak{E}_{\jmath j}^{[3](4)} + \mathfrak{E}_{\jmath j}^{[4](4)}. 
\end{equation}

The first term is obtained by injecting Eq.~\eqref{eq:Hvib_[1]_lambda4} into Eq.~\eqref{eq:Eijl[1]} and retaining the terms proportional to $\lambda^4$ which gives
\begin{equation}
     \mathfrak{E}_{\jmath j}^{[1](4)} = \mathcal{E}_i^{(6),\textrm{elph}} + \mathcal{E}_i^{(6),\textrm{vec1}} + \mathcal{E}_i^{(6),\textrm{anh1}}+ \mathcal{E}_i^{(6),\textrm{elm}}, 
\end{equation}
where 
\begin{align}
    \mathcal{E}_i^{(6),\textrm{elph}} \equiv & \langle \vartheta_{\jmath j} | E^{\rm Born \{ \mathbf{2} \}} \cdot \boldsymbol{\rho}_j^2 | \vartheta_{\jmath j} \rangle_{\boldsymbol{\rho}_j} \label{eq:6elph}\\
    \mathcal{E}_i^{(6),\textrm{vec1}} \equiv & \langle \vartheta_{\jmath j} | \hat{T}^{\rm vec(1) \{ \mathbf{3} \}} \cdot \boldsymbol{\rho}_j^3 + \hat{T}^{\rm vec(2) \{ \mathbf{2} \}} \cdot \boldsymbol{\rho}_j^2 | \vartheta_{\jmath j}  \rangle_{\boldsymbol{\rho}_j} \nonumber \\
    \mathcal{E}_i^{(6),\textrm{anh1}} \equiv &  \langle \vartheta_{\jmath j} | E^{\rm BO \{ \mathbf{6} \}} \cdot \boldsymbol{\rho}_j^6 | \vartheta_{\jmath j} \rangle_{\boldsymbol{\rho}_j}  \nonumber\\
    \mathcal{E}_i^{(6),\textrm{elm}} \equiv &  \langle \vartheta_{\jmath j} | \sum_{\substack{\kappa\alpha\\\kappa'\alpha'}} \frac{1}{2}\hat{\pi}_{\kappa\alpha}  \Delta m^{\rm elm,-1}_{\substack{\kappa\alpha\\\kappa'\alpha'}} \hat{\pi}_{\kappa'\alpha'} | \vartheta_{\jmath j} \rangle_{\boldsymbol{\rho}_j}.  \label{eq:6el}
\end{align}

The second term comes from the fourth order of Eq.~\eqref{eq:Eijl[2]} and gives:
\begin{equation}\label{eq:24}
     \mathfrak{E}_{\jmath j}^{[2](4)} = \mathcal{E}_i^{(6),\textrm{mix1}} + \mathcal{E}_i^{(6),\textrm{vec2}} + \mathcal{E}_i^{(6),\textrm{anh2}}, 
\end{equation}
with the first contribution being
\begin{align}
    \mathcal{E}_i^{(6),\textrm{mix1}} \equiv &  \Big(\langle \vartheta_{\jmath j} |  (E^{\rm BO \{ \mathbf{3} \}} \!\cdot\! \boldsymbol{\rho}_j^3) \hat{\mathfrak{G}}_{\perp \jmath j}^{\textrm{ho}}[\boldsymbol{\rho}_j] (E^{\rm Born \{ \mathbf{1} \}} \!\cdot\! \boldsymbol{\rho}_j) \nonumber  \\
    & \quad +  (E^{\rm BO \{ \mathbf{3} \}} \!\cdot\! \boldsymbol{\rho}_j^3) \hat{\mathfrak{G}}_{\perp \jmath j}^{\textrm{ho}}[\boldsymbol{\rho}_j] (\hat{T}^{\rm vec(1) \{ \mathbf{2} \}} \!\cdot\! \boldsymbol{\rho}_j^3) \nonumber    \\
    & \quad + (E^{\rm BO \{ \mathbf{4} \}} \!\cdot\! \boldsymbol{\rho}_j^4) \hat{\mathfrak{G}}_{\perp \jmath j}^{\textrm{ho}}[\boldsymbol{\rho}_j] (\hat{T}^{\rm vec(1)\{ \mathbf{1} \}} \!\cdot\! \boldsymbol{\rho}_j) | \vartheta_{\jmath j} \rangle_{\boldsymbol{\rho}_j} \nonumber \\
    &+  \textrm{(c.c.)} \Big),
\end{align}
where (c.c.) indicates the complex conjugate of the preceding term. 
The second and third contributions in Eq.~\eqref{eq:24} are 
\begin{align}
    \mathcal{E}_i^{(6),\textrm{vec2}} \equiv &
    \langle \vartheta_{\jmath j} | (\hat{T}^{\rm vec(1) \{ \mathbf{1} \}} \cdot \boldsymbol{\rho}_j) \hat{\mathfrak{G}}_{\perp \jmath j}^{\textrm{ho}}[\boldsymbol{\rho}_j] \nonumber \\
    & \,\,\,\,\, \times (\hat{T}^{\rm vec(1) \{ \mathbf{1} \}} \cdot \boldsymbol{\rho}_j) | \vartheta_{\jmath j} \rangle_{\boldsymbol{\rho}_j}, \\
    \mathcal{E}_i^{(6),\textrm{anh2}} \equiv & \,\,\, \Big(\langle \vartheta_{\jmath j} | (E^{\rm BO \{ \mathbf{3} \}} \cdot \boldsymbol{\rho}_j^3) \hat{\mathfrak{G}}_{\perp \jmath j}^{\textrm{ho}}[\boldsymbol{\rho}_j] \nonumber \\
    & \,\,\,\,\,\,\,\,\,\,\, \times (E^{\rm BO \{ \mathbf{5} \}} \cdot \boldsymbol{\rho}_j^5)| \vartheta_{\jmath j} \rangle_{\boldsymbol{\rho}_j} + (\textrm{c.c.})\Big) \nonumber\\
    & + \langle \vartheta_{\jmath j} |(E^{\rm BO \{ \mathbf{4} \}} \cdot \boldsymbol{\rho}_j^4) \hat{\mathfrak{G}}_{\perp \jmath j}^{\textrm{ho}}[\boldsymbol{\rho}_j]  \nonumber \\
    & \,\,\,\,\,\,\,\,\,\,\,\, \times (E^{\rm BO \{ \mathbf{4} \}} \cdot \boldsymbol{\rho}_j^4) | \vartheta_{\jmath j} \rangle_{\boldsymbol{\rho}_j} .
\end{align}
Obtaining the third term in Eq.~\eqref{eq:Ei6} can be done using the $2n+1$ theorem~\cite{Gonze1995}, 
with the first-order wavefunction being obtained thanks to the Green's function $\hat{\mathfrak{G}}_{\perp \jmath j}^{\textrm{ho}}[\boldsymbol{\rho}_j]$, which gives:
\begin{widetext}
\begin{align}
     \mathfrak{E}_{\jmath j}^{[3](4)} =& \mathcal{E}_i^{(6),\textrm{mix2}} + \mathcal{E}_i^{(6),\textrm{anh3}}, \\
    \mathcal{E}_i^{(6),\textrm{mix2}} \equiv &
    \, \, \,   \Big( \langle \vartheta_{\jmath j} | (\hat{T}^{\rm vec(1) \{ \mathbf{1} \}} \cdot \boldsymbol{\rho}_j)\hat{\mathfrak{G}}_{\perp \jmath j}^{\textrm{ho}}[\boldsymbol{\rho}_j]
    (E^{\rm BO \{ \mathbf{3} \}} \cdot \boldsymbol{\rho}_j^3) \hat{\mathfrak{G}}_{\perp \jmath j}^{\textrm{ho}}[\boldsymbol{\rho}_j] (E^{\rm BO \{ \mathbf{3} \}} \cdot \boldsymbol{\rho}_j^3) | \vartheta_{\jmath j}  \rangle_{\boldsymbol{\rho}_j}  + \textrm{(c.c.)}\Big)  \nonumber \\
    &+  \langle \vartheta_{\jmath j} | (E^{\rm BO \{ \mathbf{3} \}} \cdot \boldsymbol{\rho}_j^3)\hat{\mathfrak{G}}_{\perp \jmath j}^{\textrm{ho}}[\boldsymbol{\rho}_j] (\hat{T}^{\rm vec(1) \{ \mathbf{1} \}} \cdot \boldsymbol{\rho}_j) \hat{\mathfrak{G}}_{\perp \jmath j}^{\textrm{ho}}[\boldsymbol{\rho}_j] (E^{\rm BO \{ \mathbf{3} \}} \cdot \boldsymbol{\rho}_j^3) | \vartheta_{\jmath j} \rangle_{\boldsymbol{\rho}_j} ,\\
    \mathcal{E}_i^{(6),\textrm{anh3}} \equiv & \, \, \, \Big( \langle \vartheta_{\jmath j} | (E^{\rm BO \{ \mathbf{4} \}} \cdot \boldsymbol{\rho}_j^4)\hat{\mathfrak{G}}_{\perp \jmath j}^{\textrm{ho}}[\boldsymbol{\rho}_j] (E^{\rm BO \{ \mathbf{3} \}} \cdot \boldsymbol{\rho}_j^3) \hat{\mathfrak{G}}_{\perp \jmath j}^{\textrm{ho}}[\boldsymbol{\rho}_j] (E^{\rm BO \{ \mathbf{3} \}} \cdot \boldsymbol{\rho}_j^3) | \vartheta_{\jmath j}  \rangle_{\boldsymbol{\rho}_j}  + \textrm{(c.c.)}\Big)  \nonumber \\
    &+ \langle \vartheta_{\jmath j} | (E^{\rm BO \{ \mathbf{3} \}} \cdot \boldsymbol{\rho}_j^3)\hat{\mathfrak{G}}_{\perp \jmath j}^{\textrm{ho}}[\boldsymbol{\rho}_j] (E^{\rm BO \{ \mathbf{4} \}} \cdot \boldsymbol{\rho}_j^4) \hat{\mathfrak{G}}_{\perp \jmath j}^{\textrm{ho}}[\boldsymbol{\rho}_j] (E^{\rm BO \{ \mathbf{3} \}} \cdot \boldsymbol{\rho}_j^3) | \vartheta_{\jmath j} \rangle_{\boldsymbol{\rho}_j},
\end{align}
\end{widetext}
The final term in Eq.~\eqref{eq:Ei6} is obtained by constructing first the second-order wavefunction from the first-order one, thanks to the Green's function 
$\hat{\mathfrak{G}}_{\perp \jmath j}^{\textrm{ho}}[\boldsymbol{\rho}_j]$, then combining it with the first-order Hamiltonian and the first-order wavefunction. 
We do not give the details of such calculation, but only the final result,
\begin{align}
    \mathfrak{E}_{\jmath j}^{[4](4)} =& \mathcal{E}_i^{(6),\textrm{anh4}},  \\
    \mathcal{E}_i^{(6),\textrm{anh4}} \equiv & -\langle \vartheta_{\jmath j} | \Big((E^{\rm BO\{ \mathbf{3} \}} \cdot \boldsymbol{\rho}_j^3)   \hat{\mathfrak{G}}_{\perp \jmath j}^{\textrm{ho}}[\boldsymbol{\rho}_j]\Big)^3 \nonumber \\
    & \qquad \,\,\,\, \times (E^{\rm BO\{ \mathbf{3} \}} \cdot \boldsymbol{\rho}_j^3  ) | \vartheta_{\jmath j} \rangle_{\boldsymbol{\rho}_j} \nonumber\\
    & + \langle \vartheta_{\jmath j} | (E^{\rm BO\{ \mathbf{3} \}} \cdot \boldsymbol{\rho}_j^3)
    \Big( \hat{\mathfrak{G}}_{\perp \jmath j}^{\textrm{ho}}[\boldsymbol{\rho}_j]\Big) ^2 \nonumber \\
    & \qquad  \times  (E^{\rm BO\{ \mathbf{3} \}} \cdot \boldsymbol{\rho}_j^3)| \vartheta_{\jmath j} \rangle_{\boldsymbol{\rho}_j}  \nonumber \\
    & \times  \langle \vartheta_{\jmath j} | (E^{\rm BO\{ \mathbf{3} \}} \cdot \boldsymbol{\rho}_j^3)\hat{\mathfrak{G}}_{\perp \jmath j}^{\textrm{ho}}[\boldsymbol{\rho}_j] \nonumber \\
    & \qquad \times (E^{\rm BO\{ \mathbf{3} \}} \cdot \boldsymbol{\rho}_j^3) | \vartheta_{\jmath j} \rangle_{\boldsymbol{\rho}_j}.
\end{align}

We can further group the various terms with similar origin as:
\begin{align}
    \mathcal{E}_i^{(6),\textrm{vec}} \equiv& \quad \,\mathcal{E}_i^{(6),\textrm{vec1}} +  \mathcal{E}_i^{(6),\textrm{vec2}} \label{eq:6vec}\\
    \mathcal{E}_i^{(6),\textrm{mix}} \equiv& \quad \, \mathcal{E}_i^{(6),\textrm{mix1}} +  \mathcal{E}_i^{(6),\textrm{mix2}} \label{eq:6hyb} \\
    \mathcal{E}_i^{(6),\textrm{anh}} \equiv& \quad \,\mathcal{E}_i^{(6),\textrm{anh1}}  +  \mathcal{E}_i^{(6),\textrm{anh2}}  \nonumber \\
                              &+   \mathcal{E}_i^{(6),\textrm{anh3}}  + \mathcal{E}_i^{(6),\textrm{anh4}}. \label{eq:6anh}
\end{align}

\section{Normal mode coordinates}\label{sec:normal_modes}

In this section, our goal is to obtain the lowest-order electron-phonon contribution to the total energy, 
$\mathcal{E}^{(4),\textrm{elph}}_{i}$, in terms of derivatives of the BO electronic
eigenfunctions with respect to normal mode coordinates, instead of derivatives with respect to nuclei displacements, as in Eq.~\eqref{eq:Eijl[1]2_Born_2}. 

Normal mode coordinates refer to the eigenmodes of the phononic Hamiltonian with physical masses, in which the Hamiltonian becomes separable.
In the main text, the formulation of the BO perturbation expansion refers to the auxiliary harmonic oscillator Hamiltonian Eq.~\eqref{eq:H_E_vib_first}, in which relative nuclear masses are present instead of the physical ones. 
There is therefore only a simple scaling factor between these Hamiltonians.
In this appendix, we use the standard definition of the normal modes.
The normal modes are obtained by performing a Taylor expansion of the BO energy around the BO equilibrium configuration $\{\bR^0_j \}$. 
Using the notation of Eq.~\eqref{eq:XR_Taylor} to express Eq.~\eqref{eq:EBO_0R_lambda}, the BO energy is

\begin{multline}\label{eq:EBO_0R}
E^{\rm BO}_j\{\bR\} = E^{\rm BO}_{j|\mathbf{0}} + \frac{1}{2} \sum_{\substack{\kappa\alpha \\ \kappa'\alpha'}} \frac{\partial^2 E^{\rm BO}_j}{\partial R_{\kappa\alpha}\partial R_{\kappa'\alpha'}}_{|\mathbf{0}} \!\! \Delta R_{\kappa\alpha j}\Delta R_{\kappa'\alpha'j} \\
+ \mathcal{O}( \Delta \bR^3).
\end{multline} 
We then define the phononic Hamiltonian $\hat{H}_j^{\rm ph}$ as the nuclear kinetic operator plus the second-order BO Hamiltonian
$\hat{H}^{\rm BO}$: 
\begin{multline}\label{eq:Hvib}
\hat{H}_j^{\rm ph}\{\bR\} \equiv -\frac{1}{2}\sum_{\kappa\alpha}\frac{1}{M_{\kappa}}\frac{\partial^2}{\partial R_{\kappa\alpha}^2} \\
+ \frac{1}{2} \sum_{\substack{\kappa\alpha \\ \kappa'\alpha'}} \frac{\partial^2 \hat{H}_j^{\rm BO}}{\partial R_{\kappa\alpha}\partial R_{\kappa'\alpha'}}_{|\mathbf{0}}\Delta R_{\kappa\alpha j}\Delta R_{\kappa'\alpha' j},
\end{multline} 
so that
\begin{multline}\label{eq:HwithHvib}
\hat{H} \{ \bR \}  = \hat{H}^{\rm BO}_{j|\mathbf{0}} + \sum_{\kappa\alpha}  \frac{\partial \hat{H}^{\rm BO}_j}{\partial R_{\kappa\alpha}}_{|\mathbf{0}}  \Delta R_{\kappa\alpha j} \\
  + \hat{H}_j^{\rm ph}\{\bR\}  + \mathcal{O}(\Delta \bR^3).
\end{multline} 
Note that the first-order BO Hamiltonian is not included in $\hat{H}_j^{\rm ph}\{ \bR \}$ but in $\hat{H} \{ \bR \}$.
Indeed, the expectation value of the first-order BO Hamiltonian for a reference state at the minimum of the $j$-BO hypersurface vanishes due to the Hellman-Feynman theorem.   

Next, we describe the nuclear vibrations with multi-dimensional harmonic oscillators. 
Using the secular equation and following the notation from Ref.~\onlinecite{Ponce2014a},
\begin{align}\label{eq:secular}
\sum_{\kappa'\alpha'} C_{\substack{\kappa\alpha j \\ \kappa'\alpha'}} U_{\nu \kappa'\alpha' j} =& M_{\kappa} \omega_{\nu j}^2 U_{\nu\kappa\alpha j}, \\
C_{\substack{\kappa\alpha j \\ \kappa'\alpha'}} \equiv & \frac{\partial^2 E^{\rm BO}_j} {\partial R_{\kappa\alpha}\partial R_{\kappa'\alpha'}}_{|\mathbf{0}},
\end{align}
where $C$ are the interatomic force constants while $\omega_{\nu j}^2$ is the vibrational energy of mode $\nu$ from the $j$ BO hypersurface.
The eigenfunctions of Eq~\eqref{eq:secular} satisfy the completeness and orthonormalization conditions
\begin{align}\label{eq:compl}
\sum_\nu \sqrt{M_{\kappa}M_{\kappa'}} U_{\nu\kappa\alpha j}^* U_{\nu\kappa'\alpha' j} =& \delta_{\kappa\kappa'} \delta_{\alpha\alpha'}, \\
\sum_{\kappa\alpha} M_\kappa U_{\nu\kappa\alpha j}^* U_{\nu'\kappa\alpha j} =& \delta_{\nu\nu'}, \label{eq:ortho}
\end{align}
where $U_{\nu\kappa\alpha j}$ are the eigendisplacement vectors of the harmonic equation of motion.

The transformation to normal mode coordinates $\eta_{\nu j}$ is defined as follows
\begin{align}\label{eq:deltaR}
\Delta R_{\kappa\alpha j} =& \sum_{\nu} \eta_{\nu j} U_{\nu\kappa\alpha j}, \\
\eta_{\nu j} =&  \sum_{\kappa\alpha} M_{\kappa} U_{\nu\kappa\alpha j} \Delta R_{\kappa\alpha j}. 
\end{align} 

Using the relationships Eqs.~\eqref{eq:compl} and \eqref{eq:ortho},
\begin{align}
\sum_{\kappa\alpha} \frac{1}{M_\kappa}\frac{\partial^2}{\partial R_{\kappa\alpha}^2} =& \sum_{\kappa\alpha} \frac{1}{M_\kappa}\frac{\partial}{\partial R_{\kappa\alpha}} \sum_\nu \frac{\partial \eta_{\nu j}}{\partial R_{\kappa\alpha}} \frac{\partial}{\partial \eta_{\nu j}} \nonumber \\
=& \sum_{\substack{\nu\nu' \\ \kappa\alpha}} \frac{1}{M_\kappa}  M_\kappa U_{\nu \kappa\alpha j} M_\kappa U_{\nu'\kappa\alpha j} \frac{\partial^2}{\partial \eta_{\nu' j} \partial \eta_\nu} \nonumber \\
=& \sum_{\nu}\frac{\partial^2}{\partial\eta_{\nu j}^2},
\end{align}
and similarly using Eqs.~\eqref{eq:secular}, \eqref{eq:ortho}, and \eqref{eq:deltaR},
\begin{align}
\sum_{\kappa\alpha} C_{\substack{\kappa\alpha j \\ \kappa'\alpha'}} \Delta R_{\kappa\alpha j}\Delta R_{\kappa'\alpha' j} =&  \sum_{\substack{\nu\nu' \\ \kappa\alpha}} \eta_\nu U_{\nu \kappa\alpha j} M_\kappa \omega_{\nu j}^2 U_{\nu' \kappa\alpha j} \eta_{\nu' j} \nonumber \\
 =& \sum_{\nu} \eta^2_{\nu j} \omega_{\nu j}^2. \label{eq:secularE2}
\end{align}

The phononic Hamiltonian, Eq.~\eqref{eq:Hvib}, becomes
\begin{equation}\label{eq:phnu}
\hat{H}^{\rm ph}_j[ \bbeta_j ] = \sum_{\nu}
\Big( -\frac{1}{2}\frac{\partial^2}{\partial\eta_{\nu j}^2} + \frac{1}{2}\omega^2_{\nu j} \eta_{\nu j}^2\Big).
\end{equation}

The Hamiltonian $\hat{H}^{\rm ph}_j[\bbeta_j ]$ is separable. 
The eigen-equation
\begin{equation}\label{eq:evol}
\hat{H}^{\rm ph}_j[ \bbeta_j ] \chi_{\jmath j}
[ \bbeta_j ] =  E_{\jmath j}^{\rm ph} \chi_{\jmath j} [ \bbeta_j ],
\end{equation}
has the following ground-state solution, stemming from the treatment of the quantum 
harmonic oscillator:
\begin{align} \label{eq:theta}
\chi_{0 j} [ \bbeta_j ] =& \Pi_\nu  \Big( \frac{\omega_{\nu j}}{\pi} \Big)^{1/4}e^{-\frac{1}{2} \eta^2_{\nu j} \omega_{\nu j}}, \\
 E_{0 j}^{\rm ph} =& \sum_\nu \frac{1}{2} \omega_{\nu j}. \label{eq:Eph}
\end{align}

Similarly, the fourth-order electron-phonon contribution to the total energy $\lambda_0^4 \mathcal{E}^{(4),\textrm{elph}}_{i}$, Eqs.~\eqref{eq:elph4_main}, \eqref{eq:Born_lambda} and \eqref{eq:fourthorder}
becomes
\begin{equation}\label{eq:E4elphfinal}
 E^{(4),\textrm{elph}}_{i} = \frac{1}{2} \sum_\nu  \bigg\langle \frac{\partial \phi_j }{\partial \eta_{\nu j} }
 _{|\bO} \bigg| 
 \hat{\cal{P}}^{\textrm{BO}}_{\perp j}
 \bigg|\frac{\partial \phi_j }{\partial \eta_{\nu j}}_{|\bO} \bigg\rangle
\end{equation}
in normal mode coordinates.
This result can be deduced from Appendix VII of Ref.~\onlinecite{Born1954}, Eq.~(VII.24) at page 405.

\section{Software improvements and numerical details}\label{app:improvement}

\begin{table}[t]
  \caption{\label{tableAppD}
Eigenvalue renormalization due to electron-phonon interaction using the adiabatic AHC theory within the rigid-ion approximation at the valence band maximum $\Delta \varepsilon$ of diamond. 
The density is computed on a 8$\times$8$\times$8 $\bk$-point zone-center grid with only the $\mathbf{q}=\boldsymbol{\Gamma}$ point included and a fixed 5~meV smearing. 
We are testing the impact of \textbf{G}-vector sphere truncation, which is the \texttt{boxcutmin} input variable.
Exact densities are obtained with \texttt{boxcutmin} equal to 2 or larger.  
}
\begin{tabular}{ l r r r }
  \toprule\\   
  & \multicolumn{3}{c}{$\Delta \varepsilon_{\boldsymbol{\Gamma}}^{\rm AHC}$ (meV)} \\
 \texttt{boxcutmin}  & Fan & DW & Total \\
\hline
1.1 & -552.930 & 576.220 & 23.290 \\
1.5 & -552.923 & 576.236 & 23.313 \\
2   & -554.906 & 578.712 & 23.806 \\
3   & -553.023 & 576.360 & 23.337 \\
4   & -553.313 & 576.721 & 23.408 \\
10  & -552.795 & 576.078 & 23.283  \\
  \botrule
\end{tabular}  
\end{table}

At the start of this project, at the end of 2022, by comparing the adiabatic AHC renormalization between a primitive cell and a 2$\times$1$\times$1 supercell, computed using crystal symmetry to reduce the computational cost, we found that the \textsc{Abinit} software was giving inconsistent results. 
We traced back this bug to the reconstruction of the dynamical matrices using crystal symmetries where the dimensionless real-space and reciprocal-space primitive translation were used instead of their dimensional counterpart.
This bug was corrected in \textsc{Abinit} v9.11.6.
For information, this bug gave $\Delta \varepsilon^{\rm AHC}$= 161.10~meV instead of the correct 149.24~meV for the VBM of a diamond 2$\times$1$\times$1 supercell with 8 active space bands and a 100$\times$100$\times$100  $\bq$-point interpolation, reported in Table~\ref{tab:table3}.   

We also noted the presence of a second bug which concerns the Debye-Waller term in the case where the highest energy state was degenerate and when the interpolation of the potential was used. 
In that case the Sternheimer contribution was omitted. 
This bug has been fixed in \textsc{Abinit v9.10.5}.

Another issue was linked with a missing i$\eta$ in the denominator of the Debye-Waller term on the active space, see Eq.~\ref{eq:ahceq}.  
Such factor is important to exactly cancel the Fan term on the active space. 
This missing factor was only present when using the Fourier interpolation of the perturbed potential. 
This bug was corrected in \textsc{Abinit} v9.11.6.

All three bugs did not affect previous published results~\cite{Ponce2014,Antonius2014,Ponce2015,Miglio2020,BrousseauCouture2022,BrousseauCouture2023}, as they did not rely on the interpolation of the perturbed potential.

\section{Zero-point renormalization at $\mathbf{q}=\boldsymbol{\Gamma}$} \label{sec:zpr}

To assess the eigenenergy contribution to that second-order energy, we compute the zero-point renormalization (ZPR) for all $\mathbf{k}$-point of an homogeneous 8$\times$8$\times$8 $\bk$-grid using crystal symmetry, and report their values in Table~\ref{tab:table2}.
We find a close agreement between perturbation theory and finite difference (FD), with less than 1.5~meV difference. 
From its weighted summation, we find that the eigenenergy contribution to the second-order energy is -84.214~meV, which amounts to over one-third the total second-order contribution.
This value can be approximated via density functional perturbation theory (DFPT) and the rigid-ion approximation (RIA) which gives $-100.783$~meV, overestimating by 20\% the true value. 
As mentioned earlier, we expect the RIA to improve when summing over dense $\mathbf{q}$-points grids.

\begin{table}[ht]
  \caption{\label{tab:table2}
Comparison of the sum over all occupied bands (1-4) of the ZPR, between the finite difference (FD) and Allen-Heine-Cardona (AHC) theory,
for diamond with 40~Ha cutoff.
The contribution of each  \textbf{k}-point is listed, with the corresponding weight. 
8$\times$8$\times$8 $\bk$-grid, standard scalar-relativistic PBE pseudopotential from \textsc{PseudoDojo}. 
For the FD, Richardson extrapolation with Romberg order 4, starting from a $h=2$ displacement.
}
\begin{tabular}{ r c r r r r}
  \toprule\\
$\mathbf{k}$-point & weight & $\Delta \varepsilon_{1-4,\mathbf{k}}$ & $\Delta \varepsilon_{1-4,\mathbf{k}}^{\rm NRIA}$  & Diff. & $\Delta \varepsilon_{1-4,\mathbf{k}}^{\rm AHC} $   \\
 & & (meV) & (meV)  & (meV)  & (meV) \\
\hline
 0,     0,  0                            & $\frac{ 1}{512}$ &  139.369 &  25.441  &  113.928  &   113.952  \\
$\frac{ 1}{8},   0,  0 $                 & $\frac{ 8}{512}$ &  -76.081 &  21.885  &  -97.966  &   -98.186  \\
$\frac{ 1}{4},   0,  0 $                 & $\frac{ 8}{512}$ &  -78.347 &  19.149  &  -97.496  &   -97.762  \\
$\frac{ 3}{8},   0,  0 $                 & $\frac{ 8}{512}$ &  -39.933 &  18.013  &  -57.946  &   -58.217  \\
$\frac{ 1}{2},   0,  0 $                 & $\frac{ 4}{512}$ &  -24.155 &  17.710  &  -41.865  &   -42.136  \\
$\frac{ 1}{8}, \frac{1}{8},  0 $         & $\frac{ 6}{512}$ & -100.133 &  21.769  & -121.902  &  -121.681  \\
$\frac{ 1}{4}, \frac{1}{8},  0 $         & $\frac{24}{512}$ & -100.774 &  18.400  & -119.174  &  -120.007  \\
$\frac{ 3}{8}, \frac{1}{8},  0 $         & $\frac{24}{512}$ &  -78.439 &  16.929  &  -95.368  &   -96.023  \\
$\frac{ 1}{2}, \frac{1}{8},  0 $         & $\frac{24}{512}$ &  -64.532 &  16.769  &  -81.301  &   -81.435  \\
-$\frac{3}{8}, \frac{1}{8},  0 $         & $\frac{24}{512}$ &  -67.085 &  17.484  &  -84.569  &   -84.113  \\
-$\frac{1}{4}, \frac{1}{8},  0 $         & $\frac{24}{512}$ &  -83.279 &  19.005  & -102.284  &  -101.285  \\
-$\frac{1}{8}, \frac{1}{8},  0 $         & $\frac{12}{512}$ &  -93.054 &  21.372  & -114.426  &  -113.031  \\
$\frac{ 1}{4}, \frac{1}{4},  0 $         & $\frac{ 6}{512}$ &  -97.988 &  19.000  & -116.988  &  -116.658  \\
$\frac{ 3}{8}, \frac{1}{4},  0 $         & $\frac{24}{512}$ &  -88.777 &  16.678  & -105.455  &  -105.796  \\
$\frac{ 1}{2}, \frac{1}{4},  0 $         & $\frac{24}{512}$ &  -84.746 &  15.773  & -100.519  &  -100.471  \\
-$\frac{3}{8}, \frac{1}{4},  0 $         & $\frac{24}{512}$ &  -85.456 &  16.128  & -101.584  &  -100.957  \\
-$\frac{1}{4}, \frac{1}{4},  0 $         & $\frac{12}{512}$ &  -89.319 &  17.483  & -106.802  &  -105.630  \\
$\frac{ 3}{8}, \frac{3}{8},  0 $         & $\frac{ 6}{512}$ &  -78.257 &  17.372  &  -95.629  &   -95.258  \\
$\frac{ 1}{2}, \frac{3}{8},  0 $         & $\frac{24}{512}$ &  -80.215 &  15.980  &  -96.195  &   -95.960  \\
-$\frac{3}{8}, \frac{3}{8},  0 $         & $\frac{12}{512}$ &  -84.827 &  15.692  & -100.519  &   -99.836  \\
$\frac{ 1}{2}, \frac{1}{2},  0 $         & $\frac{ 3}{512}$ &  -72.174 &  16.750  &  -88.924  &   -88.543  \\
$\frac{ 3}{8}, \frac{1}{4}, \frac{1}{8}$ & $\frac{24}{512}$ &  -95.325 &  19.478  & -114.803  &  -112.791  \\
$\frac{ 1}{2}, \frac{1}{4}, \frac{1}{8}$ & $\frac{48}{512}$ &  -89.036 &  17.192  & -106.228  &  -104.996  \\
-$\frac{3}{8}, \frac{1}{4}, \frac{1}{8}$ & $\frac{24}{512}$ &  -85.732 &  16.339  & -102.071  &  -101.185  \\
$\frac{ 1}{2}, \frac{3}{8}, \frac{1}{8}$ & $\frac{24}{512}$ &  -88.837 &  16.676  & -105.513  &  -104.352  \\
-$\frac{3}{8}, \frac{3}{8}, \frac{1}{8}$ & $\frac{48}{512}$ &  -90.944 &  15.234  & -106.178  &  -105.401  \\
-$\frac{1}{4}, \frac{3}{8}, \frac{1}{8}$ & $\frac{24}{512}$ &  -92.176 &  15.392  & -107.568  &  -106.800  \\
-$\frac{3}{8}, \frac{1}{2}, \frac{1}{8}$ & $\frac{12}{512}$ &  -86.008 &  15.087  & -101.095  &  -100.799  \\
-$\frac{1}{4}, \frac{1}{2}, \frac{1}{4}$ & $\frac{ 6}{512}$ &  -94.059 &  13.946  & -108.005  &  -107.826  \\
 & &  & &   \\
 & & E$^{\rm Allen}$ & E$^{\rm Allen,NRIA}$  & Diff.  & E$^{\rm Allen,AHC}$\\
 \hline
\multicolumn{2}{c}{\textbf{k}-integrated} &  -84.214   &   17.086  &  -101.300  &  -100.783 \\
  \botrule
\end{tabular}
\end{table}

\section{Size consistency}\label{sec:size}

In this section, we examine the size consistency of $\Delta \varepsilon_{n\bk}^{\rm AHC}$, by comparing results obtained for a primitive cell and for a supercell.

We compute the adiabatic zero-point renormalization using the rigid-ion approximation for the Debye-Waller term~\cite{Ponce2014a} and a Fourier interpolation of the perturbed potential~\cite{Eiguren2008,Gonze2020}.
We evaluate the diagonal~\cite{Lihm2020} self-energy at the DFT eigenvalues (on-the-mass shell) and use a Sternheimer expression to replace the sum-over-state expression~\cite{Gonze2011,Ponce2015}.

\begin{table}[t]
  \caption{\label{tab:table3}
Band renormalization comparison using the adiabatic AHC theory within rigid-ion approximation at the valence band maximum (VBM)   $\Delta \varepsilon$ of diamond between the primitive cell and a 2$\times$1$\times$1 supercell. 
The density of the primitive cell is computed on a 4$\times$4$\times$4 $\bk$-point zone center grid while the perturbed potential and phonons are interpolated 
from a 4$\times$4$\times$4 $\bq$-point grid to a dense 100$\times$100$\times$100  $\bq$-point grid one using quadrupoles and a fixed 5~meV smearing.
The band renormalization is computed at the relaxed lattice parameter of 6.7035~Bohr at 0~K.
The computed dynamical quadrupole $Q$ using perturbation theory~\cite{Royo2019} is reported.
We show the results when the active space is composed of valence band only or with some empty states as well and also the impact of using crystal symmetry.
The Debye-Waller (DW) term has no active space. 
}
\begin{tabular}{ l r r r r}
  \toprule\\
  & \multicolumn{2}{c}{primitive} & \multicolumn{2}{c}{supercell} \\
  & sym. & no sym. & sym. & no sym. \\
\hline
nsym & 48 &  1 & 24 &  1 \\
nkpt & 8  & 64 &  8 & 32 \\
Q (eBohr) &   2.431104 &   2.431215 &   2.431114 &   2.431116 \\
\\
 & \multicolumn{4}{c}{4$\times$4$\times$4  $\bq$-grid (meV)} \\  
Active space  & \multicolumn{2}{c}{4 bands} & \multicolumn{2}{c}{8 bands} \\
\hline
$\Delta \varepsilon^{\rm AHC}$  &   114.47 &   114.38 &   114.07 &   114.69   \\
$\Delta \varepsilon^{\rm Fan}$  & -1549.54 & -1549.64 & -1548.93 & -1549.53 \\
$\Delta \varepsilon^{\rm DW}$   &  1664.01 &  1664.02 &  1663.00 &  1664.22  \\  
\\    
 & \multicolumn{4}{c}{ 100$\times$100$\times$100  $\bq$-interpolation (meV)} \\  
Active space  & \multicolumn{2}{c}{4 bands} & \multicolumn{2}{c}{8 bands} \\
\hline
$\Delta \varepsilon^{\rm AHC}$ &   149.73 &   150.05 &   149.24 &   149.77 \\
$\Delta \varepsilon^{\rm Fan}$ & -1577.59 & -1577.71 & -1577.02 & -1577.76 \\
$\Delta \varepsilon^{\rm DW}$  &  1727.33 &  1727.75 &  1726.27 &  1727.54 \\  
Active space  & \multicolumn{2}{c}{9 bands} & \multicolumn{2}{c}{18 bands} \\
\hline
$\Delta \varepsilon^{\rm AHC}$ &   150.82 &   151.20 &   150.36 &   150.84 \\
$\Delta \varepsilon^{\rm Fan}$ & -1578.99 & -1579.27 & -1578.39 & -1579.04 \\
$\Delta \varepsilon^{\rm DW}$  &  1729.81 &  1730.46 &  1728.75 &  1729.88 \\
\\
ecut \& $\mathbf{k}$ & \multicolumn{2}{c}{Fan (meV)} & DW (meV) & AHC (meV) \\
\cline{2-5}  
Ha & active & full & full & full \\
  & \multicolumn{4}{c}{symmetry} \\ 
\hline
40 \& 4$^3$ & 173.997 & -1577.592 & 1727.326 & 149.734 \\
60 \& 8$^3$ & 176.978 & -1582.746 & 1735.074 & 152.328 \\
\\ 
  & \multicolumn{4}{c}{no symmetry} \\ 
\hline 
40 \& 4$^3$ & 174.305 & -1577.707 & 1727.755 & 150.048 \\
60 \& 8$^3$ & 176.997 & -1582.777 & 1735.082 & 152.305 \\
  \botrule
\end{tabular}  
\end{table}

We show in Table~\ref{tab:table3} the comparison between the adiabatic zero-point renormalization $\Delta \varepsilon_{n\bk}^{\rm AHC}$ for the case of the valence band maximum of diamond at the zone-center using consistent electron $\bk$ and interpolated phonon momentum grids $\bq$ between a primitive cell and a  2$\times$1$\times$1 supercell.
The total energy is exactly the same in all cases. 
Importantly, in the adiabatic case, the results should be independent of the active space size. 
As shown in Table~\ref{tab:table3}, this is not exactly respected. This was traced back to the 
presence of the finite lifetime $i\eta$ in the active space, affecting the results by up to 1.3~meV on the ZPR. 
When doing these tests, we noted the presence of a bug for the Debye-Waller term in the case where the highest energy state was degenerate and the interpolation of the potential was used.  
The bug has been fixed in \textsc{Abinit v9.10.5}as detailed in Appendix~\ref{app:improvement}.

Next, we look at the impact of crystal symmetry in the primitive cell case. 
We find a small impact, lower than 0.4~meV, see Table~\ref{tab:table3}.
The agreement between the case where symmetries are used and the case where they are not used can be systematically improved with more converged momentum grids and increasing planewave energy cutoff. 
We then confirm the size consistency between the primitive cell and a 2$\times$1$\times$1 supercell when disabling all symmetries, with a difference lower than 0.4~meV.
Finally, we look at the supercell case using symmetry. 
In that case, there is a larger difference of about 10~meV, due to \textsc{Abinit} being unable to detect symmetries with transformation matrices not based on integers for the supercell geometry. 
We note that this problem would not be present in a isotropically generated supercell such as 2$\times$2$\times$2 the primitive cell.
Fixing this issue would require a profound change of the \textsc{Abinit} core routines and is left for further work. 

We finish by presenting in Table~\ref{tab:table3} the values of the ZPR without interpolation of the perturbed potential and conclude that interpolation is not the source of the problem, but in fact symmetry detection.  
We also remark that exclusively for Table~\ref{tab:table3}, the data were computed with \textsc{Abinit v9.10.5} and a PBE pseudopotential without NLCC from \textsc{PseudoDojo}~\cite{Setten2017}, without spin-orbit coupling. 

\bibliography{Bibliography}